\documentclass[manuscript,linenumbers]{aastex701}

\newcommand\betaapp{{\beta}_{app}}
\newcommand\vapp{{v}_{app}}
\newcommand\Ho{{H}_{0}}
\newcommand\Omegam{{\Omega}_{m}}
\newcommand\Omegal{{\Omega}_{\Lambda}}
\newcommand\masyr{{mas~${\rm yr}^{-1}$}}

\graphicspath{{./}{figures/}}

\usepackage{graphicx}
\begin{document}

\title{A Three-Decade VLBI Study of the Nucleus in the Lobe-Dominated Quasar 3C207}
\shorttitle{Three-Decade VLBI Study of 3C207}

\correspondingauthor{T.~A. Rector}
\email{tarector@alaska.edu}

\author{D.~H. Hough} 
\affiliation{Department of Physics \& Astronomy, Trinity University, 
San Antonio, TX  78212}
\email[]{lucia.a.hough@gmail.com}  

\author[0000-0002-7675-4752]{J.~P. Linick} 
\affiliation{Department of Physics \& Astronomy, Trinity University, 
San Antonio, TX  78212}
\email[]{jlinick@gmail.com}  

\author[0000-0003-4072-7768]{E.~L. Danielson} 
\affiliation{Department of Physics \& Astronomy, Trinity University, 
San Antonio, TX  78212}
\email[]{ibndaniel@gmail.com}  

\author{S.~M. Escobedo} 
\affiliation{Department of Physics \& Astronomy, Trinity University, 
San Antonio, TX  78212}
\email[]{stephen.escobedo@swri.org}  

\author[0009-0008-2373-9373]{H.~D. Ibaroudene} 
\affiliation{Department of Physics \& Astronomy, Trinity University, 
San Antonio, TX  78212}
\email[]{hakima.ibaroudene@swri.org}  

\author{B.~D. Sadler} 
\affiliation{Department of Physics \& Astronomy, Trinity University, 
San Antonio, TX  78212}
\email[]{benjamin.sadler@gmail.com}  

\author{N.~A. Polito} 
\affiliation{Department of Physics \& Astronomy, Trinity University, 
San Antonio, TX  78212}
\email[]{npolito@smhall.org}  

\author[0000-0003-0287-4126]{R.~C. Vermeulen}
\affiliation{Netherlands Foundation for Research in Astronomy, Postbus 2, 
7990 AA Dwingeloo, The Netherlands}
\email[]{rvermeulen@astron.nl}  

\author{C.~E. Aars}
\affiliation{Department of Physics, Collin College, Central Park Campus, 
2200 W. University Dr., McKinney, TX  75070}
\email[]{caars@collin.edu}  

\author{C.~L. Newton}
\affiliation{Department of Physics \& Astronomy, Texas Christian University, 
Box 298840, Ft. Worth, TX  76129}
\email[]{fakeemail1@google.com}  

\author[0000-0001-8164-653X]{T.~A. Rector}
\affiliation{Department of Physics and Astronomy, University of Alaska Anchorage, Anchorage, AK 99508, USA}
\email[show]{tarector@alaska.com}  






\begin{abstract}

We present results from very-long-baseline-interferometry (VLBI) 
observations of the nucleus in the lobe-dominated quasar 3C207.  These observations were completed at 8.4 or 
10.7~GHz (X-band) from 1981 to 2010, spanning 29 years. The nucleus of 3C207 is the strongest and 
most variable in the 3CR complete sample of LDQs, which is under study to test 
relativistic jet models over a wide range of jet orientation angles. Images 
have typical resolutions of $\sim$0.5-1.0 milliarcseconds (mas) and 
sensitivities of $\sim$0.1-0.2 mJy ${\rm beam}^{-1}$. The VLBI 
core region has flux density outbursts at mean intervals of $\sim$7~yr; two of 
these are multiple outbursts from a stationary ``true'' core that feeds a 
``swinging component'' $\sim$0.5~mas to the east. The position angle (PA) of 
the swinging component shows a long-term increase of $\sim$40\arcdeg , with a 
short-term reversal of $\sim$10\arcdeg . A one-sided, curved VLBI jet extends 
$\sim$25~mas eastward, with components spanning a PA range of 
$\sim$25\arcdeg . The jet components have average apparent transverse 
velocities $\vapp \approx10$c. One component shows apparent acceleration from 
7$c$ to 14$c$ at 2-3~mas from the true core, where the flow is redirected 
toward PA $\sim$90\arcdeg. Another component shows marginal evidence 
for apparent deceleration. Individual jet components expand until reaching 
the recollimation zone. Our results are consistent with a physical model in 
which 3C207 has quasi-periodic outbursts, jet precession by ballistic 
components on a conical surface with a small opening angle, and a 
recollimation process that modifies component motions and narrows the conical 
geometry on a scale of $\sim$100 pc.    

\end{abstract}







\section{Introduction} \label{sec:intro}

This work is part of a larger, long-term very-long-baseline-inteferometry (VLBI) 
survey to test models of relativistic jets in active galactic nuclei (AGN) and 
AGN unification \citep{1989AJ.....98.1208H}. In contrast to the various highly productive 
surveys of bright, compact AGN \citep[e.g.,][]{1978Natur.276..768R,2009AJ....137.3718L}, we have selected 
a complete sample of 25 lobe-dominated quasars (LDQs) from the revised 3CR 
catalog \citep{1983MNRAS.204..151L}. Unlike the bright, compact AGN, the LDQs have been 
selected by their extended, steep-spectrum radio emission on kpc scales, which minimizes 
orientation bias due to highly-beamed structures and permits tests of jet 
models over a wide range of orientations. To define the complete sample, 
\citet{1989AJ.....98.1208H} considered all quasars in the 3CR catalog, used a strict 
flux density cutoff for the extended emission (10~Jy at 178~MHz), and examined 
source structure and spectra to select all the ``double-lobed quasars'' (DLQ) 
with two steep-spectrum lobes straddling a nucleus coincident with the optical 
identification. We also used the definition of the quantity $R$ to be the 
ratio of arcsecond nucleus flux density to extended flux density at 5~GHz, following \citet{1982MNRAS.200.1067O}. \citet{2002AJ....123.1258H} then adopted the term 
``lobe-dominated quasar'' and defined an LDQ to have $R < 1$. All 25 DLQs were 
found to have $R < 1$, so they all qualify as LDQs. The 3CR LDQ 
nuclei have flux densities $S$ that span a vast range, from $\sim$1 
millijansky (mJy) to $\sim$1~Jy at frequencies $\nu$ from 5~GHz to 15~GHz. 

The key results from our previous VLBI studies include: (1) All 25 LDQ 
nuclei are detected; (2) 22 have one-sided jets (3 have faint cores only); 
(3) the inner jets typically have apparent bends of a few degrees; (4) in 15 
sources with multiple-epoch observations, the jet components have apparent 
transverse speeds ranging from stationary to at least $\sim$10$c$; (5) there are some 
indications that outer components can be faster than inner ones, but thus far 
only one component in one source -- 3C207 -- has been followed long enough to 
actually observe a positive acceleration; (6) four sources have 
low-polarization core/inner jet regions with significant rotation measure, two 
of which -- including 3C207 -- have long ($\sim$25-40~mas), moderately-polarized 
jets with longitudinal magnetic fields \citep[e.g.,][]{2002AJ....123.1258H,2007AAS...210.0211D,2008ASPC..386..274H,2010AAS...21543417P}. 


We report here on a nearly three-decade X-band VLBI study of the nucleus in the LDQ \object[3CRR]{3C207} (1Jy~0838+133), with redshift $z$ = 0.6805 
\citep{2026AJ....171..285D}, projected linear size of 98~kpc and J2000 coordinates of the nucleus: 
${08}^{\rm h} {40}^{\rm m} {47.587}^{\rm s}$, 
$13\arcdeg 12\arcmin 23.54\arcsec$ \citep{2002AJ....123.1258H}. 
\citet{2012ApJS..201...38T} estimate 3C207 to be powered by a supermassive black hole with mass log $M_{BH} = 8.92$ ($M_{\odot})$.  It has also been detected at $\gamma$-ray energies, which is unusual for FRII radio galaxies \citep[e.g., ][]{2024ApJ...976..120P} and- as discussed below- speaks to the beamed nature of this source.
The nucleus of 3C207 is the brightest (maximum $\sim$1.5~Jy), most variable, 
and best-studied in the 3CR LDQ complete sample. 

Observations were made 
at 10.7~GHz from 1981.76 to 1991.15, and at 8.4~GHz from 1991.16 to 2010.81. 
While our previous results over a shorter time range showed that jet speeds are 
faster further from the core in 3C207 \citep{2008ASPC..386..274H}, the picture over 29 years is 
more complex, with a swinging component at nearly constant radius next to 
the core, variable direction of component ejection, component acceleration and 
possible deceleration, and recollimation on the $\sim$100 pc scale.    

For context, the kiloparsec-scale VLA structure of 3C207 shows 
a prominent one-sided jet to the east \citep{2010ApJ...710..764K,2014ApJS..212...19F}, roughly aligned with the parsec-scale jet studied herein. The closest jet knot, at a 
distance $d = 0.78$\arcsec\ from the arcsecond nucleus, has position angle 
PA = 94\arcdeg; the two brightest knots also lie along this PA. A third jet 
knot, and also the terminal hot spot at $d = 6.4$\arcsec, lie along PA = 
91\arcdeg; and the overall mean jet axis is at PA = 93\arcdeg. The outer 
lobes are not collinear with the nucleus; the eastern hot spot - nucleus - 
western hot spot misalignment angle is 23\arcdeg. 

We assume a standard $\Lambda$-CDM cosmology with 
$\Ho$ = 70 km ${\rm s}^{-1}$ ${\rm Mpc}^{-1}$, $\Omegam$ = 0.3, and $\Omegal$ = 0.7. 


\section{Observations} \label{sec:obs}

The nucleus of the lobe-dominated quasar 3C207 was observed with VLBI 
arrays for 32 epochs between 1981.76 and 2010.81. The VLBI arrays included 
telescopes of the European VLBI Network, the US VLBI Network, the NASA Deep 
Space Network, and the National Radio Astronomy Observatory's (NRAO)\footnote{Through 2015, the VLBA was operated by the National Radio Astronomy 
Observatory as a facility of the National Science Foundation operated under 
cooperative agreement by Associated Universities, Inc.} Very Long Baseline 
Array (VLBA) and Very Large Array (VLA). The observing frequency was 8.4 or 
10.7~GHz (i.e., ``X-band"). The ({\it u,v}) coverage for each source ranged from 
a few ``pilot survey'' scans of several minutes' duration to imaging tracks 
with many scans over a $\sim$10-12 hour period. For example, in 1995.67, 
fourteen scans of duration 6-7 minutes spread over $\sim$10 hours were made. 
To look for any rapid variations, three sets of multiple-epoch observations 
were each conducted as a series of six closely-spaced epochs, from 2001.14 to 
2002.21, from 2002.96 to 2003.77, and from 2005.02 to 2005.86. We will refer 
to these as the 2001, 2003, and 2005 series, respectively. A journal of early (1981-1999) VLBI 
observations is given in Table~\ref{tbl:obs}.  All observations  after this time were completed with the full NRAO VLBA array\footnote{Nominally 10 antennas, although see the Table~\ref{tbl:obs} notes for exceptions.} at 8.4~GHz with a bandwidth of 32~MHz, 2-bit sampling, RCP polarization, and total integration times $\sim$25-40~min, with the exceptions of observations on 2009~Dec~2 and 2010~Oct~22, which used bandwidths of 64 and 128~MHz respectively.


\startlongtable
\begin{deluxetable}{ccccccc}
\tablecaption{Journal of VLBI Observations \label{tbl:obs}}
\tablecolumns{7}
\tablenum{1}
\tablewidth{0pt}
\tablehead{ 
\colhead{Date}     & \colhead{Stations\tablenotemark{a}} & 
\colhead{$\nu$ (GHz)\tablenotemark{b}}  & \colhead{$\Delta\nu$ (MHz)\tablenotemark{c}} & 
\colhead{${N}_{bits}$\tablenotemark{d}} & \colhead{Polarization\tablenotemark{e}} &
\colhead{$\tau$ (min)\tablenotemark{f}}
}
\colnumbers
\startdata
1981 Oct 4  &B K G F (O)        &10.7 &56 &1 &LCP & $\sim$25 \\
1983 Jun 19 &B K G F O          &10.7 &56 &1 &LCP & $\sim$30 \\
1984 Aug 5  &B K G F O          &10.7 &28 &1 &LCP & $\sim$125 \\
1988 Mar 2  &B (L) K (G) F O    &10.7 &56 &1 &LCP & $\sim$20 \\
1989 Apr 7  &B L (N) (K) G PT O &10.7 & 2 &1 &LCP & $\sim$185 \\
1991 Feb 24 &B L K G PT O       &10.7 &28 &1 &LCP & $\sim$250 \\
1991 Feb 27 &T B L K G Y PT KP O&8.4  &28 &1 &RCP & $\sim$325 \\
1995 Sep 2  &VLBA               &8.4  &32 &2 &RCP & $\sim$80 \\
1998 Jan 30 &VLBA               &8.4  &32 &2 &RCP & $\sim$120 \\
1999 Aug 22 &VLBA               &8.4  &32 &1 &RCP/LCP & $\sim$90 \\
\enddata
\tablenotetext{a}
{Station legend: 
T = Onsala Space Observatory, Onsala, Sweden (20 m);
B = Max-Planck-Institut f\"ur Radioastronomie, Effelsberg, Germany (100 m); 
L = Istituto di Radioastronomia, Medicina, Italy (32 m);
N = Istituto di Radioastronomia, Noto, Italy (32 m); 
K = Haystack Observatory, Westford, MA (37 m); 
G = National Radio Astronomy Observatory (NRAO), Green Bank, WV (43 m); 
F = George R. Agassiz Station, Fort Davis, TX (26 m);   
Y = NRAO Very Large Array, Socorro, NM (27 x 25 m); 
Y1 = NRAO Very Large Array, Socorro, NM (1 x 25 m);
O = Owens Valley Radio Observatory, Big Pine, CA  (40 m);  
VLBA = NRAO Very Long Baseline Array, all ten 25 m telescopes at 
Brewster, WA (BR), Fort Davis, TX (FD), Hancock, NH (HN), 
Kitt Peak, AZ (KP), Los Alamos, NM (LA), Mauna Kea, HI (MK), North 
Liberty, IA (NL), Owens Valley, CA (OV), Pie Town, NM (PT), and 
Saint Croix, Virgin Islands (SC). Y1 replaced PT at 2003.45.
Stations that did not provide (useful) data at pre-VLBA epochs are 
in parentheses; for VLBA epochs, these stations were OV at 2002.21, 
BR at 2002.96, NL at 2003.12, HN and NL at 2005.02, 
PT at 2005.22, KP at 2005.70, and NL at 2010.81.}
\tablenotetext{b}
{Observing frequency.}
\tablenotetext{c}
{Bandwidth.}
\tablenotetext{d}
{Number of bits per sample.}
\tablenotetext{e}
{Right and/or left circular polarization.}
\tablenotetext{f}
{Approximate total on-source integration time.}
\end{deluxetable}


The data from 1981 to 1988 and from 1991 were recorded with the Mark 
III/IIIA VLBI system and processed on the Mark III correlator at the MIT 
Haystack Observatory \citep{1983Sci...219...51R}. The 1989 data were recorded with the 
Mark II VLBI system and processed on the CIT-JPL correlator at the California 
Institute of Technology \citep{1973IEEEP..61.1242C}. The data from 1995 to 2010 were obtained 
with the NRAO VLBA; data from 1995 to 2009 were processed on the VLBA hardware 
correlator \citep{1994IEEEP..82..658N}, and the data from 2010  were processed on 
the VLBA-DiFX software correlator \citep{2007PASP..119..318D}, both in Socorro, NM.   



\begin{figure}[ht]
\plotone{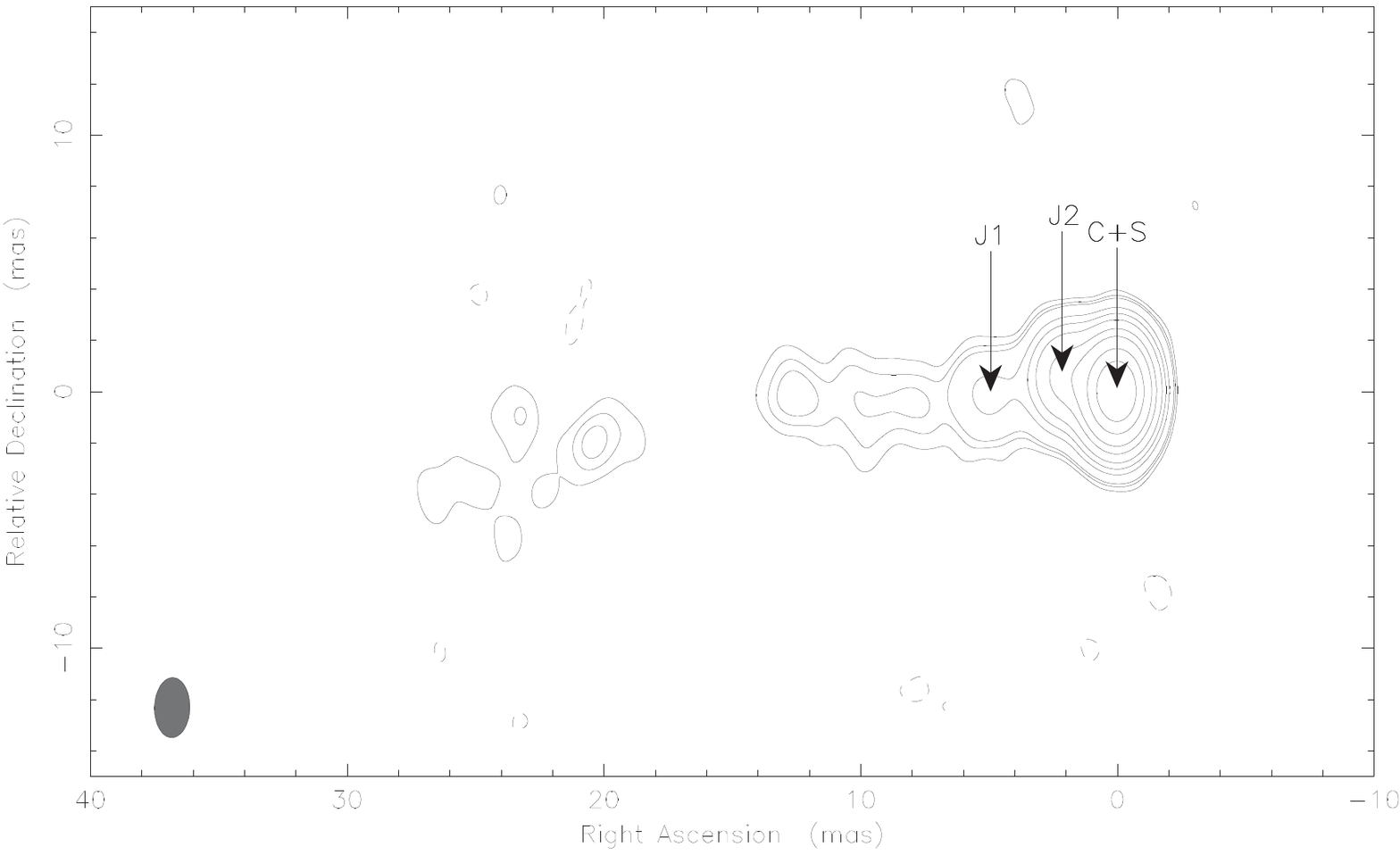}
\caption{8.4~GHz VLBA map of the overall 3C207 jet at epoch
1998.08. The ``true" core (C) and ``swinging" component (S), as well as two jet components (J1 and J2) are labeled. Additional jet components were identified in later epochs. The peak intensity is 0.615~Jy~{beam}$^{\rm -1}$, with contours of -0.05\%, 0.05\%, 0.1\%, 0.15\%, 0.5\%, 1\%, 2\%, 5\%, 10\%, 25\%, and 50\% of the peak. The CLEAN beam is shown shaded in the lower left-hand corner, with a FWHM of $2.34 \times\ 1.38$~mas at $-1.24\arcdeg$. 
\label{fig:1998_map}}
\end{figure}

Imaging and modelfitting were done using the Caltech DIFMAP software 
\citep{1994BAAS...26..987S,1997ASPC..125...77S}. The images all have rms noise levels comparable to or 
slightly above theoretical noise estimates; see Figure~\ref{fig:1998_map} for a representative example made at conventional 
resolution with natural weighting. Modelfits to all core and jet 
features for all epochs from 1981.76 to 2010.81 are given in 
Table~\ref{tbl:mod_sng}. In some cases, one or two elliptical Gaussian components 
in the core region would give slightly improved modelfits. However, these 
usually collapsed to ``line'' sources of zero width. So our Gaussian modelfits 
presented here use circular Gaussians almost exclusively, except in the one 
case -- epoch 2002.02 -- for which the core region is best modeled by a single 
elliptical Gaussian. Several tests comparing one set of modelfits using 
circular Gaussians with another set using elliptical Gaussians showed that all  
component parameters common to the two sets agreed within the uncertainties. 
The jet component labeled J2 usually displays a double substructure, J2a and 
J2b; in a few cases, this also occurs for J1 and J3. The modelfits for these 
double substructures are given in Table~\ref{tbl:mod_dbl}. 

Modelfit uncertainties were determined, during the process of seeking the 
highest-quality final image, from the range of parameters that 
gave acceptable fits for a variety of different imaging attempts that employed 
different schemes for starting models, initial sets of baselines, 
{\it a priori} amplitude corrections, windowing, phase and amplitude 
self-calibration, cleaning, etc. We excluded models for which the initial clean 
box on the dominant VLBI core component produced spurious symmetrization on 
scales below the synthesized beam width. 

We note that some of our results were previously reported in \citet{2002AJ....123.1258H} 
and \citet{2013EPJWC..6108009H}. These are updated and expanded upon here. 


\section{Results} \label{sec:res}
 
\subsection{Flux Density Variability in the Milliarcsecond Core Region} 
\label{subsec:var1}

\begin{figure}[ht]
\plotone{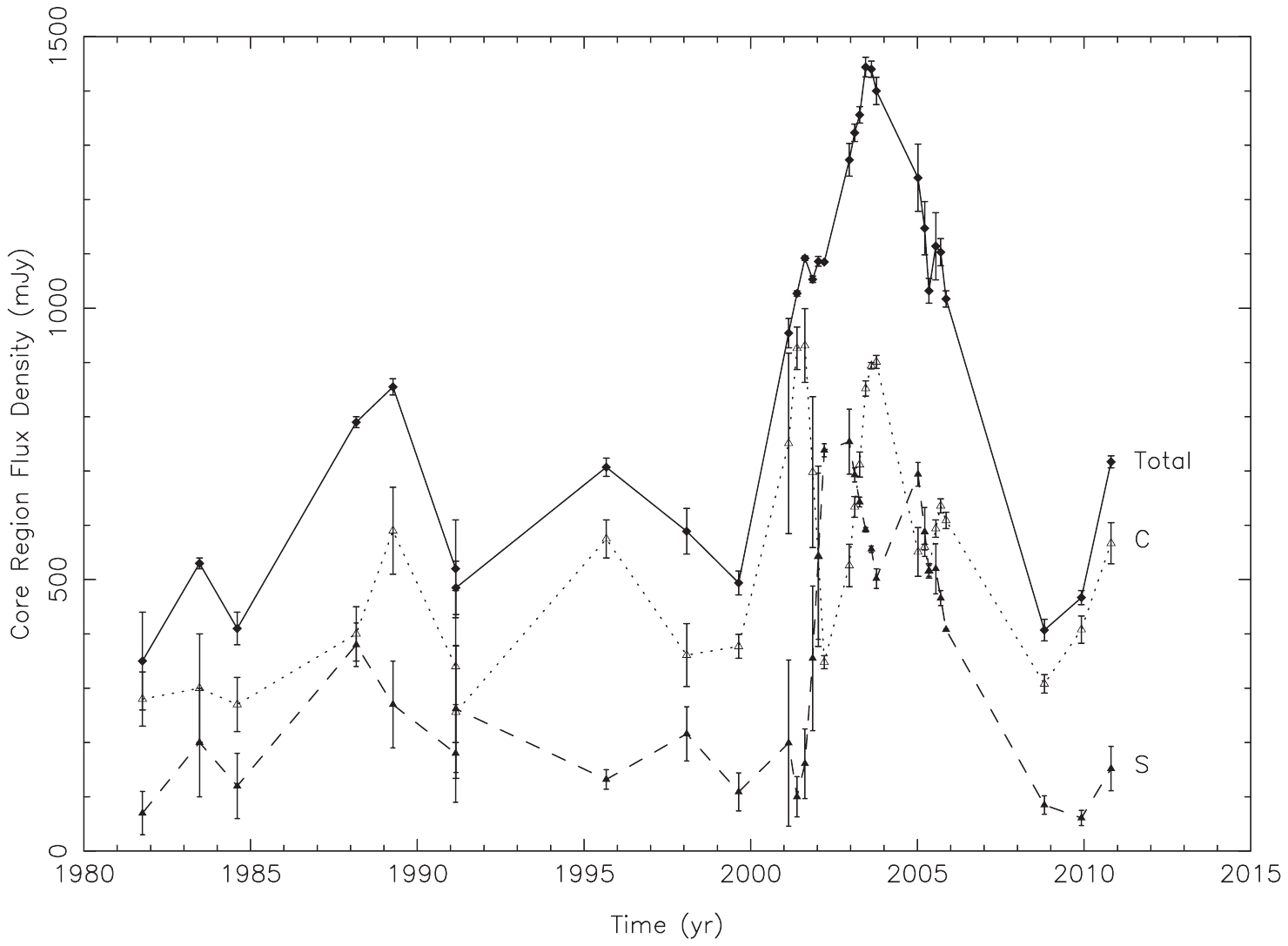}
\caption{X-band VLBI flux densities in the core region of 3C207 as a function of time. The flux is shown for the true core C, the swinging component S, and their combined (Total) emission.  The 1981.76--1991.15 data were measured at 10.7~GHz, and the 1991.16--2010.81 data at 8.4~GHz.
\label{fig:fluxtime}}
\end{figure}

As the best means of tracking variability, we plot the X-band VLBI flux 
densities of the milliarcsecond core region as a function of time as determined by modelfitting 
(Figure~\ref{fig:fluxtime}). Super-resolved imaging and modelfitting both show that the 
core region consists of a close double: the ``true'' core C (the base of the 
jet) to the west and a ``swinging'' component S to the east that appears to 
maintain a nearly constant radial distance from the core while undergoing 
significant PA changes; these are discussed further in Section~\ref{subsec:morph1}. The only 
exception is epoch 2002.03, for which a single elliptical Gaussian component 
was fitted to the core region. Five ``outbursts'' are evident, and while the 
time sampling is not sufficient to define them all precisely, we can 
characterize the outbursts in terms of their typical spacing, rise and decay 
times, and rise and decay amplitudes. We identify these outbursts by their 
highest observed flux densities at 1983.47, 1989.27, 1995.67, 2003.45, and the last epoch
2010.81, where the flux density was still rising.  This is suggestive of quasi-periodic events with a mean 
spacing of $6.9 \pm 0.9$~yr. The rise/decay times and amplitudes are simply defined 
 by adjacent local minima 
and maxima.   With only one observation prior to the 1983.47 maximum, we take 
the minimum rise time for this event to be 1.7~yr. Including the large 
pre-outburst gaps, the maximum rise times for the 1989.27 and 1995.67 events 
are 4.7 and 4.5~yr, respectively. The rise time for the 2003.45 event is 
3.8~yr, and the minimum rise time for the 2010.81 event is 2.0~yr. These data 
are consistent with a typical rise time of $\sim$3-4~yr. By similar methods, 
the typical decay time is also $\sim$3-4~yr. The rise and decay amplitudes 
range over a factor of $\sim$10, at $\sim$100~mJy for 1983.47, $\sim$400~mJy 
for 1989.27, $\sim$200~mJy for 1995.67, and $\sim$1000~mJy for 2003.45; the 
rise amplitude is at least $\sim$300~mJy for 2010.81. This indicates that the 
outbursts may alternate regularly between weaker and stronger amplitudes. 

The well-sampled 2003.45 outburst provides the best evidence for nearly 
symmetrical rise and decay profiles for the overall core region. However, it is 
very interesting that a closer examination of the separate profiles for C and S 
reveals that the relatively simple, smooth overall profile~masks the fact that 
C actually experienced {\it three} more closely-spaced ``sub-outbursts.'' We 
estimate the peaks of these events to be at 2001.5, 2003.7,  and 2005.7; the 
first two peaks are between each event's two highest and nearly equal flux 
densities (all $\sim$900~mJy), while the last peak is captured near the end of 
the 2005 series.  Thus there is an observed interval of $\sim$2.1~yr between 
the C sub-outbursts. Further, S mirrors the first two sub-outbursts at later 
times; a presumed third S event would have occurred after the 2005 series.  
During the first S event, its two highest (and nearly equal) flux densities 
($\sim$750~mJy) occur at 2002.21 and 2002.96, on either side of a 0.75~yr gap, 
so that a reasonable estimate of the time of the first S peak is in the gap at 
$\sim$2002.6. During the second S event, the highest flux density occurs after 
a gap of 1.25~yr at 2005.02. Assuming symmetry in the profile, the $\sim$500~mJy levels at 2003.77 (just before the gap) and $\sim$2005.45 yield an estimate 
of $\sim$2004.6 for the time of the second S peak. Therefore the two S peaks 
are $\sim$2.0~yr apart, and lag the first two C peaks by $\sim$1.0~yr. Although 
the peaks for S were not directly observed, the S profiles suggest that these 
peaks may have approached the $\sim$900~mJy level of  the C peaks. The maxima 
of S roughly overlap the minima of C, which leads to the flux density of S 
exceeding that of C when C is near its minima.

     The other outbursts, which have weaker amplitudes, coarser sampling, and 
sometimes larger flux density uncertainties, did not show any clear evidence 
for multiple sub-outbursts. The 1989.27 event shows the lone S peak preceding 
the lone C peak by $\sim$1~yr. However, it is possible that a double outburst
occurred, with the first peak for C in the $\sim$1985-1988 gap with no data. 
This is consistent with the two components that are 
associated with this event. The 1995.67 event shows the S peak following the C 
peak by $\sim$2~yr; again, the possibility of a double event cannot be ruled 
out, although it appears that only one component was ejected during this event 
(see Section~\ref{subsec:morph1}).


Due to the unprecedented strength of the 2003.45 outburst, we check here to 
see if 3C207 still satisfies the $R < 1$ criterion for a LDQ at the peak of the 
outburst (as shown in Figure~\ref{fig:fluxtime}). In our earlier works, we assumed a nuclear spectral 
index $\alpha_{\rm nuc} = 0.2$ for 3C207 ($S \propto {\nu}^{-\alpha}$). Here, 
we can try to calculate the spectral index from nearly simultaneous 8.4~GHz and 
10.7~GHz VLBI model flux densities in 1991 (see Table~\ref{tbl:mod_sng}), but the resulting 
value has a very large uncertainty ($\alpha_{\rm nuc} = -0.2 \pm 0.9$). We turn 
next to epoch 1999.64, for which we have simultaneous 5.0~GHz and 8.4~GHz VLBI 
model flux densities of $526\pm15$~mJy (unpublished) and $644\pm34$~mJy 
(Table~\ref{tbl:mod_sng}). These data yield $\alpha_{\rm nuc,5-8} = -0.39 \pm 0.12$. The peak 
8.4~GHz flux density occurs at epoch 2003.45, for which we have simultaneous 
8.4~GHz and 15.4~GHz VLBI model flux densities of $1493\pm19$~mJy (Table~\ref{tbl:mod_sng}) and 
$1836\pm16$~mJy (unpublished). These data give $\alpha_{\rm nuc,8-15} = 
-0.34 \pm 0.03$. We will assume an inverted spectrum using the average value 
$\alpha_{\rm nuc} = -0.37 \pm 0.06$, which leads to an estimate of the nuclear 
flux density that would be observed at 5~GHz $S_{\rm nuc,5} = 1232\pm40$~mJy. 
The extended flux density at 5~GHz is $S_{\rm ext,5} = 730\pm54$~mJy 
\citep{1974MNRAS.169..477P}, and the spectral index of the extended emission is 
$\alpha_{\rm ext} = 0.79\pm0.03$ \citep{1980MNRAS.190..903L}. To compute $R$, we multiply 
the observed nuclear-to-extended flux density ratio by $(1+z)^{\Delta\alpha}$, 
where $\Delta\alpha = \alpha_{\rm nuc} - \alpha_{\rm ext}$. The result is 
$R = 0.92 \pm 0.08$, and thus 3C207 still (barely!) satisfies the $R < 1$ LDQ requirement even 
at the peak of its strongest outburst. 


\subsection{Jet Morphology}  \label{subsec:morph1}

The VLBI structure of 3C207 consists of the dominant core region and a 
curved one-sided jet. Here we first briefly discuss the jet structure from 
pre-VLBA imaging. We then describe in detail the jet features as seen in VLBA 
images.  These maps were made in several ways: at conventional resolution, with ``super-resolution" to probe the core 
region, with light tapering to enhance features further from the core, and with heavy tapering to detect faint, distant jet components.   

\subsubsection{Pre-VLBA Images, 1991}  \label{subsubsec:jet1}

\begin{figure}[ht]
\plottwo{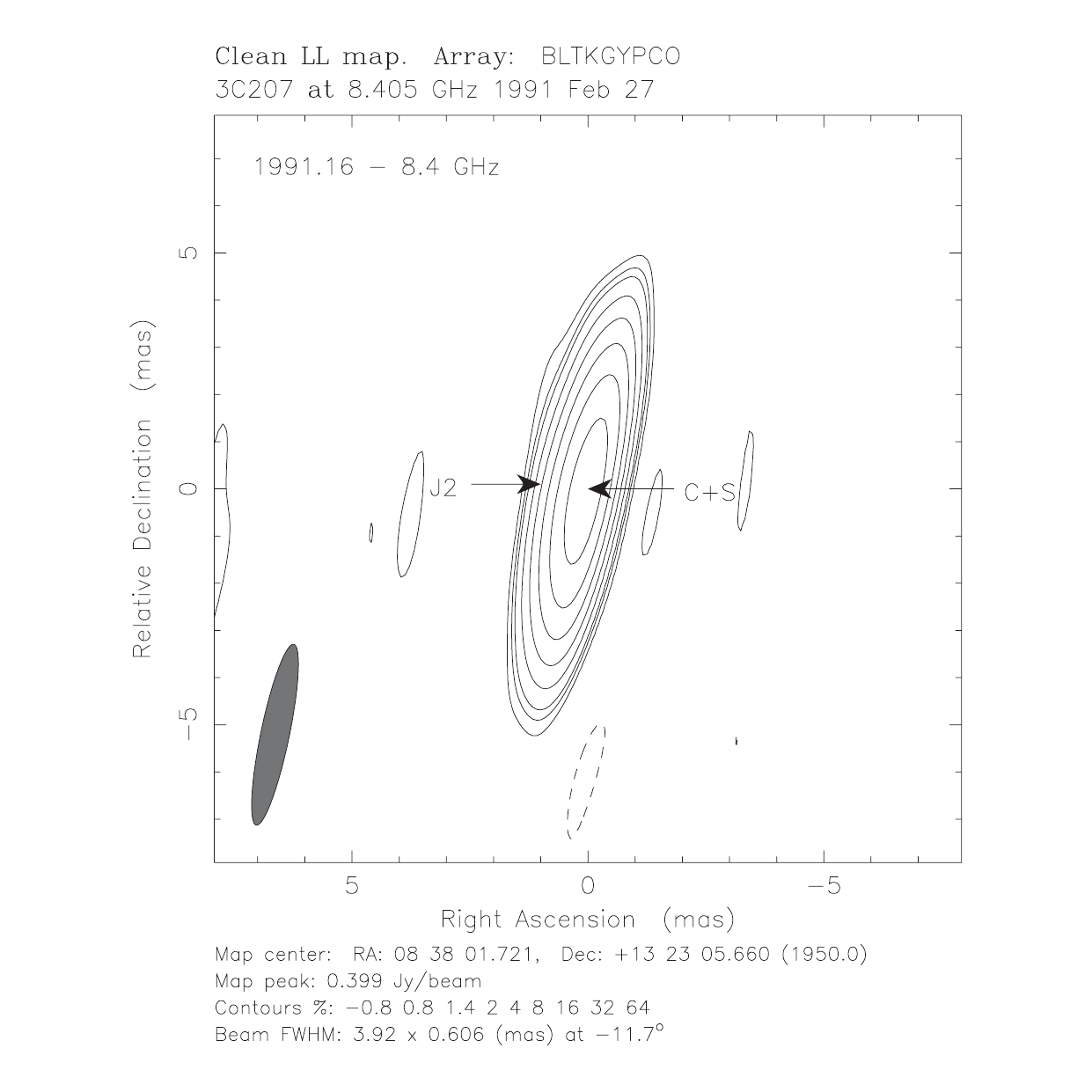}{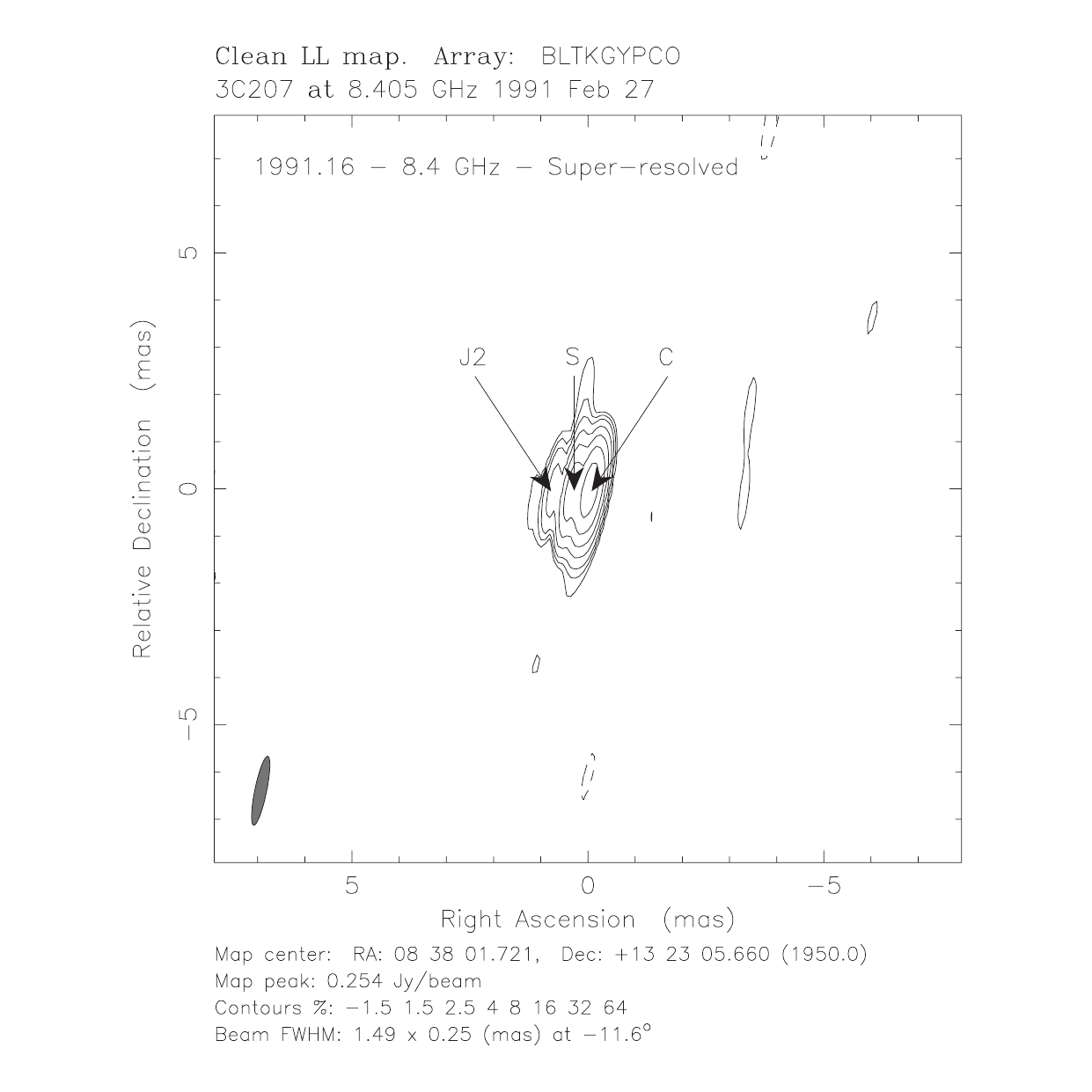}
\caption{8.4~GHz CLEAN VLBI maps of the nucleus of 3C207 at epoch 1991.16, shown at conventional resolution (left) and 2x super resolution (right). The core (C), swinging component (S), and jet component J2 are labeled. The peak intensity, contour levels, and beam parameters are given at the bottom of each image.
\label{fig:1991_map}}
\end{figure}

\begin{figure}[ht]
\plotone{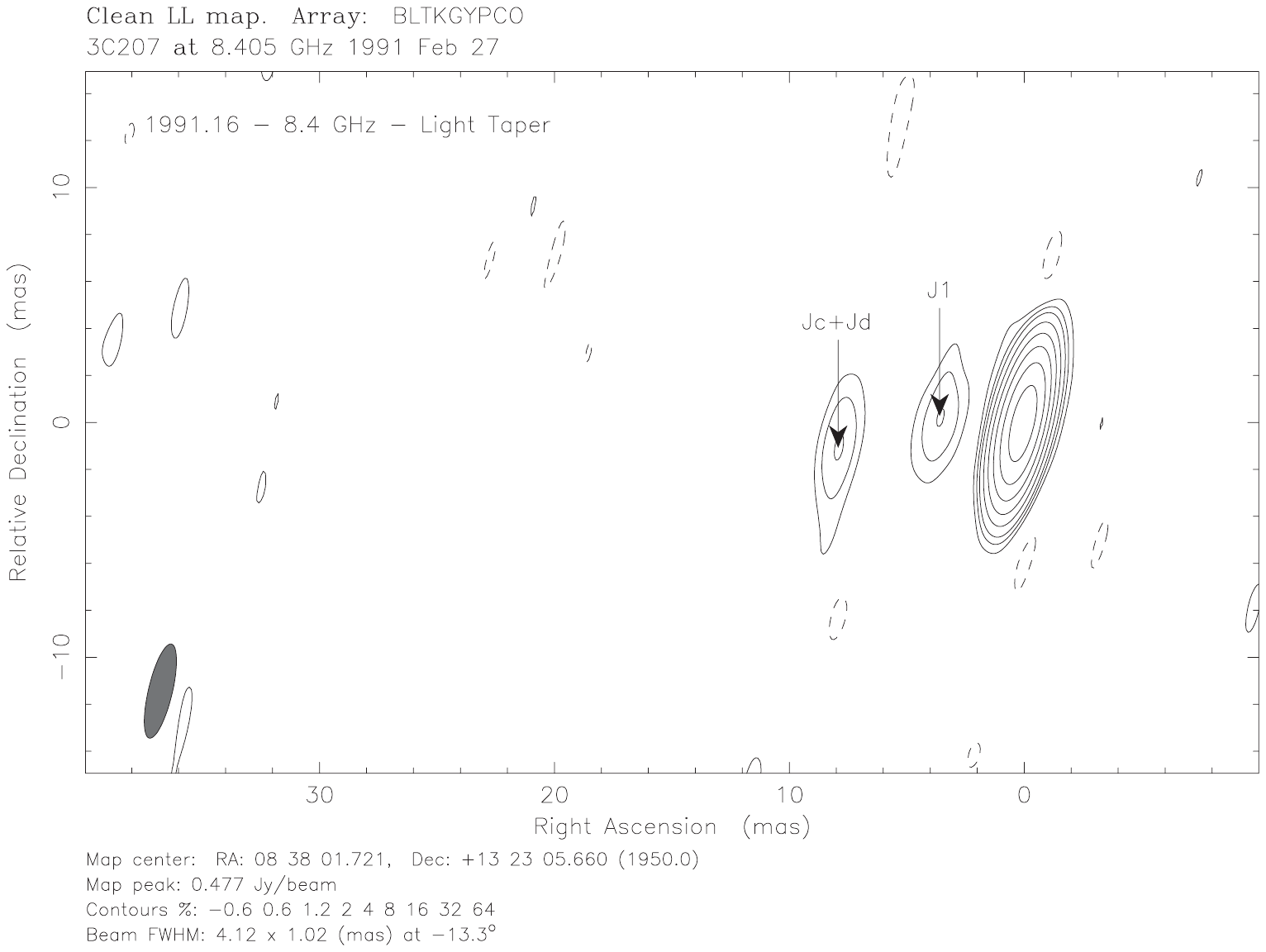}
\caption{``Wide-field" 8.4~GHz CLEAN VLBI map at epoch 1991.16, with a light taper to enhance the detection of faint components further from the nucleus. Jet components J1 as well as Jc+Jd are labeled. 
\label{fig:1991_map_lt}}
\end{figure}

In Figure~\ref{fig:1991_map} we display our earliest VLBI images of 3C207 that show 
both resolved structure in the core region and extended structure in the jet. 
These are pre-VLBA images from 1991.16. 8.4~GHz images are shown at conventional resolution with natural 
weighting and super-resolved by a factor of two with uniform weighting.  Observations at 10.7~GHz were also obtained on 1991.15.  At conventional resolution, the core 
region -- consisting of the true core C and the swinging component S -- has an
extension to the east we identify as jet component J2 or its sub-component J2a 
(as discussed in Section~\ref{subsubsec:jet2} below). Super-resolution reveals all three components 
distinctly. Modelfit parameters agree well for the two epochs, except for the 
PA of J2, which is obviously more southerly at epoch 1991.16. The 
discrepancy is at the $\sim2\sigma$ level. Since both the 10.7~GHz and 8.4~GHz images are pre-VLBA, 
they have inferior {\it (u,v)} coverage. The north-south/east-west beam ratios 
of 13:1 at 10.7~GHz and 6.5:1 at 8.4~GHz for natural weighting lead to 
poor constraints on position angles for east-west structure. With no reason to 
favor one epoch over the other, we include J2 modelfit results for both epochs 
1991.15 and 1991.16 in our analyses below, but also consider how their 
omission affects our results. In addition, Figure~\ref{fig:1991_map_lt} shows an 8.4~GHz image  
with a light Gaussian taper that downweights the {\it (u,v)} data by a factor 
of two at a radius of 125 mega-wavelengths (M$\lambda$). This reveals components that we identify 
as J1 and the combination Jc+Jd (also see Section~\ref{subsubsec:jet2}), located $\sim$4~mas and 
$\sim$8~mas east of the core, respectively. While there were hints of faint 
structure east of the core region at 10.7~GHz, no robust features were detected 
due to limited sensitivity.  

Two other 10.7~GHz pre-VLBA images (not presented here), at epochs 
1984.60 and 1989.27, show hints of faint jet features east of the core region, 
but there are no robust structures that appear consistently for different 
runs of DIFMAP.

 
\subsubsection{VLBA Images, 1995-2010}  \label{subsubsec:jet2}

\begin{figure}[ht]
\plotone{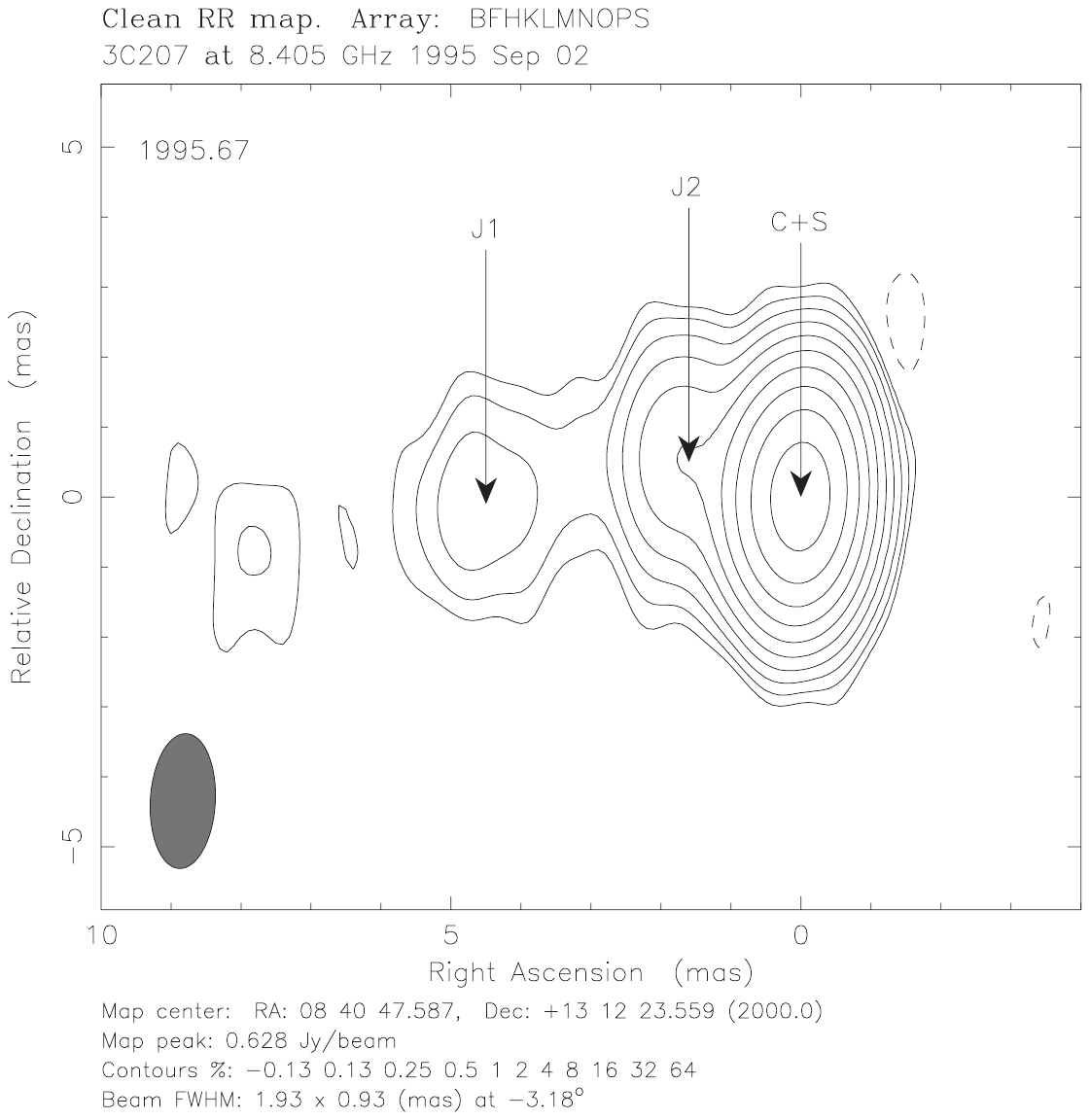}
\caption{Example 8.4~GHz VLBA map of the inner jet at epoch
1995.67. The true core (C), swinging component (S), and jet components (J1--J4) are labeled. 
The complete figure set (9 images) is available in the online journal. (*The other 8 images are also temporarily in the appendix.*)
\label{fig:vlba_maps}}
\end{figure}

Figure~\ref{fig:vlba_maps} shows representative VLBA images of 3C207 from 1995.67 to 2010.81, made at conventional 
resolution with natural weighting. 
Thanks to the stability and reliability of the VLBA, all of the images obtained during this time are of similar quality.  However, for epochs 2001.63, 
2003.27, and 2005.55, we used a light Gaussian taper that downweights the 
{\it (u,v)} data by a factor of two at a radius of 100~M$\lambda$ to 
enhance the fainter jet features. For each of the three series with six 
closely-spaced observations -- 2001, 2003, and 2005 -- we select one 
representative image. On these images, we label the true core C and the 
swinging component S, and also the multiple jet knots J1--J4 in 
order of their emergence from the core region and subsequent motion to the 
east. The jet always appears continuous through the last visible knot (either 
J1 or J2) at this resolution. By comparison of adjacent components, it is 
clear that each component follows its own independent path 
(see Section~\ref{subsec:traj}). From the very first epoch, 1995.67, it is obvious that 
the jet is curved, with a pronounced difference in position angles for J1 and 
J2. The eastward motions of J1, J2, and J3 are easily visible in this series 
of images, with J1 and J2 following different trajectories, and with J2 moving 
to larger position angles with increasing core distance. We note that 
\citet{2002AJ....123.1258H} did not consider the swinging component scenario, and in that work 
misidentified component S at $d\sim0.3$-0.5~mas as J1 at epochs 1981.76 and 
1983.47; at epochs 1988.17 and 1995.67, S is most likely blended with J2 and 
J3, respectively. The 1995.67 image clearly shows J1 and J2 well separated 
from the core region. These components were correctly identified in 
\citet{2002AJ....123.1258H}, who also correctly stated that J3 was created in an outburst 
around this time. By 2001.63, J3 is emerging beyond 1~mas from the peak of the 
core region; J4 does so by 2008.81. With a few exceptions, J1 was visible 
through epoch 2009.92. Note that data quality did not permit reliable imaging 
of jet components at 2005.02. 

\begin{figure}[ht]
\plotone{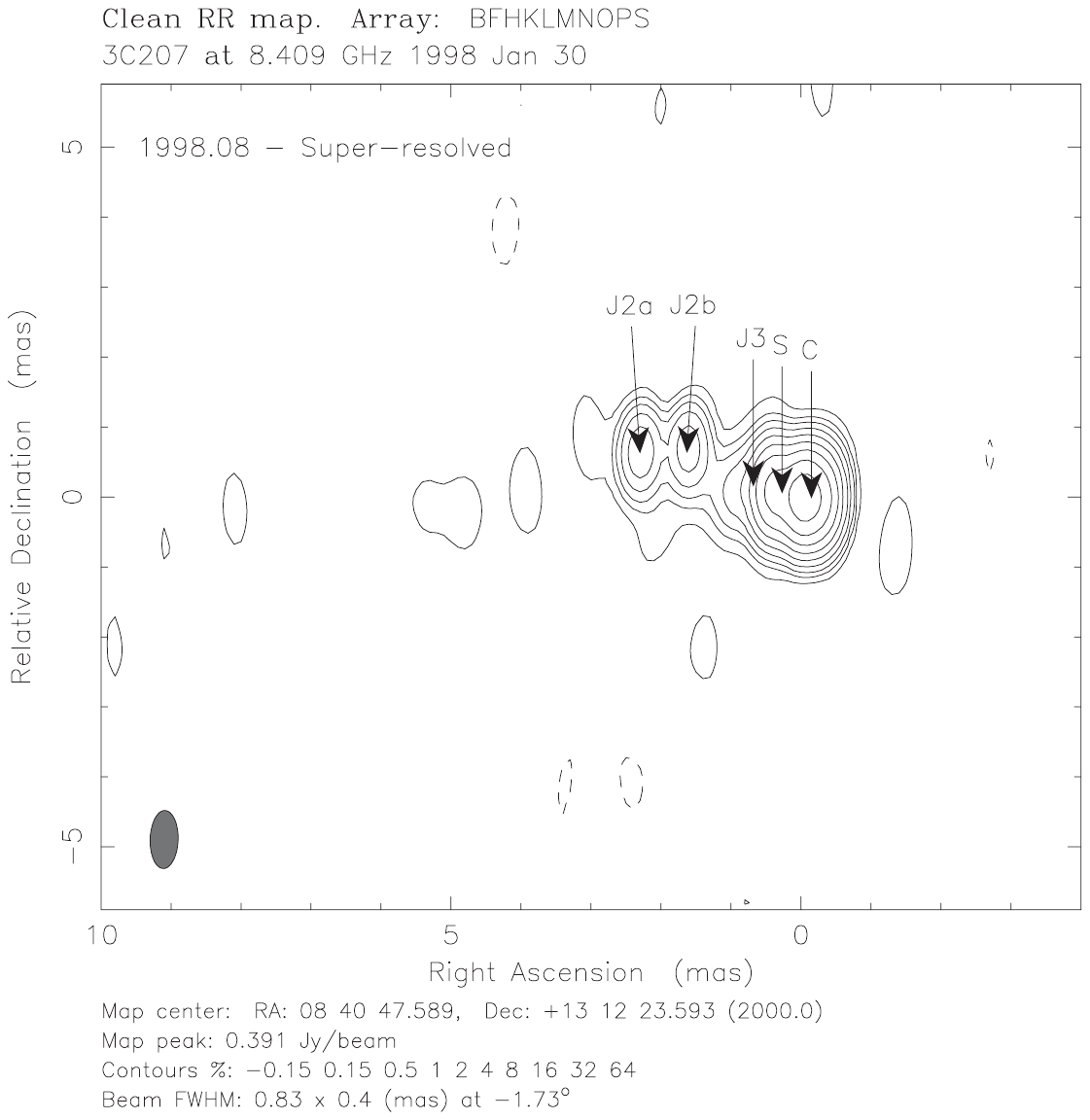}
\caption{Example 8.4~GHz super-resolved VLBA map of the inner  jet at epoch
1998.08. The true core (C), swinging component (S), and jet components (J1--J4) are labeled. 
The complete figure set (5 images) is available in the online journal. (*The other 4 images are also temporarily in the appendix.*)
\label{fig:vlba_maps_sr}}
\end{figure}

Figure~\ref{fig:vlba_maps_sr} shows selected VLBA images of 3C207 for five epochs between 1998.08 and 2009.92, made with uniform weighting and super-resolved by a factor of two. 
While overinterpretation of details must be avoided, 
areas of high signal-to-noise ratio on the images can give reliable 
indications of structure, and generally the clean components are a good 
guide for fitting Gaussian components to the data. The epochs shown in Figure~\ref{fig:vlba_maps_sr} were chosen to 
alternate between the core component C and the swinging component S 
having the highest flux density, and to show changes in PA for component S. 
At epoch 1998.08, C is clearly dominant, with S at PA = 
82\arcdeg. But by epoch 2002.96, the broad 2003 
outburst is underway, near the first peak of S around 2002.6, and its PA has increased 
to 86\arcdeg. Then at epoch 2003.77, we are at the second peak 
of C-- and S is still at PA = 86\arcdeg. Next at epoch 2005.22, we are between the second peak of S and the third peak of C, 
with S still slightly brighter and having backtracked to a decreased PA = 
81\arcdeg. At epoch 2009.92, C is once again dominant, while 
S has resumed its southerly motion and reached PA = 101\arcdeg. The 
individual jet knots J3 and J4 appear distinct throughout this sequence of 
images, and it is evident that they emerge at larger position angles than knot 
J2. The double nature of knot J2 -- with subcomponents J2a and J2b -- becomes 
apparent. 



\begin{figure}[ht]
\plotone{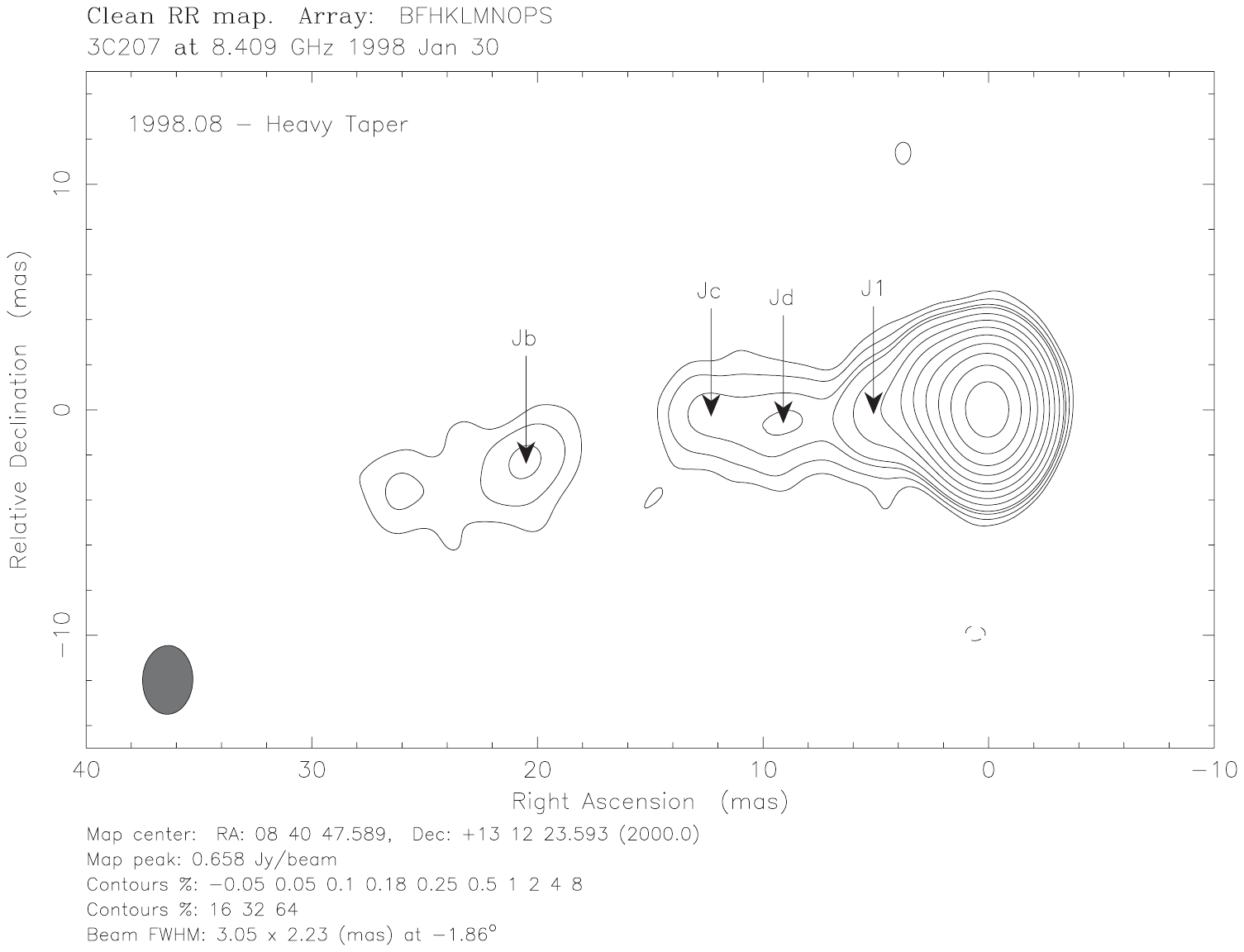}
\caption{Example 8.4~GHz VLBA map at epoch 1998.08, made with natural weighting and a heavy Gaussian taper. Jet component J1 is labeled, along with fainter features (Jb, Jc, and Jd). 
\label{fig:vlba_maps_ht}}
\end{figure}

To search for faint components in the extended jet beyond $d \sim7$-8 
mas, we used natural weighting and a heavy Gaussian taper that downweights the 
{\it (u,v)} data by a factor of ten at a radius of 100~M$\lambda$. 
Figure~\ref{fig:vlba_maps_ht} shows an example map.
Depending upon the details of the {\it (u,v)} coverage and self-calibration, 
some of the faint, outer jet features are more robust at some epochs, but not 
reliably detected in others. We require a feature to be present on all the 
clean images at a given epoch with a peak-to-rms ratio $\geq$ 5 for a reliable 
detection. 
In Tables~\ref{tbl:mod_sng} and \ref{tbl:mod_dbl} these fainter
components are labeled Ja, Jb, Jc, and Jd inward along the jet, in the 
presumed order of their ejection from the core region. We give these components 
special labeling because they appear only a few times, in cases where longer 
or wider-bandwidth observations were made, or when data calibration was 
uncomplicated.

    
It was also possible to detect component J1 at several more epochs through 
2009.92. Our best heavily-tapered image with natural weighting is for epoch 
1998.08. This shows faint jet features out to a distance $d \sim$26~mas from 
the milliarcsecond core. The inner jet structure lies along PA 
$\sim75$\arcdeg. The jet knots are located as follows: J1 at $d\sim5$~mas  
and PA $\sim92$\arcdeg, Jd at $d\sim9$~mas and PA $\sim93$\arcdeg, Jc at
$d\sim12$~mas and PA $\sim91$\arcdeg, Jb at $d\sim21$~mas and PA 
$\sim96$\arcdeg\ with distinct southerly displacement.
While most of our proper motion studies will be confined to the inner 
$\sim$7-8~mas of the jet, the outer components Jc and Jd do show outward motion 
on the heavily-tapered images, so we do include some motion analyses for these 
faint outer components below.          

As suggested by the images and discussed above, and confirmed by 
modelfitting at all epochs but one, the core region consists of a close double: 
the component C to the west that represents the true core, and a swinging 
component S to the east that appears to maintain a quasi-stationary radial 
position while undergoing significant position angle swings. The one exception 
is 2002.03, when the only acceptable fit was a single elliptical Gaussian for 
the core region. We do not consider the core region to be composed literally 
of two discrete components; rather, the double model is a representation of the 
core region structure that provides excellent fits at the $\sim$1~mas 
resolution of the VLBA at 8.4~GHz.


As mentioned above, the jet knot J2 also shows a double substructure 
(J2a and J2b) on most of super-resolved images, in particular when its peak is at 
$d\sim2.0-2.5$~mas. Modelfitting at these epochs confirms the need for two 
components to obtain good fits. At other epochs, superresolved images suggest
a double-- which can be successfully fitted to the data, though with only small
improvements to the fits.   While the images do not show it as clearly for jet 
knot J4, modelfitting requires three components (J4a, J4b, and J4c) for good 
fits when its peak is at $d\sim1.2$-1.5~mas. For J2, we suspect that it was 
spawned by a double event (Section~\ref{subsec:var1}), while for J4, we presume that its 
triple structure is due to the triple nature of the 2003.45 outburst. Since J2 and J4 most often appear as single components on images at conventional 
resolution, and can often be well modeled with single components, we conduct 
two separate analyses for them below: (1) we treat J2 and J4 as single 
components, either fitting a single component or locating them by the 
coordinates of their brightness centroids and summing their flux densities; and 
(2) we treat them as multiple components at those epochs when they had multiple 
substructure, and consider the impact on the results of the single-component 
analysis.


\begin{figure}[ht]
\plotone{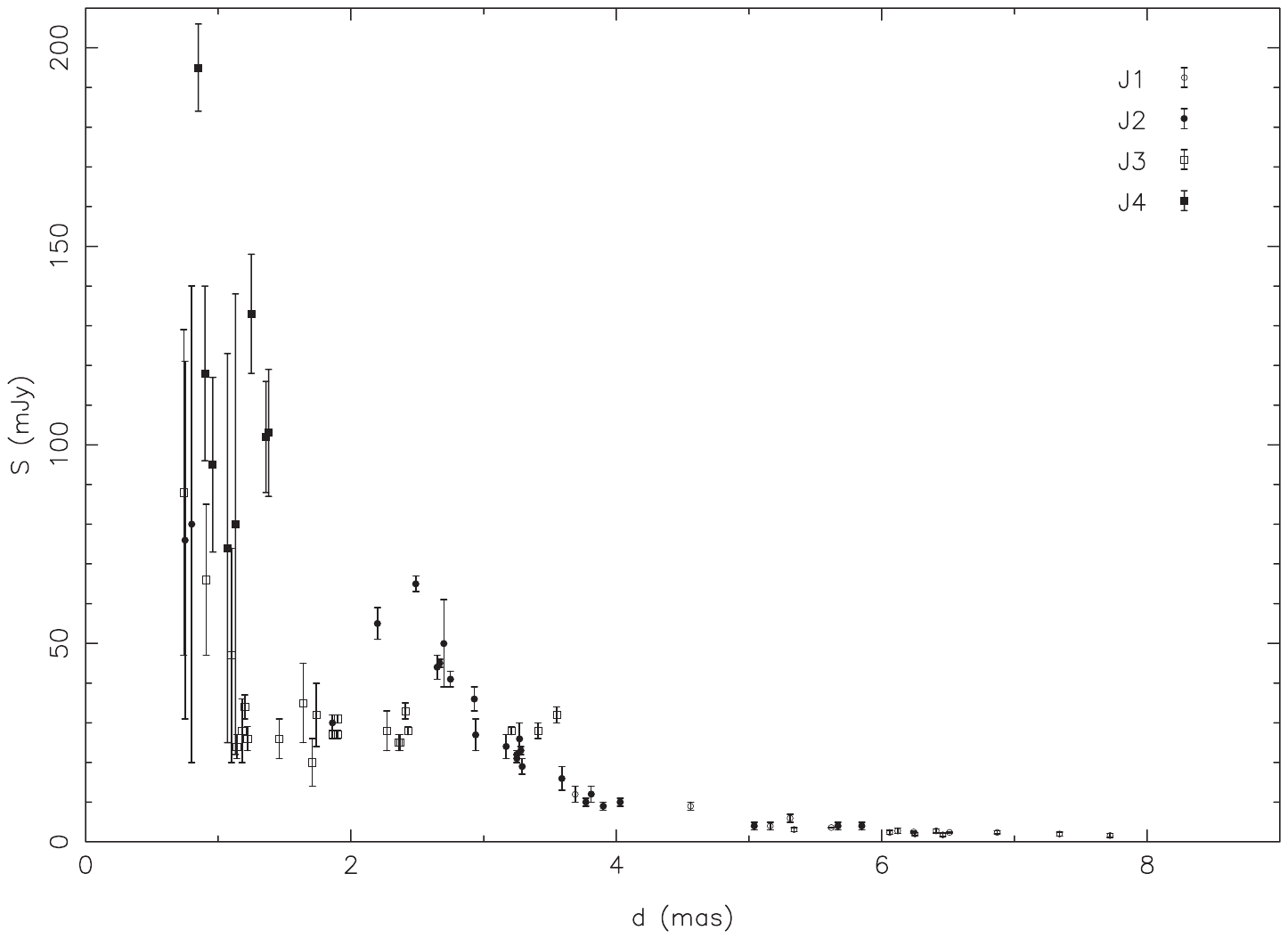}
\caption{The flux density of jet components J1--J4 as a function of distance $d$ from the core.  Error bars on $d$ have been omitted to permit a better view of the flux density trends; the uncertainties in $d$ are typically $<$ 0.1~mas. The large uncertainties for components within $\sim$1~mas are due to the proximity of the swinging component S.
\label{fig:jetflux}}
\end{figure}

\subsection{Jet Component Flux Densities}   \label{subsec:flux}

The flux densities of the four inner jet components (J1--J4) were analyzed as a function of distance from the core (Figure~\ref{fig:jetflux}).
As their large uncertainties indicate, the jet 
component flux densities can be difficult to disentangle from the swinging 
component S within $\sim$1~mas of the core component C, especially for the 
multiple-component substructures in J2 and J4. A related issue near the core 
region is uncertainty whether or not all the sub-components in J2 and 
J4 are detected; perhaps only the leading components J2a and J4a are seen.  While the overall trend is an unsurprising tendency for flux density to generally decline 
with core distance (and of course therefore with time), the individual behaviors 
are distinct in the range $d\sim1$-8~mas. The component J1 shows a gentle 
decrease between $\sim$3 and $\sim8$~mas, while J2 shows a rise followed by a 
decline between $\sim$2 and $\sim$6~mas that overlaps the J1 decline beyond 
$\sim$3~mas. From 1995.67 to 2002.96, the J2 sub-components J2a and J2b 
separately follow the rise and decline profile, but the profile for J2b is 
more sharply peaked; in addition, J2a and J2b alternate in having the stronger 
flux density.  The component J3 remains essentially constant at roughly half 
the peak flux density of J2 between $\sim$1 and $\sim$4~mas. The component J4 
starts out much stronger than both J2 and J3, but only a hint of its initial 
decline is visible between $\sim$1 and $\sim$2~mas. Both J1 and J2 display 
exponential decays with time, J1 over its full data range and J2 following a 
peak at $d\sim2.5$~mas (epoch 1999.64). The decay timescale is 8.1~yr for 
J1 and 3.5~yr for J2. As a check, for approximately-matched core distance 
ranges $d\sim3.5$-6~mas, these timescales are similar, with 7.4~yr for J1 
and 4.4~yr for J2. The limited data for the outer jet components do not reveal 
any particular trends.    


\begin{figure}[ht]
\plotone{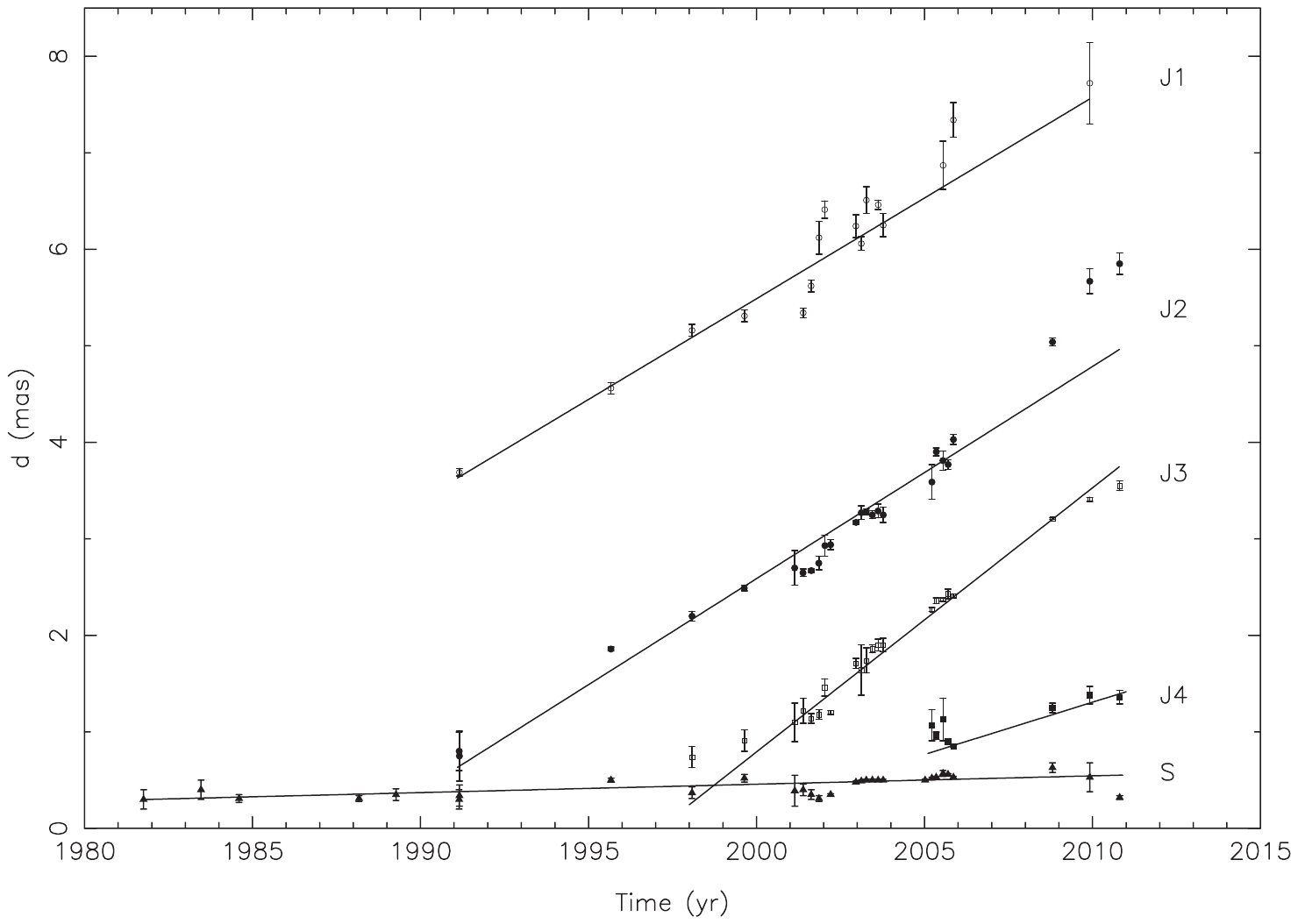}
\caption{Core distance $d$ as a function of time for the swinging component S and the jet components J1--J4. 
\label{fig:disttime}}
\end{figure}

\subsection{Component Radial Proper Motions} \label{subsec:motion1}

Component radial proper motions are of key interest. Because of the 
dominance, variability, and double structure of the core region -- which 
contains, e.g., 97\% of the flux density at the peak of the 2003.45 outburst 
-- the choice of reference points for measuring motions must be carefully 
weighed. We consider two obvious and well-defined ones: (1) the brightness 
peak on the DIFMAP image, which is the origin for cleaning and modelfitting 
(for some pre-VLBA model-fitting that instead placed the first component at 
the origin, the brightness peak was approximated as the brightness centroid of 
the core region double), and (2) the modelfit position of the western component 
in the core region that we assume is the ``true''  core C. We refer to modelfit 
component positions determined by the first method as ``unshifted'' positions, 
while those for the second method are ``shifted'' positions. Note that for 
2002.03, with the single elliptical Gaussian core region model, we used the 
average shift for the two surrounding epochs. Given that C is at the extreme 
western end of the structure, and that the 2003.45 triple ``sub-outbursts'' 
clearly occur in C (see Section~\ref{subsec:var1}), it is much more physically plausible to 
place C at the origin and use the ``shifted'' positions, and we do so 
throughout this paper, introducing the ``unshifted'' positions only when there 
are significant discrepancies. The two reference points are most likely to 
yield discrepant results for components closest to the core region, especially 
during periods when the relative brightnesses of the core C and the swinging 
component S are changing most rapidly. However, with a very small number of 
minor exceptions noted below, we find consistent quantitative results for the 
jet behaviors using both reference points.

The core distances as a function of time for the swinging component S and the four 
inner jet components J1--J4 are plotted, along with weighted 
linear fits to all the data for each component, in Figure~\ref{fig:disttime}. We now turn to 
detailed aspects of each jet component's motion. 


\subsubsection{Component S}   \label{subsubsec:S}

The weighted least-squares fit for component S
to the shifted radial positions -- those relative to the core C -- yields a  
radial proper motion ${\mu}_{\rm S} = 0.0087 \pm 0.0035$~\masyr, 
of marginal significance at the 2.5$\sigma$ level; an unweighted fit gives   
$0.0079 \pm 0.0018$~\masyr, a 4.4$\sigma$ result. An alternate 
explanation could be that the change of observing frequency from 10.7~GHz to 
8.4~GHz between epochs 1991.15 and 1991.16 introduced opacity effects that led 
to a larger core offset in the absence of motion. Separate fits for the 10.7 
GHz and 8.4~GHz data show no significant motion for either individual 
frequency. But the 8.4~GHz and 10.7~GHz positions for 1991.15 and 1991.16, 
respectively, agree well within the uncertainties at these epochs, so there is 
no significant evidence for an opacity shift. Thus we conclude that there is 
possibly a modest outward motion of S relative to C, at no more than 
$\sim$0.012~\masyr.  

For completeness, a weighted fit to the unshifted radial positions (not 
shown) is consistent with zero radial proper motion: 
${\mu}_{\rm S} = 0.0007 \pm 0.0042$~\masyr; the unweighted fit 
gives $0.0051 \pm 0.0023$~\masyr, at the 2.2$\sigma$ level. This 
lack of significant motion for the unshifted positions is not surprising, 
since the flux density and position of S help to determine the brightness 
centroid.


\subsubsection{Component J1}  \label{subsubsec:J1}

The component J1 was easily detected at three VLBA epochs with superior 
({\it u, v}) coverage: 1995.67, 1998.08, and 1999.64. Light tapering also 
picked up J1 at the pre-VLBA 1991.16 epoch, and at twelve VLBA epochs from 
2001.39 to 2009.92. It was not detected in the final epoch of observations in 2010.81, making it the only jet component to (apparently) fade from view during our survey.  A weighted linear fit gives 
${\mu}_{1} = 0.209\pm0.013$~\masyr. An unweighted fit gives
${\mu}_{1} = 0.228\pm0.015$~\masyr, which agrees with the 
weighted fit within the uncertainties. From the weighted fit, the epoch of 
zero separation is 1973.7 (or 1975.6 for the unweighted fit), so J1 must have 
been created during a flare prior to our first observations at 
epoch 1981.76. There is no evidence for significant departures from constant 
radial proper motion for J1.


\begin{figure}[ht]
\plotone{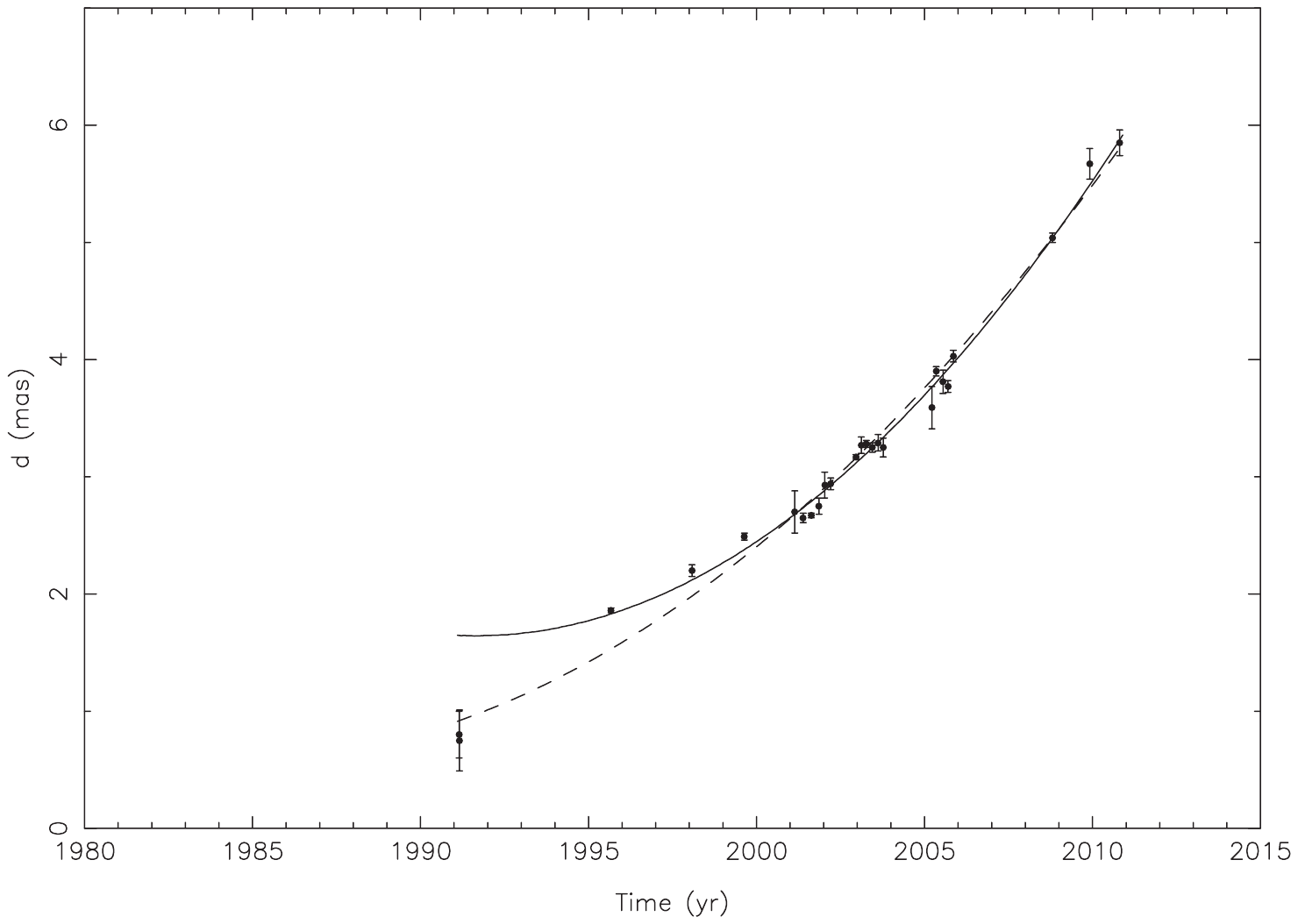}
\caption{Weighted (solid line) and unweighted (dashed line) quadractic fits for the jet component J2. The unweighted model is the most plausible as it traces back to the core region around the time of the origin of J2.
\label{fig:disttimeJ2}}
\end{figure}

\subsubsection{Component J2}  \label{subsubsec:J2}

The component J2 was continuously visible over a span of two decades, at 
10.7~GHz in 1991.15 and at 8.4~GHz from 1991.16 through the end of our survey in 2010.81. A weighted linear 
fit gives ${\mu}_{2} = 0.220\pm0.013$~\masyr; an unweighted 
fit yields ${\mu}_{2} = 0.244\pm0.012$~\masyr, which agrees 
with the weighted fit within the uncertainties. From the weighted fit, the 
epoch of zero separation is 1988.2 (or 1989.5 for the unweighted fit), so J2 
was created during the 1989.27 flux density event. 

It is evident from the weighted linear fit for J2 in Figure~\ref{fig:disttime} that 
there are significant deviations from constant radial proper motion. We 
consider the case of motion with constant acceleration. Figure~\ref{fig:disttimeJ2} shows weighted and unweighted quadratic fits.  The weighted quadratic 
fit to all the data gives an improved fit than the weighted linear fit, with a somewhat 
lower rms residual  (0.261 vs. 0.306~mas) and a 
vastly lower reduced chi-square (5.8 vs. 30.5). This fit corresponds to a 
constant radial angular acceleration 
${\alpha}_{2} = 0.0231$~mas ${\rm yr}^{-2}$, but it does not trace back to 
zero separation in the core region. An unweighted quadratic fit is much 
better, with a much lower rms residual 
(0.136 vs. 0.277~mas) and a far lower reduced chi-square (2.0 vs. 7.9), 
assigning each data point an uncertainty of 0.103~mas, which is the rms 
uncertainty for the actual data. We therefore adopt the unweighted fit as 
our best fit for the constant acceleration model. This fit corresponds to a 
radial proper motion linear in time 
${\mu}_{2} = -29.647 + 0.0014940t$~\masyr, where $t$ is in 
yr, and a constant radial angular acceleration 
${\alpha}_{2} = 0.0149$~mas ${\rm yr}^{-2}$. This suggests an acceleration 
from 0.10~\masyr\ at epoch 1991.0 to 0.38~\masyr\ at 
epoch 2010.0. This fit also nicely traces back to the core region around the time 
of the creation of J2, although with large deviations for the 1995-1999 data 
points. We note that the first two and the last three data points have large 
time gaps between the well-sampled data from 1995 to 2005, and visual 
inspection of the linear fit for J2 in Figure~\ref{fig:disttime} indicates that the last 
three data points have the largest deviations from the fit. As a further test, 
we omitted these five points and compared linear and quadratic fits to the 
1995-2005 data. Although the improvements are smaller, these quadratic fits do 
produce lower rms residuals and reduced chi-square values for both weighted 
and unweighted fits. However, these quadratic fits do not trace back well to 
the core region.  


\begin{figure}[ht]
\plotone{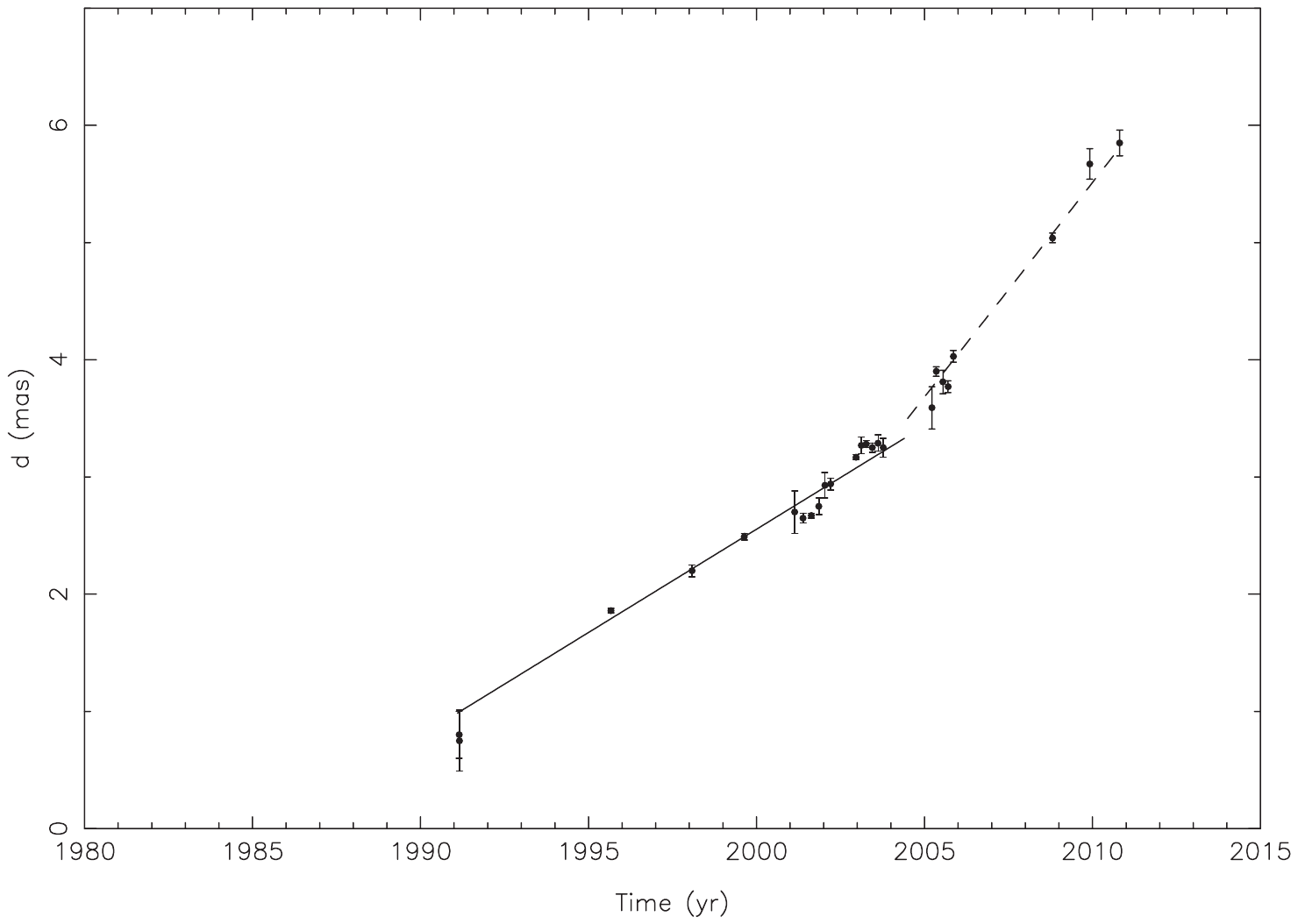}
\caption{Core distance $d$ as a function of time for jet component J2, with separate linear fits before (solid line) and after (dashed line) epoch 2004.5, yielding $\betaapp = 6.8 \pm 0.4$ beforehand and $\betaapp = 14.2 \pm 0.9$ afterwards. 
\label{fig:j2bend}}
\end{figure}

We also consider the case of a ``bend'' model for the radial proper motion 
of J2 using two separate linear fits with an abrupt change in slope. The 
``bend'' terminology applies not only to the appearance of a sharp bend on the 
proper motion graphs, but also to the fact that the jet appears to have a 
fairly sharp bend between $\sim$2 and $\sim$3~mas from the core (see Section~\ref{subsec:traj} below) at or near the point where the change in slope occurs. We selected 
four different times as potential slope break points: between 1999 and the 
2001 series at 2000.4, between the 2001 series and the 2003 series at 2002.6, 
between the 2003 series and the 2005 series at 2004.5, and between the 2005 
series and 2008 at 2007.3. We used three criteria to quantify the goodness of 
fit: the rms residual, the reduced chi-square, and the discontinuity between 
the two linear fits at the break times. Based on comparative rankings of these 
measures, the slope change at 2004.5 provides the best fit. We adopt as our 
best fit the weighted fit for 2004.5, which gives radial proper motions before 
and after the break of   
${\mu}_{2,before} = 0.176\pm0.010$~\masyr\ and 
${\mu}_{2,after} = 0.366\pm0.023$~\masyr\ (see Figure~\ref{fig:j2bend}); 
the rms residual is 0.121~mas, the reduced chi-square 
is 8.85, and the break discontinuity is 0.153~mas. Similar results are found 
for unweighted fits: ${\mu}_{2,before} = 0.195\pm0.006$~\masyr\
and ${\mu}_{2,after} = 0.396\pm0.019$~\masyr; the rms residual 
is 0.096~mas, the reduced chi-square is 10.51, and the break discontinuity is 
0.007~mas. From the other tests, the range of radial proper motions before the 
break is from 0.145~\masyr\ for 2002.6 to 
0.207~\masyr\ for 2007.3; the range after the break is from 
0.319~\masyr\ for 2002.6 to 0.431~\masyr\ for 2007.3. 
But of more importance is the range of changes in proper motion at the break, 
${\Delta \mu}_{2}$: the smallest value is $0.127\pm0.016$~\masyr\
for the unweighted 2000.4 data, while the largest value is 
$0.238\pm0.069$~\masyr\ for the weighted 2007.3 data. For our 
adopted best fit to the weighted data with a 2004.5 break, we find 
${\Delta \mu}_{2} = 0.190\pm0.025$~\masyr. Thus even for the 
smallest value of ${\Delta \mu}_{2}$, J2 experienced substantial positive 
apparent acceleration. 


Whether using the constant acceleration model or the two-slope bend 
model, it is clear that the component J2 experienced a model-dependent 
increase in radial proper motion of $\sim$0.2-0.3~\masyr\ between 
1991 and 2010.


We note here that J2 was modeled with two subcomponents, J2a and J2b, for
five epochs from 1998.08 through 2001.63, and the brightness centroid of these 
is plotted in Figures~\ref{fig:disttime} and \ref{fig:j2bend}. These data points do not seem at all 
unusual in relation to all the other epochs at which J2 was modeled with a 
single component. Within the relatively large uncertainties, fits to just these 
five epochs show that J2a and J2b separately move outward in 
lockstep with the J2 brightness centroid 
(${\mu}_{2a} = 0.13\pm0.05$~\masyr, 
${\mu}_{2b} = 0.09\pm0.05$~\masyr, and 
${\mu}_{2} = 0.130\pm0.017$~\masyr).

To determine if this result of positive apparent acceleration of J2 is 
robust, we consider some possible complicating issues. When it was first 
imaged with pre-VLBA arrays at epoch 1991.15, J2 was within $\sim$1~mas of the 
core component C. Thus it was not well distinguished from the core region, and 
it is also not clear if both J2a and J2b, or only J2a alone, had emerged by 
that time. For the last three epochs during 2008-2010, a component is clearly  
detected that we identify as J2, but it is possible that its angular size is 
smaller than expected by simple extrapolation from earlier epochs 
(Section~\ref{subsec:PA}). Therefore, we have repeated fits to the data omitting these 
four epochs. A linear fit yields 
${\mu}_{2} = 0.217\pm0.012$~\masyr; a quadratic fit 
corresponding to ${\alpha}_{2} = 0.022$~mas ${\rm yr}^{-2}$ has somewhat lower 
residuals (0.128 vs. 0.086~mas). We tested the ``bend'' model for three 
different dividing times (mid-points between two adjacent epochs) between the 
``early'' and ``late'' cases: 2000.4, 2001.7, and 2002.6. The three earlier 
fits give ${\mu}_{2,early} = 0.144\pm0.010$~\masyr, 
$0.158\pm0.012$~\masyr, and $0.168\pm0.008$~\masyr.  
The three later fits give 
${\mu}_{2,late} = 0.268\pm0.018$~\masyr, 
$0.26\pm0.03$~\masyr, and $0.275\pm0.014$~\masyr. 
Although this suggests a smaller increase in proper motion than found for 
the full data set (increase of $\sim$0.11~\masyr\ vs. 
$\sim$0.17~\masyr), the increase is substantial and leaves 
little doubt that component J2 exhibits positive apparent acceleration. 



\begin{figure}[ht]
\plotone{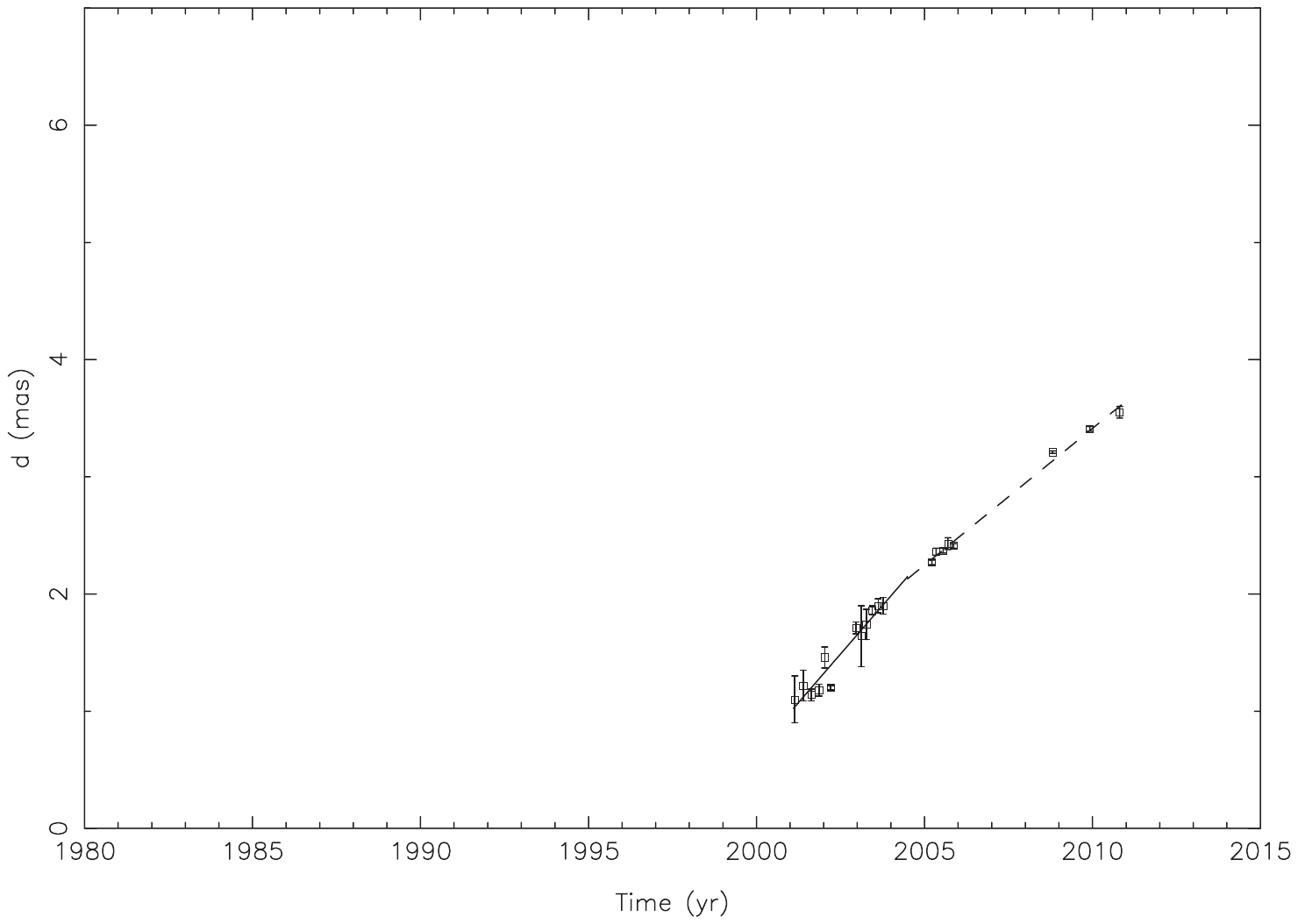}
\caption{Core distance $d$ as a function of time for jet component J3, with separate linear fits before (solid line) and after (dashed line) epoch 2004.5, yielding $\betaapp = 12.8 \pm 1.2$ beforehand and $\betaapp = 9.0 \pm 0.3$ afterwards. 
\label{fig:j3bend}}
\end{figure}

\subsubsection{Component J3}  \label{subsubsec:J3}

The component J3 was visible from 1998.08 through the end of our survey 2010.81. A weighted linear fit 
gives ${\mu}_{3} = 0.274\pm0.008$~\masyr, and it represents the 
data well, consistent with constant radial proper motion. Visual inspection of 
Figure~\ref{fig:disttime} suggests that J3 possibly emerged from the core region during the 
1995.67 flux density outburst.
      
There are mild hints of shallower slopes at the beginning and end of the 
J3 radial position as a function of time data. The initial shallower slope 
could be due to confusion between components S and J3 within $\sim$1~mas of 
component C. If we eliminate the two points with $d<$ 1~mas, a linear fit 
gives ${\mu}_{3} = 0.265\pm0.008$~\masyr, in agreement with the 
linear fit to all the data. A quadratic fit that gives a hint of constant 
deceleration with ${\alpha}_{3} = -0.016$~mas ${\rm yr}^{-2}$ produces only 
marginally better residuals, 0.072~mas vs. 0.096~mas. Even less improvement is 
seen for a fit to the unshifted data. The ``bend'' model applied to J2 (Section~\ref{subsubsec:J2}) provides similarly marginal improvements in the J3 fits, again 
omitting points with $d<1$~mas. We chose three different dividing times, i.e., mid-points between two adjacent epochs, between the ``early'' and ``late'' 
cases: 2003.4, 2004.5, and 2007.3. The three earlier fits yield
${\mu}_{3,early} = 0.31\pm0.05$~\masyr, 
$0.33\pm0.03$~\masyr, and $0.297\pm0.013$~\masyr.  
The three later fits yield
${\mu}_{3,late} = 0.238\pm0.005$~\masyr, 
$0.233\pm0.007$~\masyr, and $0.170\pm0.007$~\masyr.
The middle case, with the division at 2004.5, 
is plotted in Figure~\ref{fig:j3bend}. It  
has a residual rms of 0.071~mas. So although the early and late slopes 
formally show significant differences, hinting at a decrease in proper motion past the ``bend,"
from $\sim$0.31~\masyr\ to $\sim$0.21~\masyr, we conclude that the evidence for apparent deceleration of component J3 is marginal. 


\subsubsection{Component J4}   \label{subsubsec:J4}

The component J4 emerged in 2005.22 and was detected through the end of our survey in 2010.81. A weighted linear fit 
gives ${\mu}_{4} = 0.077\pm0.018$~\masyr. Visual inspection of 
Figure~\ref{fig:disttime} indicates that J4 emerged from the core region during the 2003.45 
flux density outburst.

Over the relatively short time span of the J4 data, only a linear fit is 
justified, with a notably lower average radial proper motion roughly one-third that of J2 and J3. Excluding points with $d<$ 1~mas, this drops slightly to 
${\mu}_{4} = 0.052\pm0.007$~\masyr. However, there are many 
complications, including the triple structure of J4, its proximity to the core 
region during all the observations, and an unusual brief period of possible 
inward motion by component S. It is not clear if the earlier observations of 
J4 represent J4a alone or both J4a and J4b. The unshifted data give a larger 
${\mu}_{4} = 0.127\pm0.019$~\masyr\ for all the data, but show 
no significant motion beyond 1~mas with 
${\mu}_{4} = 0.07\pm0.05$~\masyr. Conservatively, we conclude 
the following about the motion of J4: (1) J4 is moving outward and leaving the 
core region; and (2) it is moving more slowly than both J2 and J3 over similar 
ranges of distance, with an 
estimated ${\mu}_{4} \sim0.1$~\masyr.



\subsection{Component Position Angles}  \label{subsec:PA}

As a prelude to exploring the projected two-dimensional trajectories of 
the VLBI components in the next section, we first investigate the behavior of 
component position angles relative to the core component C.  

\begin{figure}[ht]
\plotone{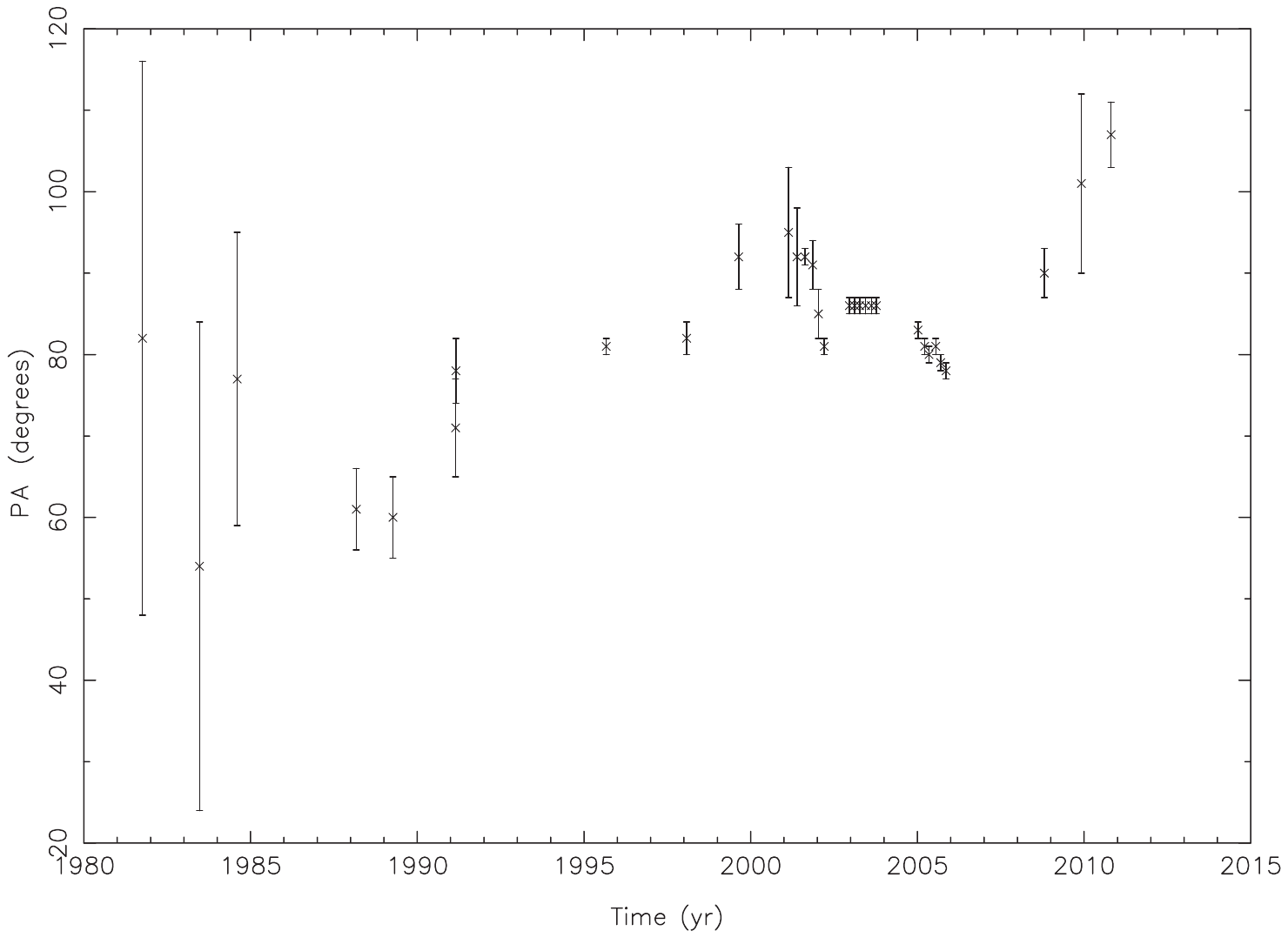}
\caption{Position angle $PA$ as a function of time for the swinging component S. 
\label{fig:pavstime}}
\end{figure}

We begin again with the swinging component S. Although the radial data 
are consistent with no or very slow radial motion, tangential motions -- 
{\it i.e.}, changes in PA -- are significant (Figure~\ref{fig:pavstime}). The measurements of PA have
large PA uncertainties during the first $\sim$5~yr, but 
for the following $\sim$25~yr shows an overall long-term increase by 
$\sim$45\arcdeg, from PA $\sim$60\arcdeg\ to PA $\sim$105\arcdeg, with 
two clear short-term reversals superimposed. Even with our long monitoring 
time, it is difficult to claim that these reversals are part of an oscillatory 
behavior, but it is evident that short-term PA swings occur on top of a 
long-term general PA increase. There is first a long, steady increase in 
PA at $\sim$2.5\arcdeg\ ${\rm yr}^{-1}$ between 1988 and early 2001. The 
first reversal, which is the best sampled, shows a fairly steady PA decrease 
of $\sim$2.6\arcdeg\ ${\rm yr}^{-1}$ during the broad 2003.45 outburst. The 
second reversal is associated with the onset of the 2010.81+ outburst and has 
the fastest PA rate-of-change, increasing at 
$\sim$5.8\arcdeg\ ${\rm yr}^{-1}$. Using this fastest PA swing and the mean 
core distance $d = 0.44$~mas, the corresponding maximum tangential proper 
motion is $\sim$0.04~\masyr. Since we have a constraint that the 
arcsecond-scale jet lies along an average PA $\sim$93\arcdeg, it seems 
unlikely that the PA would continue to show a general increase well beyond 
the observed maximum of $\sim$105\arcdeg. In fact, while the uncertainties are 
large, the mean PA for the first $\sim$5~yr is 71\arcdeg, consistent with 
part of a general decline to $\sim$60\arcdeg\ before the onset of the 
well-sampled general increase. Thus perhaps one long-term increase or decrease 
lasts at least $\sim$25~yr, suggesting a minimum period of $\sim$50~yr, with 
reversals as short as $\sim$5~yr superimposed. 


We note that for four epochs, the separation between components C and S 
was slightly below one-half the uniform beam width, at which point concerns 
might be raised about the reliability of modelfit parameters. However, the 
modelfit positions for component S at these four epochs are entirely 
consistent with epochs nearby in time, so we include these data in our 
analyses. 

\begin{figure}[ht]
\plotone{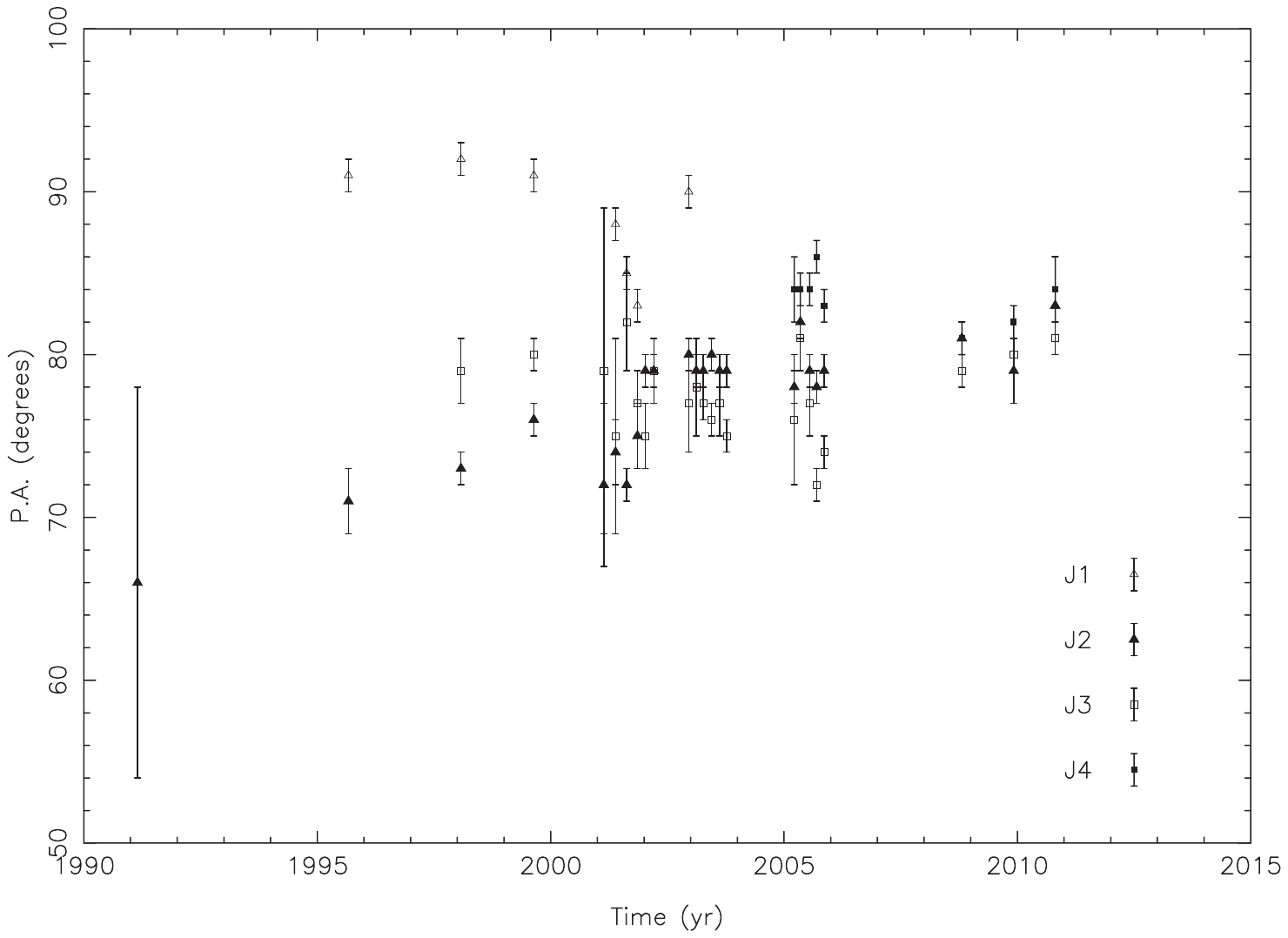}
\caption{Position angle $PA$ as a function of time for the Jet components J1--J4.
\label{fig:pavstimejet}}
\end{figure}

To search for any connection between position angles for S and the four 
jet components, we also analyzed the PAs for the jet components as a function of time (Figure~\ref{fig:pavstimejet}).
The position angles for components J1, J3, and J4 are constant (slope of PA 
as a function of time consistent with zero) during the observations: ${\rm PA}_{\rm J1} = 88.6\arcdeg \pm 1.3\arcdeg$, ${\rm PA}_{\rm J3} = 
77.5\arcdeg \pm 0.5\arcdeg$, and 
${\rm PA}_{\rm J4} = 84.0\arcdeg \pm 0.5\arcdeg$. Only J2 -- the component 
that has been observed the longest and over the largest range of separation 
from the core -- shows significant changes in PA, increasing from $66\arcdeg 
\pm 12\arcdeg$ in 1991 
and a 
better-determined $71\arcdeg \pm 2\arcdeg$ in 1995, to $\sim$80\arcdeg\ in 
2003, and remaining constant at $\sim$80\arcdeg\ thereafter. Thus it is 
evident that the four jet components travel along different position angles, 
and do {\it not} follow the same path; this will be explored in detail in the 
next section. Here, we seek to establish a link between the position angle of 
S at the peak of a flux density outburst and the ``ejection'' angle of the jet 
component created in that outburst. In the case of J2, this ejection angle is 
clearly different from later position angles. Thus we will use the innermost 
jet component position angle as the ejection angle. We do not consider J1, 
because it is only observed at large separations from the core, although its 
trajectory along PA $\sim$90\arcdeg\ traces back directly to the core 
(Section~\ref{subsubsec:J1}); and the earliest PA data for S are consistent with $PA 
= 90\arcdeg$. For J2, the innermost ${\rm PA}_{\rm J2} =  
66\arcdeg \pm 12\arcdeg$ at $d = 0.8$~mas 
agrees with 
${\rm PA}_{\rm S} = 60\arcdeg \pm 5\arcdeg$ at the peak of the 1989.27 
outburst. For J3, which first appears at $d = 0.7$~mas, its 
${\rm PA}_{\rm J3} = 79\arcdeg \pm 2\arcdeg$ agrees well with 
${\rm PA}_{\rm S} = 81\arcdeg \pm 1\arcdeg$ at the peak of the 1995.67 
outburst. And for J4, which first appears at $d\sim1$~mas, its mean 
${\rm PA}_{\rm J4} = 84\arcdeg \pm 1\arcdeg$ for its first five epochs 
during 2005 agrees well with ${\rm PA}_{\rm S} = 86\arcdeg \pm 1\arcdeg$
at the peak of the 2003.45 outburst. We note that the observed backtrack of 
${\rm PA}_{\rm S}$ by $\sim$15\arcdeg\ occurred during the flare. For 
clarity, we note the obvious fact that the position angles of J2, J3, and J4 
at any given epoch generally do not match those for S, since 
${\rm PA}_{\rm S}$ is continuously changing. Thus for the three outbursts 
that we can examine in detail, there is a strong connection between the 
position angle of S at the peaks of the outbursts and the innermost position 
angles of the jet components produced during the outbursts. We note that the 
PA values in the vicinity of the core region are generally very well 
determined, which as we saw in Section~\ref{subsec:morph1} is less often the case for inner 
radial positions, so we have confidence in these innermost jet PA values. 


\begin{figure}[ht]
\plotone{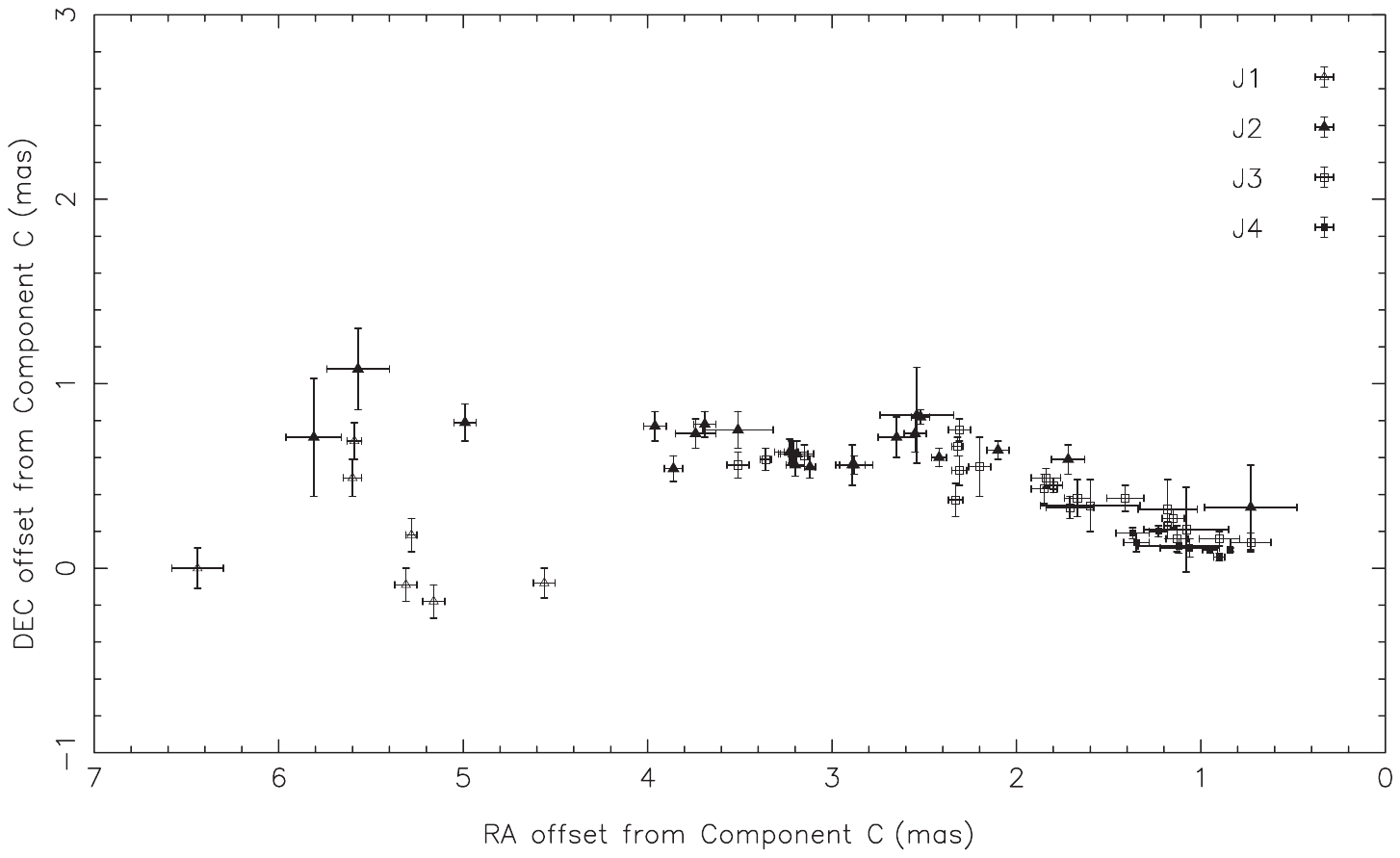}
\plotone{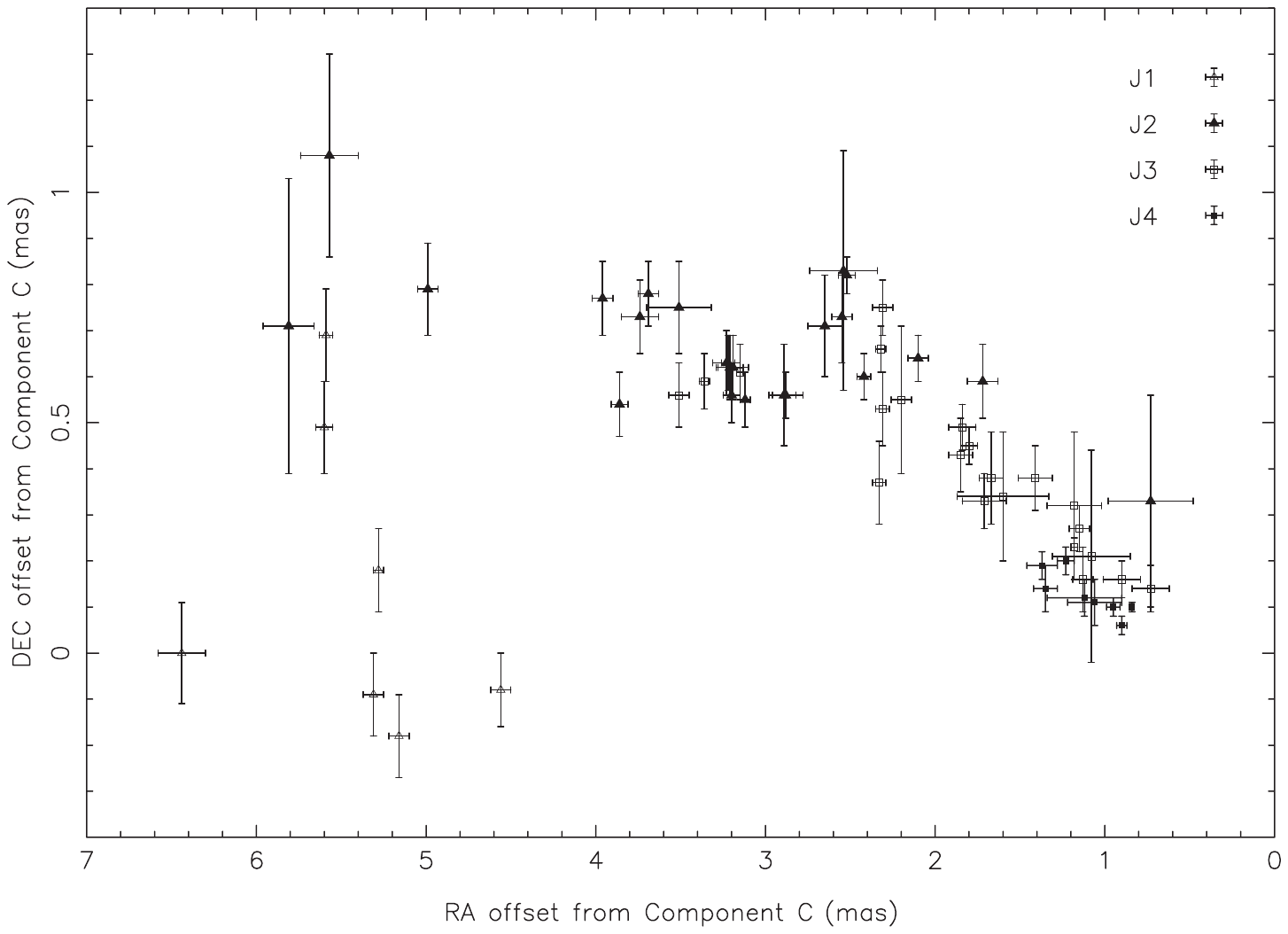}
\caption{Trajectories on the sky for the jet components J1--J4.  The true core C is at $(0,0)$ in the lower right.  The two plots contain the same data; however the top has equal scales for RA and Dec, while the bottom shows an expanded DEC scale for clarity. 
\label{fig:jettraj}}
\end{figure}

\subsection{Component Trajectories}  \label{subsec:traj}

Figure~\ref{fig:jettraj} plots the two-dimensional trajectories 
projected on the sky for all four jet components, using Right Ascension (RA) and Declination (DEC) offsets from 
the core component C at the origin. We do not exclude the 
small number of points at $d$ less than $\sim$1~mas, because these are 
critical for probing inner jet trajectories; any large fractional 
uncertainties in their radial positions will have little effect on trajectory 
determination. Uncertainties in RA and DEC were calculated by propagation of 
uncertainties in the core distances $d$ and the position angles PA At first 
glance, it may appear on Figure~\ref{fig:jettraj} that all components follow a ``noisy'' 
-- but common -- curved path. But as was already known from the significant 
PA differences between the various jet components discussed above, closer 
examination in reveals that {\it each component travels its own 
unique path.} Here we compare not the average PA values for jet components 
over their entire paths, but instead look explicitly at PA differences for 
matched ranges of distance from the core C. Between $\sim$4-7 
mas, the path of J1 is distinctly below that of J2, with a mean PA 
difference $\Delta{\rm PA}_{2-1} = -7.6\arcdeg \pm 1.8\arcdeg$. The 
paths of J2 and J3 overlap for $d$ greater than $\sim$2~mas, but the path of 
J3 is slightly below that of J2 for $d$ less than $\sim$2~mas. For all points 
with $d< 2.30$~mas, the mean PA difference $\Delta{\rm PA}_{3-2} = 
7.5\arcdeg \pm 2.2\arcdeg$. Finally, the path of J4 is slightly below that of 
J3 where both are observed out to $d\sim1.5$~mas. For all points with 
$d< 1.50$~mas, the mean PA difference $\Delta{\rm PA}_{3-2} = 
5.2\arcdeg \pm 1.0\arcdeg$. All results presented here are also found for the 
unshifted data, with very similar significance levels.     


\begin{figure}[ht]
\plotone{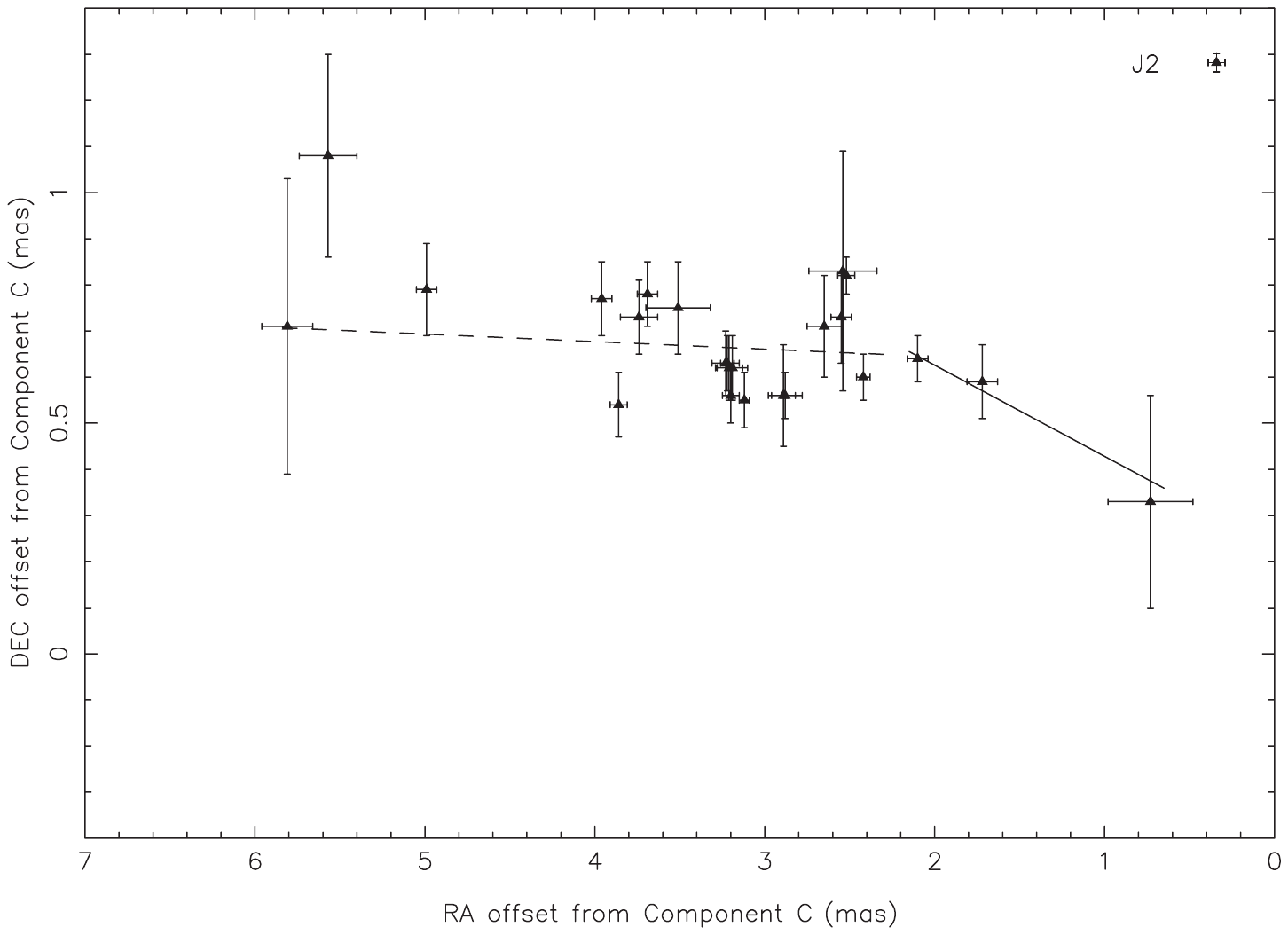}
\plotone{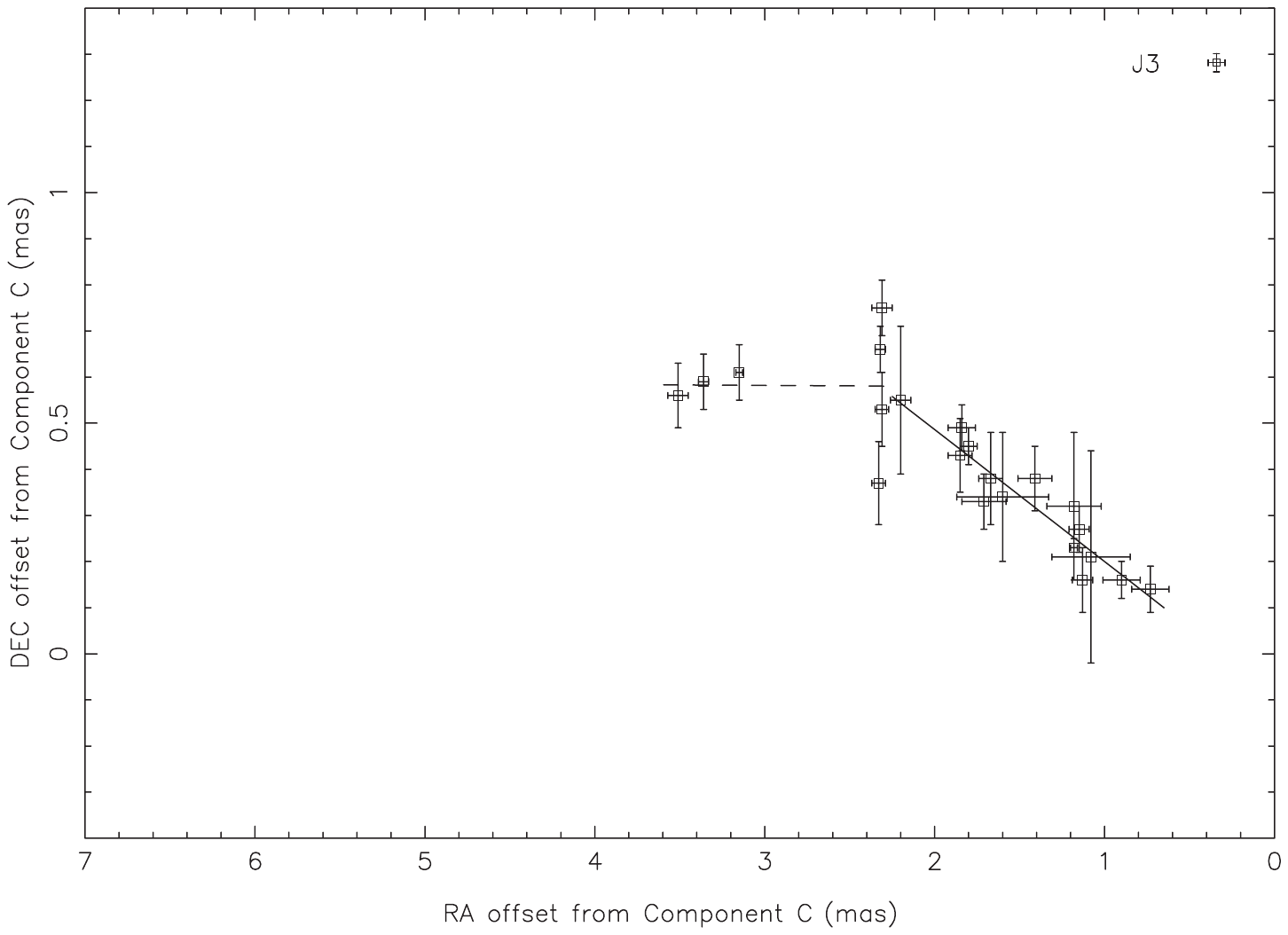}
\caption{Trajectories on the sky for the jet components J2 (top) and J3 (bottom).  The true core C is at $(0,0)$.
\label{fig:trajj2j3}}
\end{figure}

We now consider the projected trajectories for individual jet components. 
As can be seen in Figure~\ref{fig:jettraj}, there are no indications of non-linear paths 
for J1 and J4. Within the uncertainties, linear fits trace back to the origin 
for both of these components: the {\it y}-intercepts are $-0.7\pm1.4$~mas 
for J1 and $-0.08\pm0.06$ for J4. While this is also true for J3 
({\it y}-intercept $0.06\pm0.05$~mas), a single linear fit is dominated by the 
points at $d$ less than $\sim$2.5~mas; the points at larger distances give a 
definite hint of a different direction. For J2, a weighted least-squares fit 
shows that it clearly has a non-zero $y$-intercept ${b}_{J2} = 0.57 \pm 
0.10$~mas that is far north of the origin (an unweighted fit gives similar 
results, but we conservatively use the weighted fit in our analyses of J2 here 
because of the large fractional uncertainties in some of its DEC coordinates). 
 
Because the trajectories of J2 and J3 appear to ``flatten out'' past 
$d\sim2.5$~mas, this suggests trying separate linear fits for the inner and 
outer trajectories. After using several trial points for the break in slope, 
we found that $d = 2.30$~mas gives the best results that preserve  continuous 
trajectories for both J2 and J3 (Figure~\ref{fig:trajj2j3}).
For the inner jet of J2, this yields a $y$-intercept of $0.23\pm0.10$~mas; we 
will accept the improvement from a $\sim6\sigma$ to a $\sim2\sigma$ offset 
that is marginally north of the origin. For the inner jet of J3, we find a 
$y$-intercept of $-0.086\pm0.043$~mas; this actually increases the offset 
south of the origin from $\sim1\sigma$ for a single overall linear fit to 
$\sim2\sigma$ using separate inner and outer fits, and is not surprising 
given that the three outer points break towards larger PA, so we accept this 
marginal offset south of the origin. These results support a visual impression 
in Figure~\ref{fig:jettraj} that jet components travel along the angle of ejection -- as 
large as $\sim27\arcdeg$ from the mean jet axis of $\sim93\arcdeg$ on 
arcsecond scales -- within $\sim2.5$~mas of the core, but then tend to move in 
a more easterly direction closer to PA $\sim90\arcdeg$ at larger 
distances. 


To quantify the direction of motion at these larger distances, the slope 
of the linear fit for J2 beyond $d\sim2.5$~mas is $-0.117 \pm 0.035$, 
which corresponds to motion along PA $\sim83\arcdeg$. However, if the 
dividing point between inner and outer linear fits is moved slightly inward, a 
slope consistent with zero results. Therefore, we consider 
PA $\sim83\arcdeg$ to be the northernmost PA for the direction of jet 
component motion that is allowed by the data. 
We have detected this knot at two epochs only, with no reliable motion measured, 
so we will simply  assume that this PA is the maximum allowed by the data if 
there has been no redirection toward the mean jet axis. Thus the constraints 
available to us indicate that jet motion beyond $d\sim2.5$~mas occurs on 
both sides of the mean axis of the large-scale jet at ${\rm PA} = 92\arcdeg$, and is always directed within $\sim9\arcdeg$ of this axis.   


\subsection{Component Angular Diameters}  \label{subsec:diam}


     We have also investigated the time behavior of the angular 
diameters of the core region and jet components. 
Specifically, we analyzed the FWHM $D$ of the circular Gaussian components  
representing the core component C and the quasi-stationary component 
S as a function of time. For component C, a weighted least-squares fit shows 
remarkably stable core diameters, with no evidence for any long-term  
trend (slope of $0.0018 \pm 0.0027$~\masyr\ including 
upper limits for $D$, $-0.0044 \pm 0.0037$ excluding these). 
Of particular note is the stability of the core diameter throughout 
the strong 2003.45 outburst. The component S does not show any 
obvious long-term trends, either, but it does show distinct 
short-term variations connected to flux outbursts. At the 
peak of the 1995.67 outburst, it had one of its larger 
precisely-measured diameters, with $D = 0.30 \pm 0.04$~mas. 
This is followed by six consecutive upper limits (all consistent 
with $D = 0$) prior to, and in the early phases of, the 2003.45 
outburst. Throughout the rest of this outburst, there is a 
sharp rise, peak, and then decline in $D$, ranging from 
$0.16 \pm 0.02$~mas, up to $0.37 \pm 0.02$~mas at 2005.02,
and then down to $0.13 \pm 0.03$~mas.    

\begin{figure}[ht]
\plotone{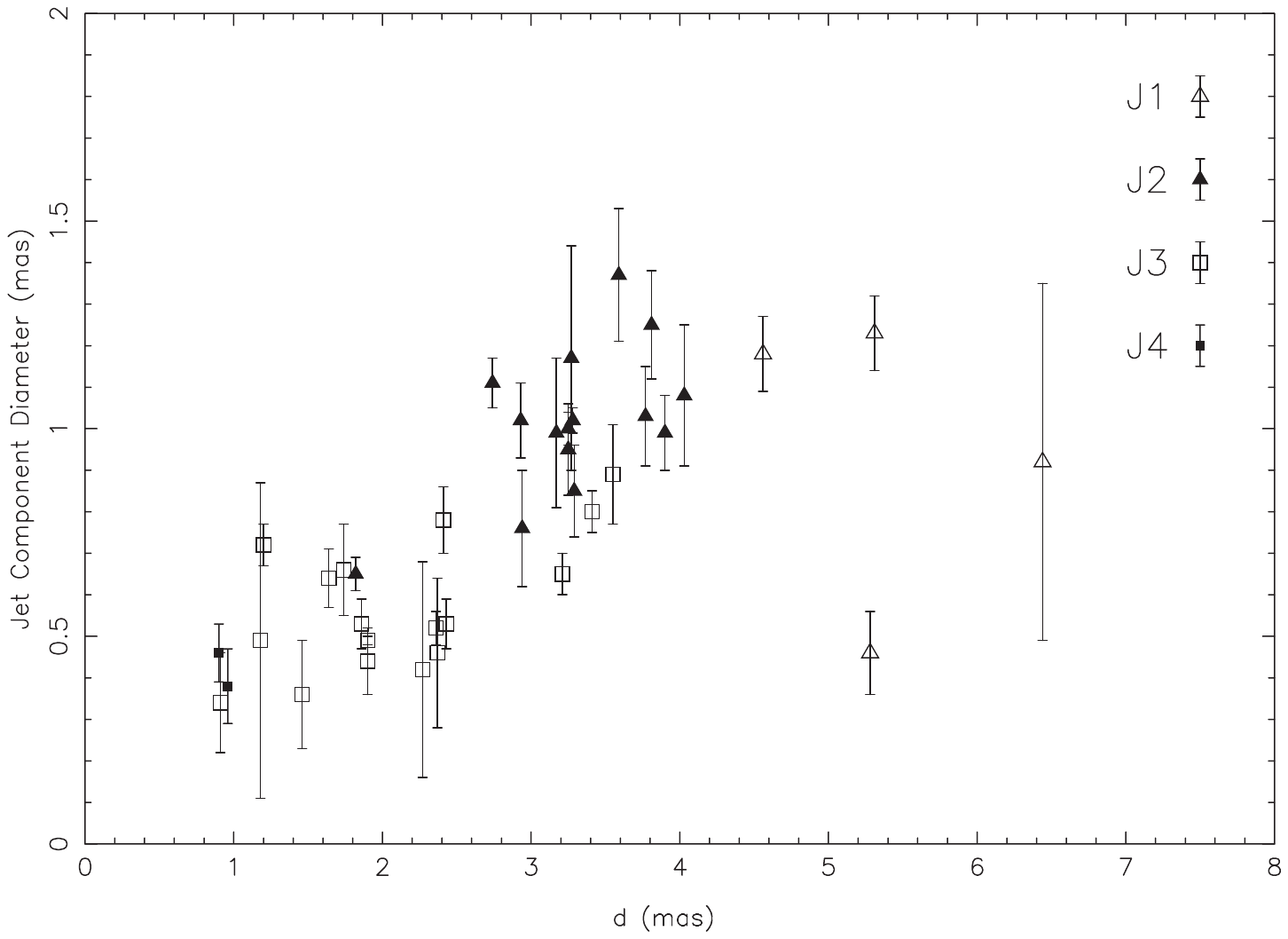}
\caption{Jet component diameter as a function of core distance $d$ for the jet components J1--J4.  The diameter is the FWHM of a single Gaussian component. 
\label{fig:jetdia}}
\end{figure}

As for the jet components, they show evidence of increasing 
diameter as they move outward -- regardless of direction of motion.
This is shown in Figure~\ref{fig:jetdia}, 
where we plot  $D$ as a function of core distance $d$ for all jet components; cases where 
only upper limits to $D$ could be estimated are not included. 
There is a strong correlation 
between FWHM and $R$ out to $\sim$4~mas, which the components J2, J3, and 
J4 also follow individually over their respective ranges of $R$. The few J1 
measurements at larger distances do not follow the correlation; the jet dims 
rapidly at these distances, and it may be that we can only measure the 
narrower inner regions of these faint features before losing them.
But over the range of the correlation, 
a linear fit yields a slope of $0.22 \pm 0.03$~mas ${\rm~mas}^{-1}$ and a 
$y$-intercept of $0.24 \pm 0.07$~mas. The latter is comparable to the typical 
diameters of the components C and S, while the former corresponds to an 
apparent jet opening half-angle of $\sim$6\arcdeg.



\section{Discussion} \label{sec:dis}

\subsection{Variability} \label{subsec:var2}

     The quasi-periodic outbursts in the core region of 3C207 have a mean 
spacing in the co-moving frame of $4.1 \pm 0.5$~yr, with rise and decay times 
of $\sim$2~yr. Despite nearly three decades of observations, the coarse 
sampling of most of the outbursts does not allow us to make any strong claims 
regarding periodicity. We do note that periods of order ten years in flux 
density monitoring data, when well-determined, have been interpreted as 
evidence for binary black holes in, e.g., OJ287 \citep{2006ApJ...643L...9V}. The minima 
between outbursts suggest a quiescent level of $\sim$400~mJy. If we define the 
fractional variability amplitude $f_v = ({S}_{max} - {S}_{min})/{S}_{min}$, 
the outbursts range from $f_v \sim$0.25 for 1983.47 to $f_v\sim$2.5 for 
2003.45. Thus the outbursts range over an order of magnitude in strength. This 
range could of course be intrinsic. If it is due to redirection of an inner 
relativistic jet component in the core region, we can estimate the range of 
angles $\theta$ to the line of sight using the Doppler factor 
$\delta = {[\gamma (1 - \beta {\rm cos} \theta)]}^{-1}$, where $\gamma$ is 
the Lorentz factor and $\beta$ is the speed parameter (assumed constant for 
this calculation). If we treat the core region as a single flat-spectrum 
component, then a ${\delta}^{3}\sim10$ flux density range only requires 
$\delta$ to range over a factor of $\sim$2. For $\gamma = 10$, this is easily 
accomplished if $\theta$ varies from, e.g., 15\arcdeg\ to 10\arcdeg.

     The details of shorter-term variations in flux density and component 
diameters are of particular interest during the 2003.45 outburst. As we have 
already seen, there are actually three sub-outbursts of C only $\sim$1.3~yr 
apart in the co-moving frame. For the first two sub-outbursts, the component S 
mirrors these with a lag of $\sim$0.6~yr. The diameter of C is remarkably 
stable during the entire 2003.45 outburst. The diameter of S begins at 
less than $\sim$0.2~mas.  It first has a distinctly non-zero diameter of 0.16~mas 
at 2002.2 just prior to the peak of the first S sub-outburst, rises to a 
maximum of 0.37~mas at 2005.0 just past the peak of the second S sub-outburst, 
and then eventually drops to 0.13~mas. 


     These behaviors for C and S support an interpretation in which three 
closely-spaced sub-outbursts in C each create a new component of small 
diameter (less than $\sim$0.2~mas) that continually expands as it moves out. 
As the first component passes through the S region and blends with it,  the 
``S blend'' increases in diameter. As the second component later does the 
same, the diameter of the S blend reaches a larger maximum value. This could 
be because the first new component also remains at least partly blended with 
S, having not yet completely cleared the S region, at least as can be 
determined given the lower quality of the 2005.02 data noted previously. By 
2005.22, the first new component is seen to be separately resolved from S, and 
the diameter of S decreases. We note that the 2002.03 epoch -- at which the 
core region could only be modeled with a single elliptical Gaussian component 
-- is halfway between the first maxima for C and S. If the first new component 
was about midway between C and S at this time, it is plausible that the 
resulting quasi-continuous line of flux required an elliptical Gaussian 
representation at our resolution. 

     We identify the newly-resolved component at 2005.22 as J4. The emergence 
of the second new component is not observed during the 2005 observations, but 
for 2008 through 2010, J4 has to be represented as a double 
(sub-components J4a and J4b) or triple (third sub-component Jc). This is 
fairly convincing evidence that the triple structure of J4 has its origins in 
the triple sub-outbursts in the core C. It also leads us to the reasonable 
speculation that the 1989.27 outburst consisted of two sub-outbursts, 
responsible for the wide diameter of J2 at its first VLBA epoch in 1995.67 and 
its double nature (J2a and J2b) at all succeeding epochs through 2001.63. It is 
not clear why J2 thereafter presented as a single component from 2001.86 through 2010.81.
Component J3 has always appeared as a single component; this raises the possibility that 
the 1995.67 outburst was a single event producing only one component. 
Component J1 has only been observed as a single component at large distances 
from the core, and we have no detailed observations of the 1983.47 outburst 
(and no observations at all of any earlier events) that may have produced it, 
so we can say nothing about the nature of its associated outburst.

%
%

\subsection{Component Motions}  \label{motion2}

     For the swinging component S, the unshifted positions are consistent with 
no radial motion, but the proper motion using the shifted positions 
($0.008 \pm 0.003$~\masyr) corresponds to a subluminal apparent 
transverse velocity (in units of $c$) $\betaapp = 0.3 \pm 0.1$. Thus at 
most, S is undergoing mildly relativistic outward motion. Its angular motions, 
however, are quite significant. The largest angular motion of 
$\sim$5\arcdeg \ ${\rm yr}^{-1}$, at the mean $d = 0.44$~mas, yields a 
modestly superluminal $\betaapp$ $\sim$1.5. This feature must therefore be 
associated with changes in direction of the mechanism that is ejecting new jet 
components, and likely represents -- in the radial direction -- a 
(nearly-)stationary shock in a (quasi-)continuous flow. 


The linear fits to core separation $d$ as a function of time for the four jet 
components yield the following apparent superluminal speeds: 
${\betaapp}_{J1} = 7.7 \pm 1.8$, ${\betaapp}_{J2} = 9.9 \pm 0.5$, 
${\betaapp}_{J3} = 9.7 \pm 0.3$, and ${\betaapp}_{J4} = 3.0 \pm 0.7$.  As discussed in Section~\ref{subsubsec:J2}, J2 is better modeled with a two-component linear fit that yields ${\betaapp}_{J2} = 6.8 \pm 0.4$ before 2004.5 and ${\betaapp}_{J2} = 14.2 \pm 0.9$ afterwards.  Likewise, J3 is better modeled with a two-component linear fit that yields $\betaapp = 12.8 \pm 1.2$ before 2004.5 and $\betaapp = 9.0 \pm 0.3$ afterwards. The 
slower speed for component J4 is consistent with our earlier 15~GHz results 
\citep{2007AAS...210.0211D,2010AAS...21543417P} and the 15~GHz results of \citet{2013AJ....146..120L}. This may be due to 
both blending with the swinging component S at small radial separations and a 
lower apparent transverse velocity for J4. 
We note that despite mistaking S for other components,  \citet{2002AJ....123.1258H} did 
find reasonable speed estimates for J1 ($\betaapp\sim10$) and J2 
($\betaapp\sim7$), because these components had been recently created at 
the times of misidentification and were then essentially still coincident with 
S.


\subsection{Model for Jet Cone Opening Angle with Recollimation Zone} 
\label{subsec:cone}

The time gaps between the outbursts that spawned J2 and J3, and then 
between those that gave rise to J3 and J4, are roughly equal, and all three 
outbursts are quite clear in the VLBI flux density record (see Figure~\ref{fig:fluxtime}). 
However, it is possible that the gap between the outbursts that formed J1 and 
J2 is considerably longer.  Our identification of an outburst peaking at 
1983.47 is only tentative; in fact, within the uncertainties, the core region 
flux density could have been constant between 1981.76 and 1984.60. Moreover, 
the time gap between J1 and J2 is clearly the largest one on the core distance 
plot in Figure~\ref{fig:disttime}. 

This large time gap is important in attempting to model the projected 
trajectories of the four jet components in terms of an orderly progression of 
ejection angle, as in, {\it e.g.}, a precessing jet. We are compelled to do 
so by the orderly progression of position angles for the successive components 
J2, J3, and J4. The trajectories in Figure~\ref{fig:jettraj} could possibly be explained 
by precession-- e.g., from a binary black hole or jet/disk misalignment-- if the time gap between ejection of J1 and J2 is long enough. 
For example, J1 may have been ejected when 
the launching mechanism was directed along the back side of the precession 
cone with PA $\sim$90\arcdeg. The ejection of J2 may have then have occurred 
when the launch direction was at or near its smallest (i.e., most northerly) 
position angle. Then J3 and J4 may have followed, closer in time and at 
successively larger position angles, as the precessing jet axis moved back 
toward PA $\sim$90\arcdeg\ along the front side.  
Beyond $\sim$2.5~mas, the trajectory of J2 strongly suggests a 
recollimation zone where the opening angle of the jet precession cone is 
reduced from its value in the inner jet. Finally, we see that individual jet
components expand until reaching the recollimation zone
(Figure~\ref{fig:jetdia}).

We now have several pieces of evidence that allow us to develop a working model that might explain much of the rich
VLBI phenomenology in 3C207. The model requires a quasi-regular
motion (perhaps precession) of the swinging component S
around the axis of a cone. Deprojection of the observed
PA range for S based on the superluminal jet speeds suggests a cone with true half-angle of $\sim$5-6\arcdeg\ or less. The jet components follow straight paths along this cone, so there is
no evidence for helical motion related to magnetic fields.
It is interesting that J4, with the slowest motion, is associated with the most powerful -- and therefore perhaps
most highy beamed -- outburst in 2003. This could be explained by motion along the cone at a very small angle to
our line-of-sight. Beyond 2-3 mas, there is a narrowing of
the opening angle in a recollimation zone, which accounts
for the redirection of J2 and J3, and seems to be associated
with their apparent acceleration and possible deceleration,
respectively. While bending of the jet must play some role
for both components, it is difficult to produce an increase
in J2 from $\betaapp \approx 7$ to $\betaapp \approx 14$ without intrinsic acceleration.

\section{Conclusions and Future Work}

After thirty years, we have a working model for 3C207
in which jet components are ejected at different angles
around a cone, with the flow recollimating at 2-3 mas downstream, of order $\sim$100~pc deprojected distance. But the
jet in 3C207 is most likely aimed close to our line-of-sight. So the question remains: How well does our model
apply to the other LDQs, taking larger orientation angles, beaming, and projection effects into account?  Other sources have evidence of swinging innermost components, but projection effects -- and thus observed PA variations -- are smaller. Good determinations of the amplitude of PA variations in several objects would help to constrain the intrinsic cone angle. Other sources show evidence of acceleration, but we still need to confirm trends observed thus far. It is still unclear the circumstances in which apparent jet acceleration/deceleration occurs, and whether it is explained best by jet bending or intrinsic acceleration. Other sources show bending or misaligned jet features, but it is not clear yet when this may be associated
with a recollimation zone as it is for 3C207.   

Future VLBI monitoring of 3C207 and other LDQs can help to resolve these questions.  We note that two additional epochs of VLBA observations of 3C207 were obtained by D.H.H. in 2014.45 and 2018.64 for this project but were not modeled.   It would be interesting to know if jet components J2-J4 were detected in these epochs, and if new components appeared on the expected time interval of $\sim$7~yr. However it might be challenging to match components because of the 4-year gaps between these and prior observations.

Other VLBI observations of 3C207 exist as well.  Using 15~GHz VLBA observations between 1994 and 2016, \citet{2019ApJ...874...43L} identify seven moving features in the core of 3C207, with median and maximum proper motions of $0.245\pm0.04$~\masyr\ ($\betaapp = 9.49\pm0.14$) and $0.275\pm0.10$~\masyr\ ($\betaapp = 10.66\pm0.39$) respectively.  These motions are consistent with the components reported in this paper.  It would be worthwhile to try and match the observed components from their survey to ours, however doing so is tricky because, as mentioned earlier, different observation frequencies can be sensitive to different parts of the jet due to resolution and opacity effects.

Our 3C207 analysis methods can be applied to the five other LDQs in our sample: 3C208, 3C212, 3C245, 3C249.1, and 3C263.  
Their VLBI cores are much fainter, ranging in brightness from $\sim$50 mJy to $\sim$500 mJy; and they do not display strong variability like
3C207. Their jets are likely oriented at larger angles to our
line-of-sight. Although they display milder eﬀects than
3C207, they do show some similar qualitative behaviors.
The best example is 3C263, with a VLBI flux density of
$\sim$100 mJy. It has a typical core-jet structure, with
four jet components that show $\betaapp \approx$ 3-7 \citep{2013EPJWC..6108009H}.
\citet{2008ASPC..386..274H} reported a preliminary trend for the overall sample suggesting that lobe-dominated quasars show positive apparent acceleration on scales of tens of parsecs, and the data for 3C207 at the time were a key part of this study.  There remains much work that D.H.H. wished to do that will remain for the next generation of radio astronomers.

\begin{acknowledgments}

The National Radio Astronomy Observatory and Green Bank Observatory are facilities of the U.S. National
Science Foundation operated under cooperative agreement by Associated Universities, Inc.

This paper was submitted posthumously for D.H.H.  David Hans Hough died on July 14, 2021, from complications of polymyositis. 
As a child Hough wanted to be an astronaut; he became an astronomer. After earning his Ph.D. at CalTech, he served as a postdoctoral fellow first at NASA JPL and then at MPIfR in Bonn.
In 1989, Hough joined Trinity University as an assistant professor. He made full professor in 2001.
Hough won the Astronomical League Award for his contributions to astronomy. 
Hough retired professor emeritus from Trinity in 2014, continuing to pursue his research and teach one or two classes a year for several years after.

He loved classical music, wiffle ball, baseball, hockey, Texas history, and most of all his family. He never forgot a friend or student. He was an active member of Christ Lutheran Church, serving on the Church Council until his death.  All of the co-authors on this paper (except for R.C.V.) are his former students.  T.A.R, his very first research mentee at Trinity, considers it an honor to finish this manuscript on his behalf.  {\it Ad astra} my friend.
\end{acknowledgments}



%

\vspace{5mm}
\facility{NRAO(VLBA)}


\appendix

Full versions of Figure~\ref{fig:vlba_maps} and Tables are below.



\begin{figure}
\plotone{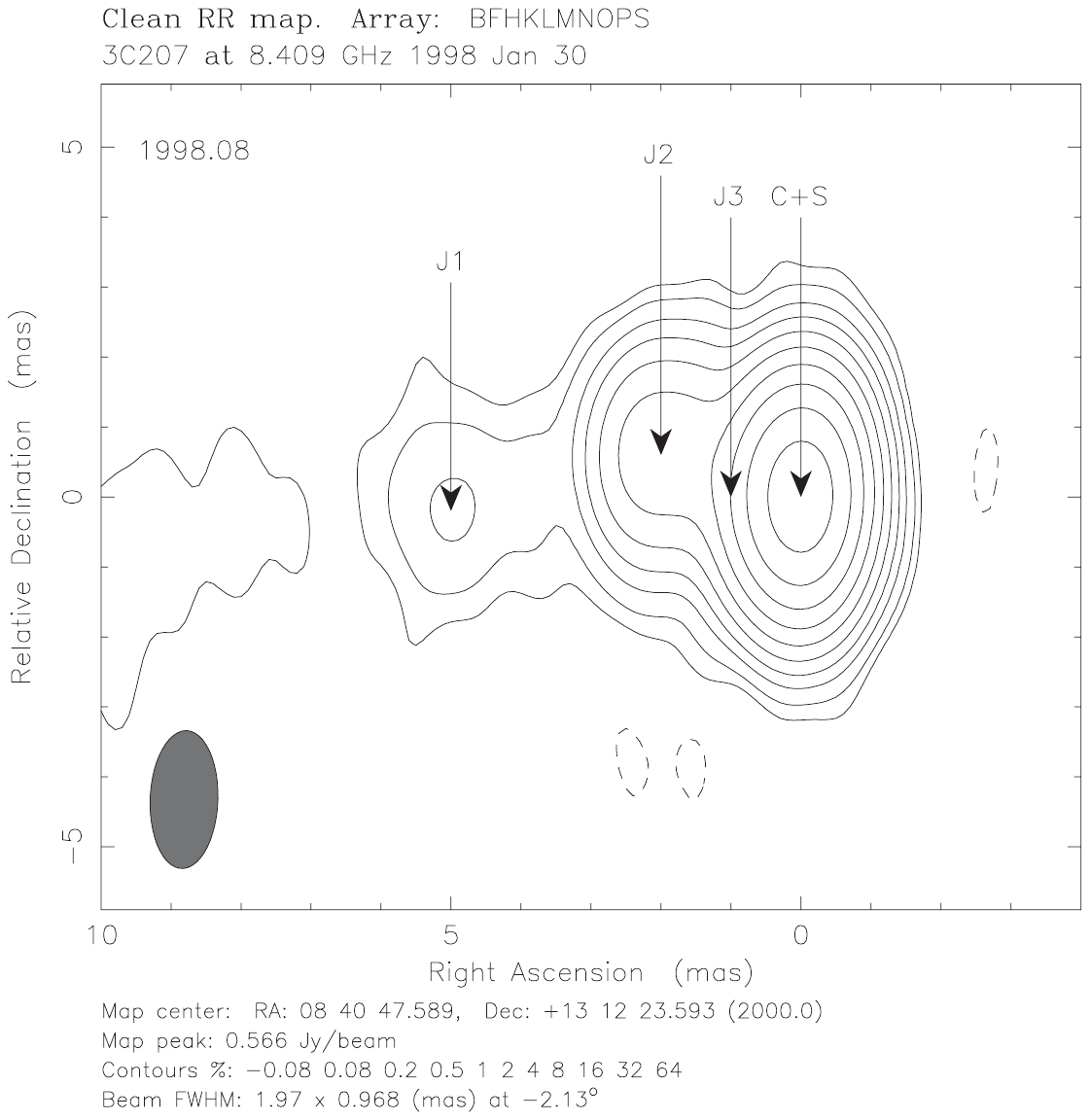}
\plotone{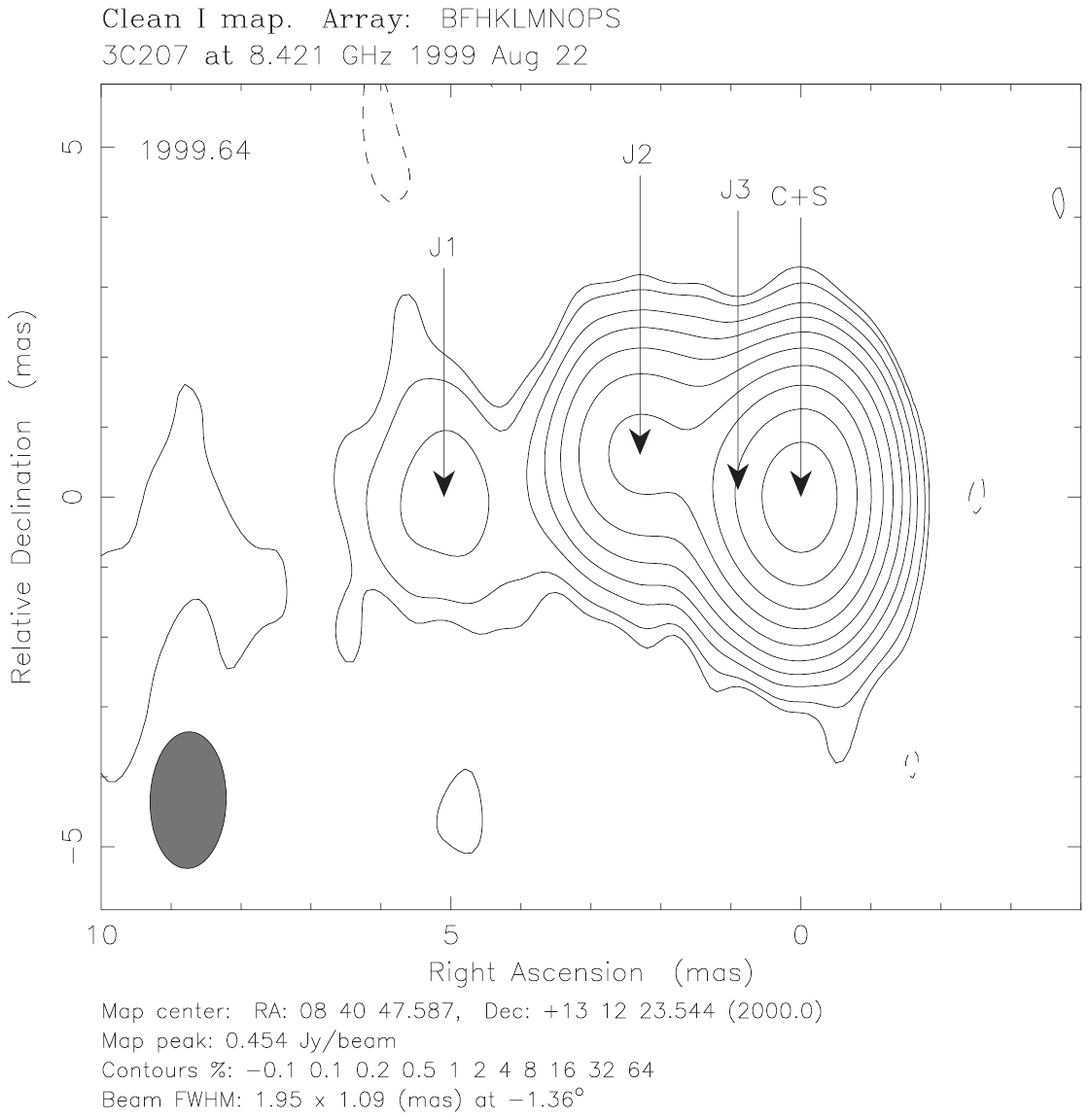}
\end{figure}
\begin{figure}
\plotone{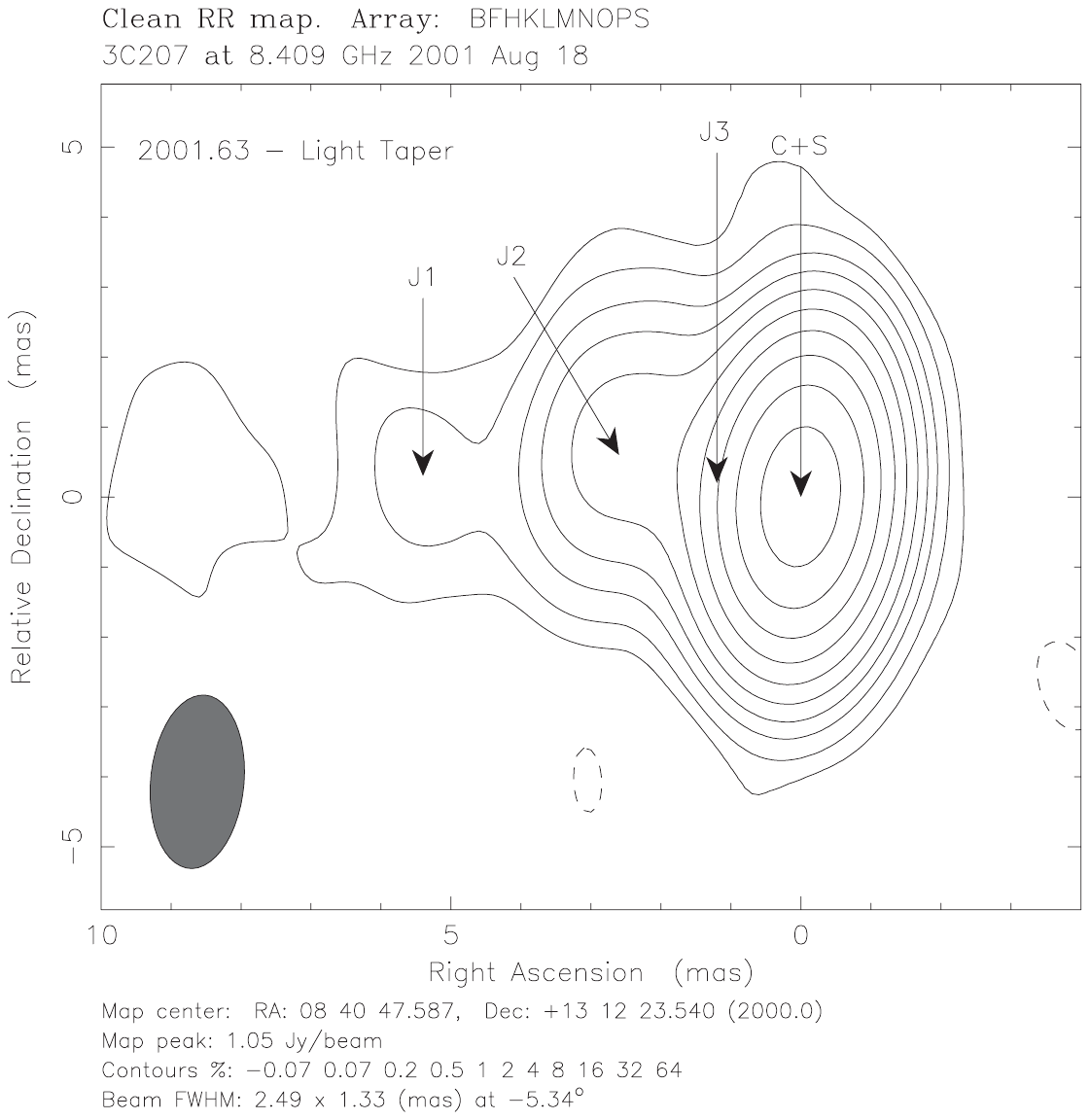}
\plotone{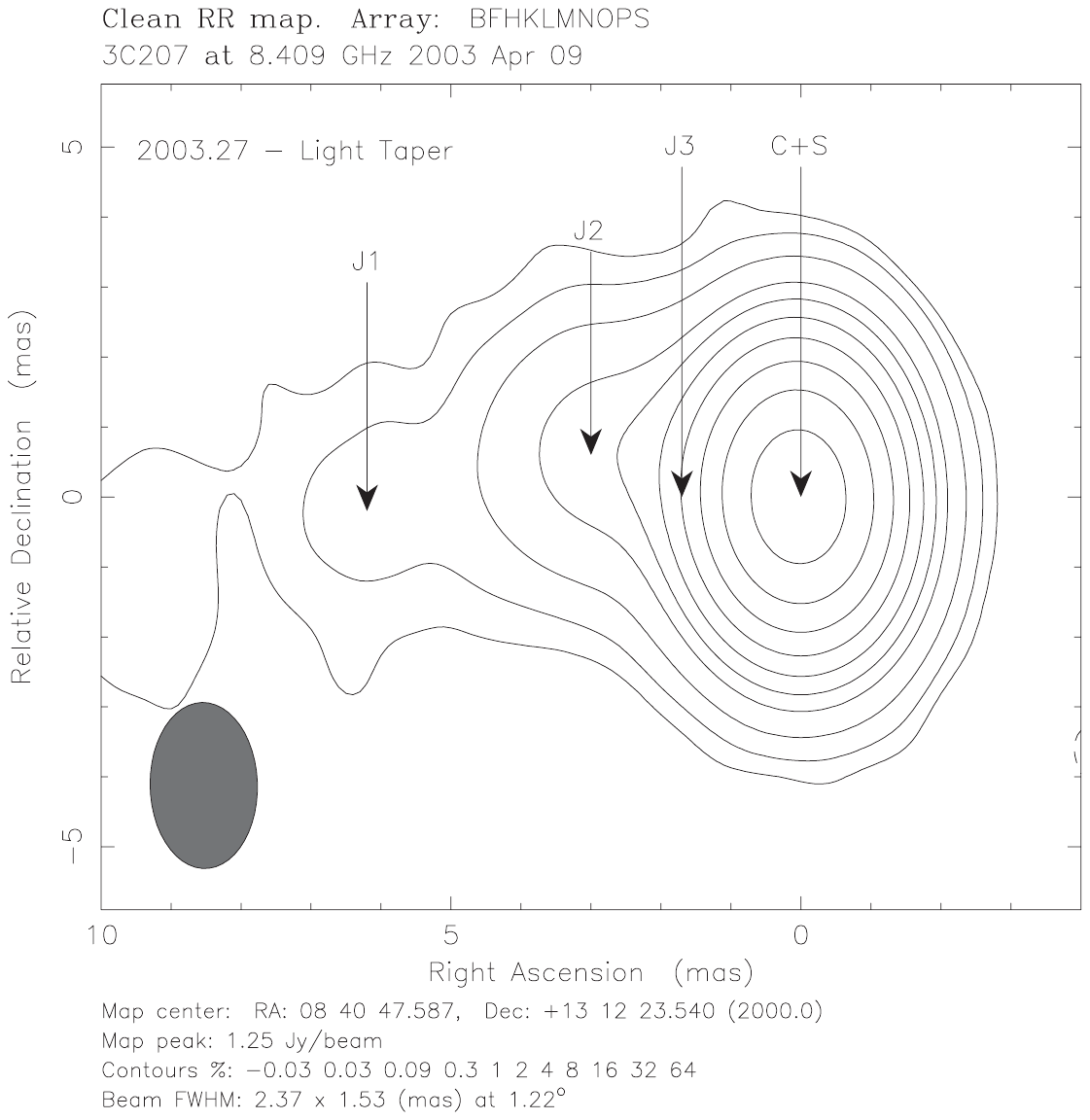}
\end{figure}
\begin{figure}
\plotone{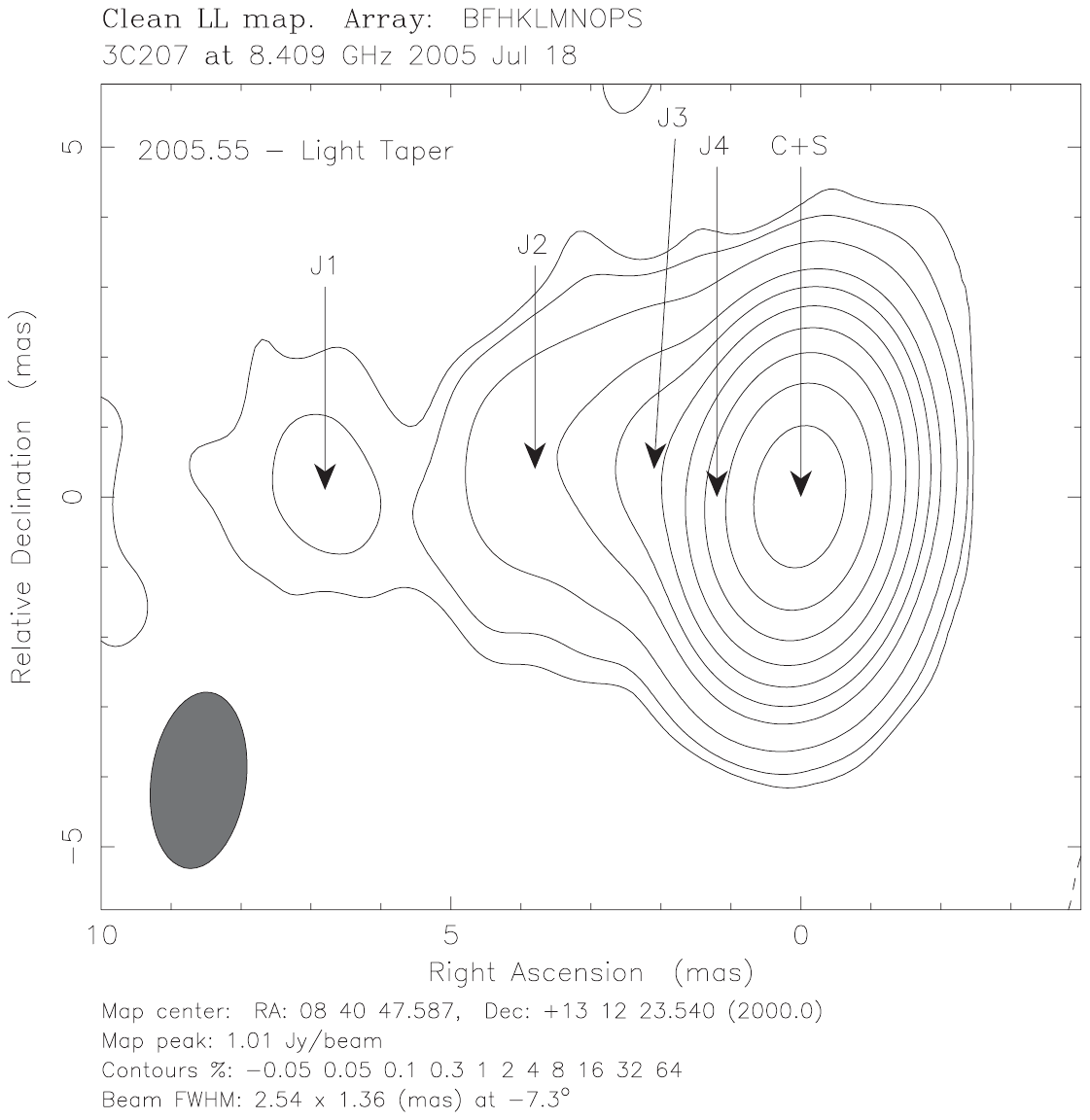}
\plotone{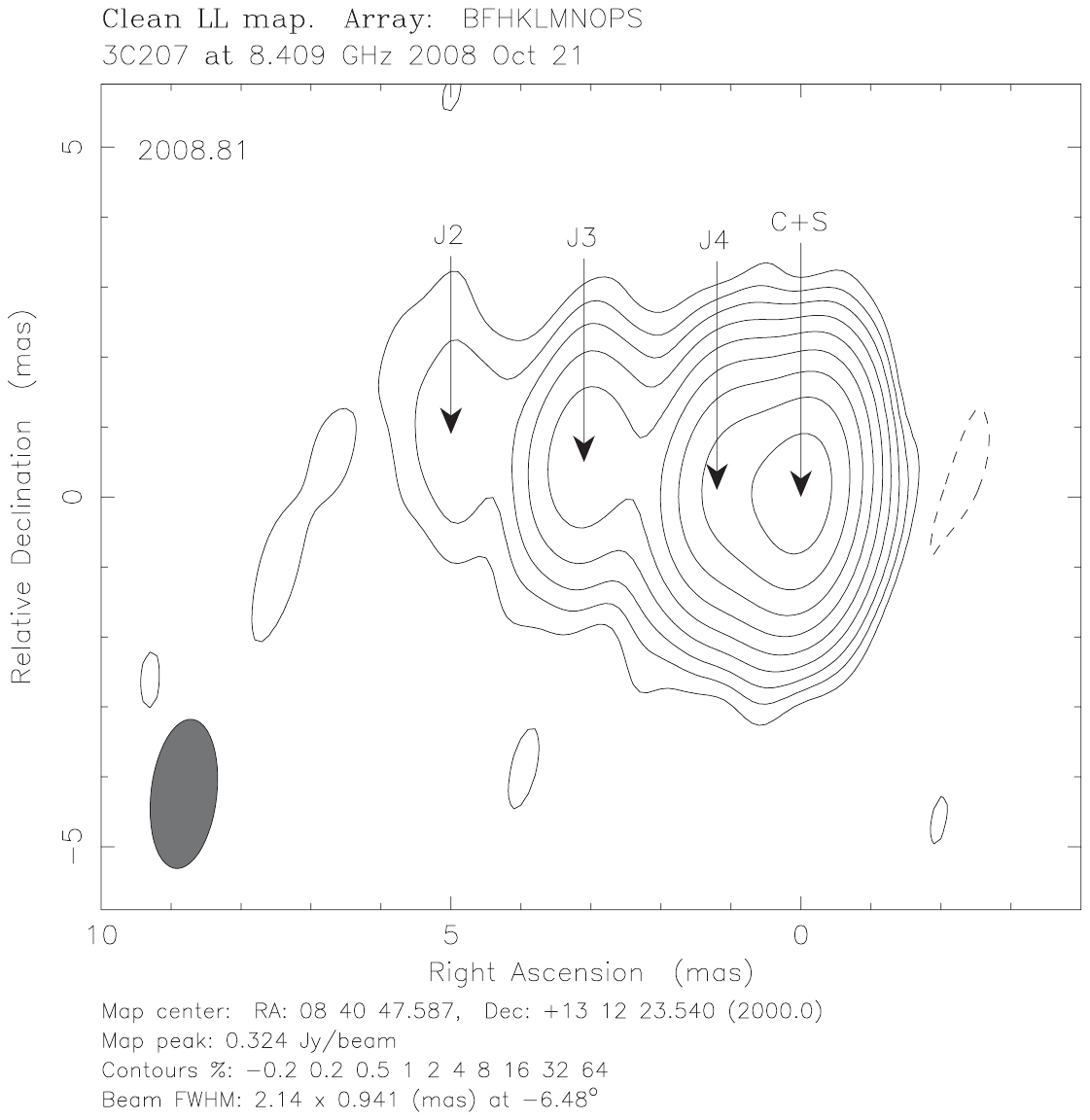}
\end{figure}
\begin{figure}
\plotone{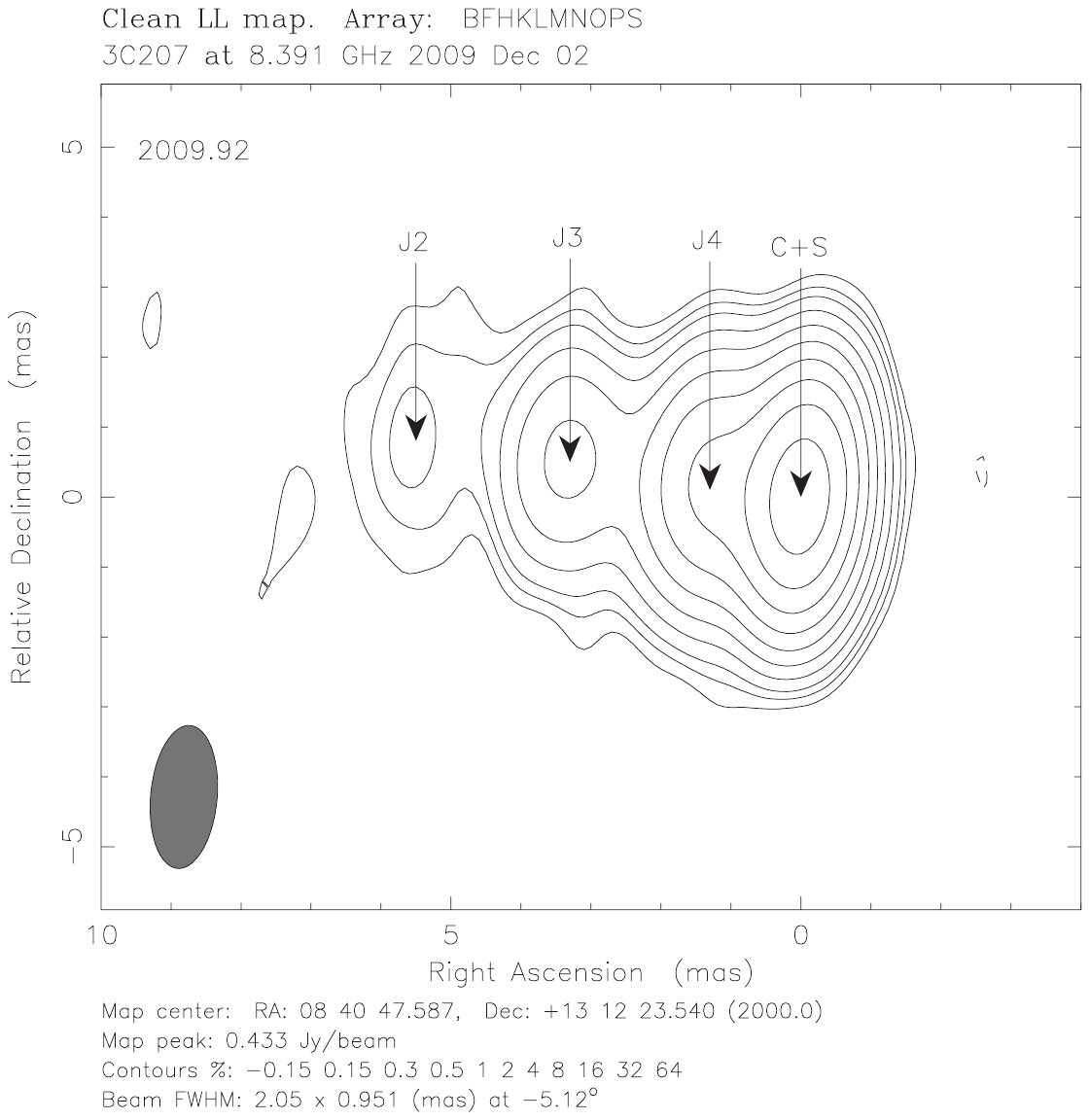}
\plotone{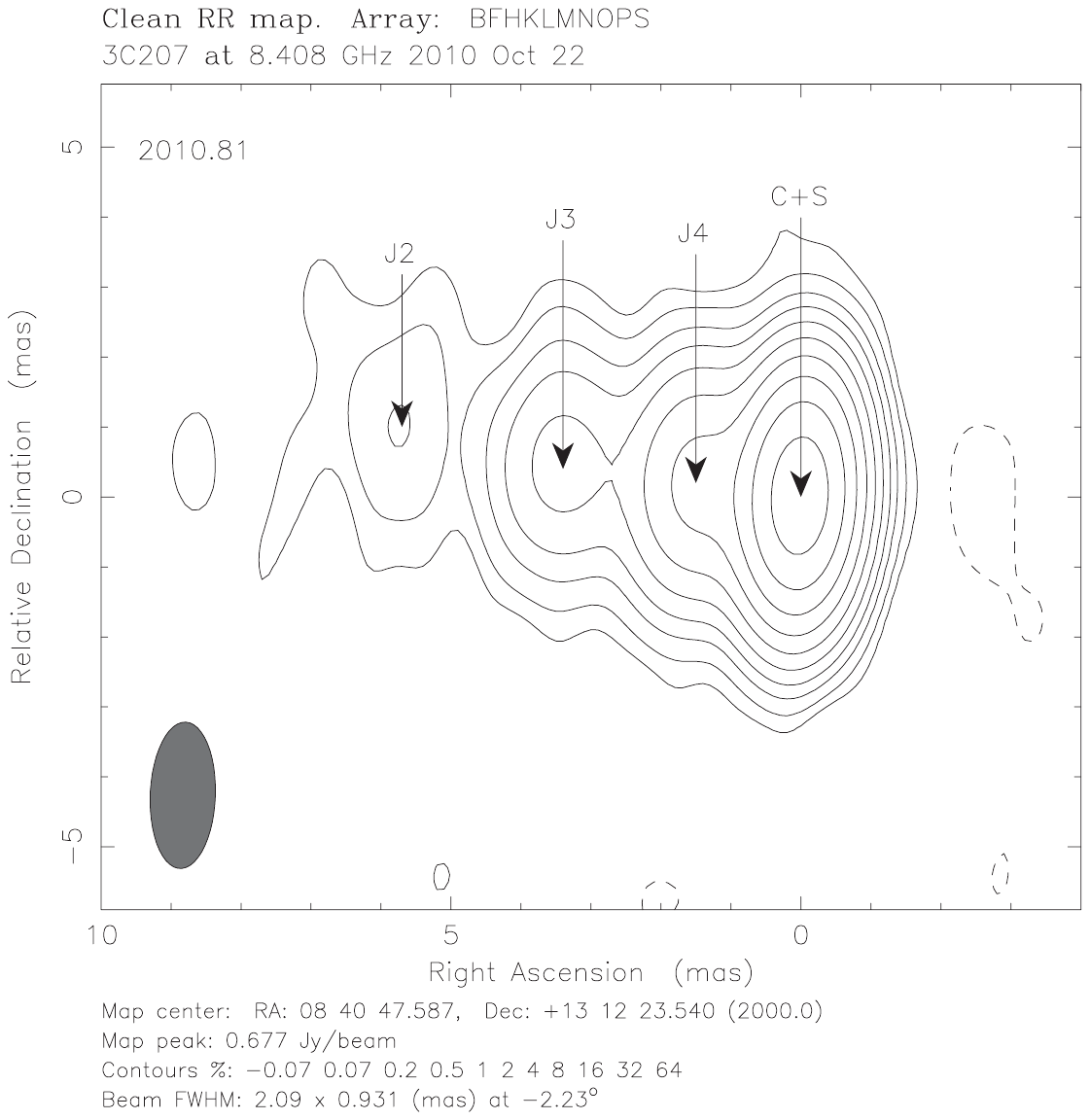}
\end{figure}

\begin{figure}
\plotone{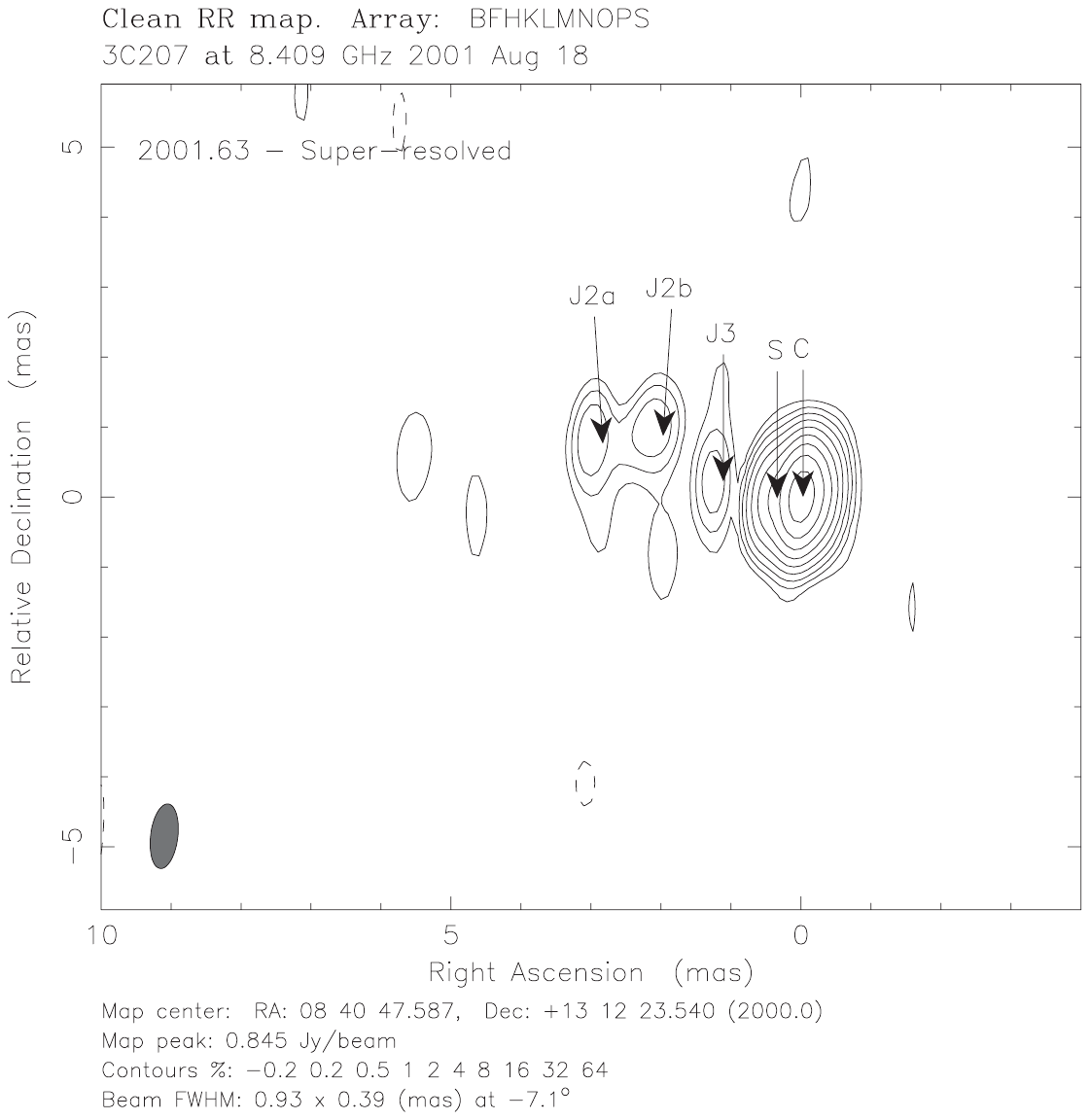}
\plotone{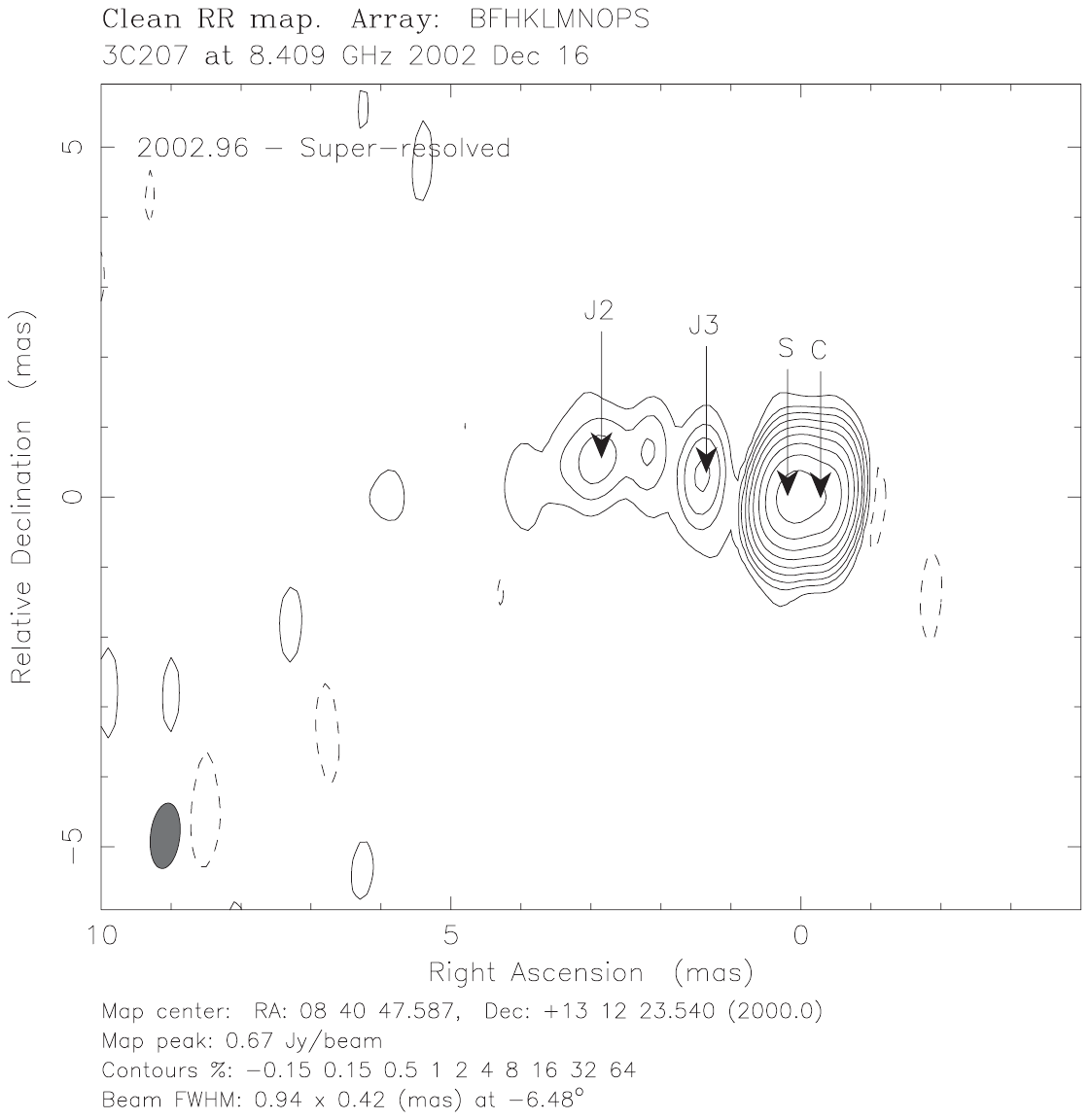}
\end{figure}
\begin{figure}
\plotone{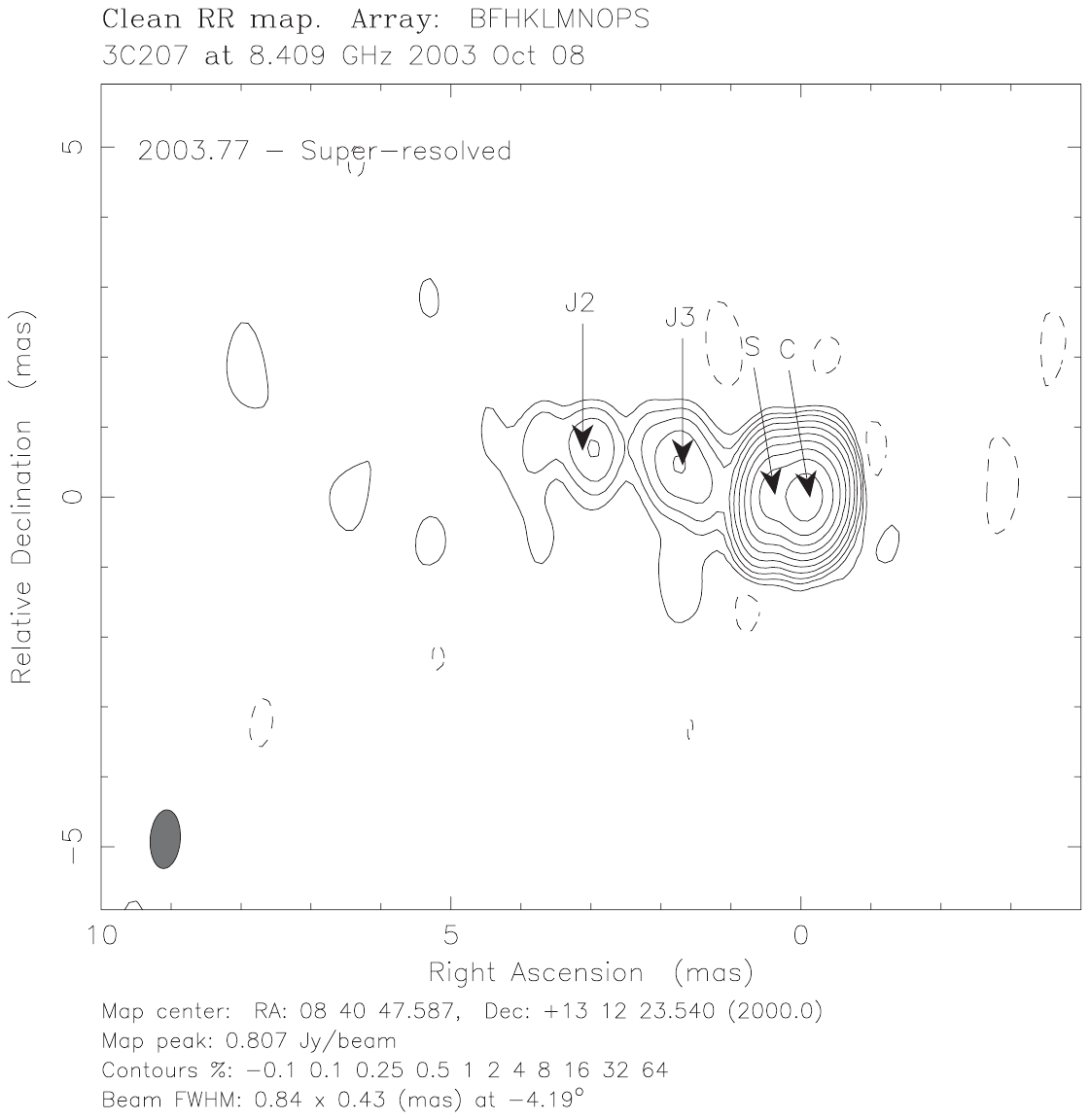}
\plotone{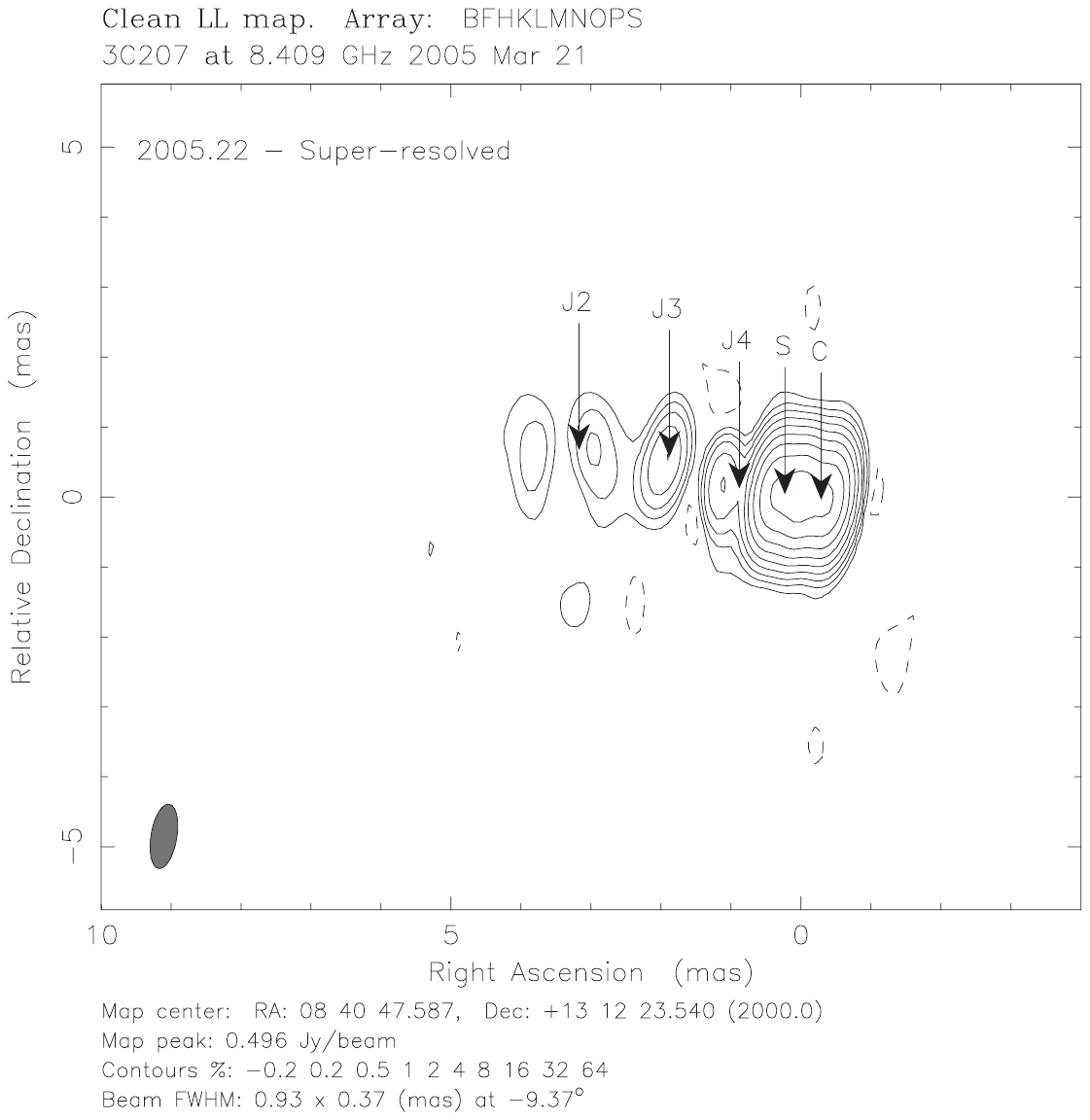}
\end{figure}


\clearpage

%

\startlongtable
\begin{deluxetable}{cccccc}
\tablecaption{Model Fitting Results for 3C207: Single Components\tablenotemark{a} \label{tbl:mod_sng}}
\tablecolumns{6}
\tablenum{2}
\tablewidth{0pt}
\tablehead{
\colhead{Epoch/Taper\tablenotemark{b}}    & \colhead{Component}                    &
\colhead{$S$ (mJy)\tablenotemark{c}}      & \colhead{$d$ (mas)\tablenotemark{d}}   &
\colhead{P.A. (\arcdeg)\tablenotemark{d}} & \colhead{$D$ (mas)\tablenotemark{e}}  
}
\colnumbers
\startdata
1981.76 & C+S\tablenotemark{f} & 360$\pm$10   & \nodata & \nodata & \nodata \\
        & C    &  300$\pm$30  & 0             &  \nodata   & $\leq$0.2      \\
        & S    &   60$\pm$20  & 0.34$\pm$0.06 &  77$\pm$22 & $\leq$0.2      \\
1983.47 & C+S  &  530$\pm$10  &  \nodata      &  \nodata   & \nodata        \\
        & C    &  300$\pm$50  & 0             &  \nodata   & $\leq$0.2      \\
        & S    &  220$\pm$60  & 0.36$\pm$0.06 &  49$\pm$7  & 0.4$\pm$0.1    \\
1984.60 & C+S  &  410$\pm$30  &  \nodata      &  \nodata   & \nodata        \\
        & C    &  270$\pm$50  & 0             &  \nodata   & $\leq$0.3      \\
        & S    &  120$\pm$60  & 0.31$\pm$0.04 &  77$\pm$18 & $\leq$0.5      \\
1988.17 & C+S  &  770$\pm$20  &  \nodata      &  \nodata   & \nodata        \\
        & C    &  360$\pm$40  & 0             &  \nodata   & $\leq$0.2      \\
        & S    &  400$\pm$50  & 0.33$\pm$0.04 &  52$\pm$9  & 0.3$\pm$0.1    \\
1989.27 & C+S  &  855$\pm$15  &  \nodata      &  \nodata   & \nodata        \\
        & C    &  590$\pm$80  & 0             &  \nodata   & $\leq$0.2      \\
        & S    &  270$\pm$80  & 0.35$\pm$0.06 &  61$\pm$6  & $\leq$0.4      \\
1991.15 & C+S  &  520$\pm$90  &  \nodata      &  \nodata   & \nodata        \\
        & C    &  340$\pm$140 & 0             &  \nodata   & $\leq$0.3      \\
        & S    &  180$\pm$90  & 0.3$\pm$0.1   &  71$\pm$6  & $\leq$0.3      \\
        & J2   &   80$\pm$60  & 0.8$\pm$0.2   &  66$\pm$12 & $\leq$0.5      \\ 
1991.16 & C+S  &  485$\pm$49  &  \nodata      &  \nodata   & \nodata        \\
        & C    &  256$\pm$122 & 0             &  \nodata   & $\leq$0.26     \\
        & S    &  257$\pm$112 & 0.32$\pm$0.09 &  78$\pm$12 & $\leq$0.48     \\
        & J2   &   71$\pm$48  & 0.68$\pm$0.20 &  91$\pm$3  & $\leq$0.70     \\
  LT    & J1   &   12$\pm$2   & 3.69$\pm$0.04 &  85$\pm$3  & $\leq$1.2      \\ 
  LT    & Jc+Jd &  11$\pm$2   & 8.07$\pm$0.10 &  97$\pm$2  & 0.86$\pm$0.08  \\
1995.67 & C+S  &  713$\pm$10  &  \nodata      &  \nodata   & \nodata        \\
        & C    &  578$\pm$32  & 0             &  \nodata   & 0.21$\pm$0.08  \\
        & S    &  135$\pm$21  & 0.49$\pm$0.03 &  82$\pm$2  & 0.26$\pm$0.08  \\
        & J2   &   34$\pm$5   & 1.76$\pm$0.13 &  72$\pm$1  & 0.65$\pm$0.04  \\
        & J1   &    9$\pm$1   & 4.53$\pm$0.10 &  91$\pm$1  & 1.18$\pm$0.09  \\
  HT    & Jd   &  3.2$\pm$0.2 & 7.89$\pm$0.05 &95.9$\pm$0.3& 1.21$\pm$0.23  \\
  HT    & Jc   &  2.2$\pm$0.5 &11.11$\pm$0.09 &91.1$\pm$0.6& $\leq$1.3      \\
1998.08 & C+S  &  589$\pm$42  &  \nodata      &  \nodata   & \nodata        \\
        & C    &  361$\pm$58  & 0             &  \nodata   & $\leq$0.24     \\
        & S    &  216$\pm$50  & 0.38$\pm$0.06 &  82$\pm$2  & $<$0.1         \\
        & J3   &   88$\pm$41  & 0.74$\pm$0.11 &  79$\pm$2  & $\leq$0.33     \\ 
        & J2   &   56$\pm$3   & 2.21$\pm$0.02 &  73$\pm$1  & \nodata        \\
        & J1a  &    4$\pm$1   & 5.21$\pm$0.06 &  92$\pm$1  & $\leq$1.2      \\
  HT    & Jd   &  3.3$\pm$0.8 & 8.51$\pm$0.40 &93.3$\pm$0.9& 2.40$\pm$0.42  \\
  HT    & Jc   &  2.4$\pm$0.9 &12.19$\pm$0.56 &91.3$\pm$0.7& $\leq$2.8      \\
  HT    & Jb   &  1.5$\pm$0.1 &20.81$\pm$0.06 &96.0$\pm$0.3& 0.58$\pm$0.35  \\
1999.64 & C+S  &  494$\pm$22  &  \nodata      &  \nodata   & \nodata        \\
        & C    &  377$\pm$22  & 0             &  \nodata   & 0.22$\pm$0.10  \\
        & S    &  109$\pm$35  & 0.52$\pm$0.04 &  92$\pm$4  & $<$0.1         \\
        & J3   &   71$\pm$24  & 0.91$\pm$0.11 &  80$\pm$1  & 0.34$\pm$0.12  \\
        & J2   &   65$\pm$2   & 2.50$\pm$0.02 &  76$\pm$1  & \nodata        \\
        & J1   &    6$\pm$1   & 5.31$\pm$0.06 &  91$\pm$1  & 1.19$\pm$0.12  \\ 
  HT    & Jd   &  2.5$\pm$0.8 & 9.00$\pm$0.12 &  97$\pm$2  & $\leq$2.9      \\
  HT    & Jc   &  2.4$\pm$0.5 &12.88$\pm$0.12 &89.8$\pm$0.6& 1.91$\pm$0.79  \\
  HT    & Jb   &  2.7$\pm$0.4 &20.74$\pm$0.05 &96.2$\pm$0.2& 1.29$\pm$0.54  \\
2001.14 & C+S  &  954$\pm$27  &  \nodata      &  \nodata   & \nodata        \\
        & C    &  754$\pm$162 & 0             &  \nodata   & 0.20$\pm$0.07  \\
        & S    &  201$\pm$155 & 0.39$\pm$0.16 &  95$\pm$8  & $\leq$0.27     \\
        & J3   &   40$\pm$20  & 1.11$\pm$0.19 &  79$\pm$10 & $\leq$0.65     \\
        & J2   &   51$\pm$10  & 2.66$\pm$0.14 &  72$\pm$5  & \nodata        \\
2001.39 & C+S  & 1027$\pm$5   &  \nodata      &  \nodata   & \nodata        \\
        & C    &  926$\pm$39  & 0             &  \nodata   & 0.26$\pm$0.03  \\
        & S    &  100$\pm$37  & 0.40$\pm$0.06 &  92$\pm$6  & $\leq$0.24     \\
        & J3   &   26$\pm$3   & 1.22$\pm$0.13 &  75$\pm$6  & $\leq$0.46     \\
        & J2   &   44$\pm$3   & 2.64$\pm$0.04 &  74$\pm$1  & \nodata        \\
  LT    & J1   &  3.1$\pm$0.5 & 5.34$\pm$0.05 &  89$\pm$2  & $\leq$0.69     \\
2001.63 & C+S  & 1092$\pm$4   &  \nodata      &  \nodata   & \nodata        \\
        & C    &  931$\pm$68  & 0             &  \nodata   & 0.28$\pm$0.02  \\
        & S    &  161$\pm$64  & 0.35$\pm$0.05 &  92$\pm$1  & $<$0.1         \\
        & J3   &   24$\pm$3   & 1.14$\pm$0.05 &  80$\pm$4  & $<$0.38        \\    
        & J2   &   45$\pm$1   & 2.66$\pm$0.02 &  73$\pm$1  & \nodata        \\
  LT    & J1   &  3.6$\pm$0.1 & 5.62$\pm$0.06 &  87$\pm$1  & $\leq$0.75     \\
2001.86 & C+S  & 1054$\pm$9   &  \nodata      &  \nodata   & \nodata        \\
        & C    &  623$\pm$214 & 0             &  \nodata   & 0.24$\pm$0.04  \\
        & S    &  439$\pm$216 & 0.31$\pm$0.03 &  90$\pm$3  & $\leq$0.21     \\
        & J3   &   28$\pm$7   & 1.13$\pm$0.09 &  85$\pm$9  & 0.43$\pm$0.32  \\
        & J2   &   45$\pm$4   & 2.69$\pm$0.06 &  74$\pm$1  & \nodata        \\
  LT    & J1   &  2.8$\pm$0.7 & 6.12$\pm$0.17 &  88$\pm$2  & 2.01$\pm$0.18  \\ 
2002.03 & C+S\tablenotemark{g} & 1086$\pm$9 & 0.19$\pm$0.08 & 84$\pm$3  & 0.44$\pm$0.02 \\
        & J3   &   26$\pm$7   & 1.47$\pm$0.08 &  73$\pm$2  & $\leq$0.49     \\
        & J2   &   37$\pm$3   & 2.97$\pm$0.11 &  78$\pm$1  & \nodata        \\
  LT    & J1   &  2.8$\pm$0.5 & 6.42$\pm$0.10 &  88$\pm$3  & 0.77$\pm$0.18  \\
2002.21 & C+S  & 1085$\pm$1   &  \nodata      &  \nodata   & \nodata        \\
        & C    &  345$\pm$15  & 0             &  \nodata   & 0.18$\pm$0.02  \\
        & S    &  741$\pm$15  & 0.35$\pm$0.01 &  81$\pm$1  & 0.16$\pm$0.02  \\
        & J3   &   34$\pm$3   & 1.20$\pm$0.02 &  79$\pm$1  & 0.72$\pm$0.05  \\
        & J2   &   27$\pm$4   & 2.95$\pm$0.05 &  79$\pm$2  & 0.75$\pm$0.13  \\  
2002.96 & C+S  & 1273$\pm$30  &  \nodata      &  \nodata   & \nodata        \\
        & C    &  574$\pm$25  & 0             &  \nodata   & 0.29$\pm$0.04  \\
        & S    &  704$\pm$11  & 0.48$\pm$0.01 &  87$\pm$1  & 0.24$\pm$0.03  \\
        & J3   &   21$\pm$6   & 1.73$\pm$0.05 &  77$\pm$3  & 0.52$\pm$0.23  \\
        & J2   &   23$\pm$3   & 3.19$\pm$0.14 &  80$\pm$2  & \nodata        \\
  LT    & J1   &  2.5$\pm$0.2 & 6.24$\pm$0.12 &88.9$\pm$0.6& $\leq$1.2      \\
  HT    & Jc+Jd &  3.9$\pm$0.5 &12.61$\pm$0.31 &87.8$\pm$0.5& 3.10$\pm$0.26 \\
2003.12 & C+S  & 1325$\pm$18  &  \nodata      &  \nodata   & \nodata        \\
        & C    &  620$\pm$24  & 0             &  \nodata   & 0.24$\pm$0.05  \\
        & S    &  700$\pm$20  & 0.49$\pm$0.01 &  86$\pm$1  & 0.28$\pm$0.04  \\
        & J3   &   28$\pm$3   & 1.79$\pm$0.10 &  78$\pm$3  & 0.63$\pm$0.09  \\
        & J2   &   21$\pm$3   & 3.32$\pm$0.12 &  79$\pm$1  & \nodata        \\
  LT    & J1   &  2.4$\pm$0.5 & 6.06$\pm$0.07 &  92$\pm$1  & 0.88$\pm$0.08  \\
  HT    & Jc+Jd &  4.0$\pm$0.1 &11.26$\pm$0.24 &  91$\pm$3  & 3.25$\pm$0.43 \\
  HT    & Ja   &  3.0$\pm$0.7 &22.93$\pm$0.17 &99.2$\pm$0.5& 1.57$\pm$1.18  \\
2003.27 & C+S  & 1356$\pm$15  &  \nodata      &  \nodata   & \nodata        \\
        & C    &  710$\pm$23  & 0             &  \nodata   & 0.20$\pm$0.03  \\
        & S    &  644$\pm$10  & 0.50$\pm$0.01 &  86$\pm$1  & 0.27$\pm$0.02  \\
        & J3   &   29$\pm$4   & 1.83$\pm$0.06 &  76$\pm$1  & 0.62$\pm$0.14  \\
        & J2   &   21$\pm$2   & 3.36$\pm$0.07 &  79$\pm$1  & \nodata        \\
  LT    & J1   &  2.4$\pm$0.1 & 6.51$\pm$0.14 &91.6$\pm$0.3& 1.10$\pm$0.21  \\
  HT    & Jd   & 1.13$\pm$0.03& 9.38$\pm$0.06 &  95$\pm$4  & $<$0.1         \\
  HT    & Jc   &  2.6$\pm$0.2 &13.45$\pm$0.08 &91.7$\pm$0.8& 1.45$\pm$0.39  \\
2003.45 & C+S  & 1445$\pm$19  &  \nodata      &  \nodata   & \nodata        \\
        & C    &  851$\pm$15  & 0             &  \nodata   & 0.20$\pm$0.01  \\
        & S    &  593$\pm$3   & 0.50$\pm$0.01 &  86$\pm$1  & 0.27$\pm$0.02  \\
        & J3   &   28$\pm$1   & 1.88$\pm$0.03 &  75$\pm$1  & 0.50$\pm$0.04  \\
        & J2   &   20$\pm$1   & 3.35$\pm$0.04 &  80$\pm$1  & \nodata        \\
2003.62 & C+S  & 1437$\pm$13  &  \nodata      &  \nodata   & \nodata        \\
        & C    &  891$\pm$6   & 0             &  \nodata   & 0.23$\pm$0.01  \\
        & S    &  550$\pm$13  & 0.50$\pm$0.01 &  86$\pm$1  & 0.30$\pm$0.02  \\
        & J3   &   28$\pm$1   & 1.92$\pm$0.06 &  75$\pm$1  & 0.41$\pm$0.10  \\
        & J2   &   17$\pm$2   & 3.30$\pm$0.12 &  79$\pm$1  & \nodata        \\
  LT    & J1   &  1.8$\pm$0.4 & 6.46$\pm$0.05 &93.1$\pm$0.8& $\leq$1.2      \\
2003.77 & C+S  & 1409$\pm$17  &  \nodata      &  \nodata   & \nodata        \\
        & C    &  901$\pm$12  & 0             &  \nodata   & 0.26$\pm$0.01  \\
        & S    &  511$\pm$8   & 0.50$\pm$0.01 &  86$\pm$1  & 0.31$\pm$0.01  \\
        & J3   &   32$\pm$1   & 1.87$\pm$0.02 &  75$\pm$1  & 0.51$\pm$0.04  \\
        & J2   &   19$\pm$1   & 3.39$\pm$0.03 &  78$\pm$1  & \nodata        \\
  LT    & J1   &  2.0$\pm$0.4 & 6.25$\pm$0.12 &  87$\pm$3  & $\leq$1.2      \\
2005.02 & C+S  & 1240$\pm$62  &  \nodata      &  \nodata   & \nodata        \\
        & C    &  551$\pm$45  & 0             &  \nodata   & 0.32$\pm$0.05  \\
        & S    &  694$\pm$22  & 0.50$\pm$0.01 &  83$\pm$2  & 0.37$\pm$0.02  \\
2005.22 & C+S  & 1046$\pm$15  &  \nodata      &  \nodata   & \nodata        \\
        & C    &  532$\pm$6   & 0             &  \nodata   & 0.23$\pm$0.01  \\
        & S    &  516$\pm$11  & 0.49$\pm$0.01 &  81$\pm$1  & $<$0.1         \\
        & J4a? &  167$\pm$11  & 0.85$\pm$0.02 &  83$\pm$1  & 0.36$\pm$0.03  \\
        & J3   &   32$\pm$3   & 2.31$\pm$0.05 &  75$\pm$2  & 0.60$\pm$0.11  \\
        & J2   &   11$\pm$2   & 3.81$\pm$0.08 &  79$\pm$1  & \nodata        \\
2005.35 & C+S  & 1031$\pm$23  &  \nodata      &  \nodata   & \nodata        \\
        & C    &  513$\pm$14  & 0             &  \nodata   & 0.17$\pm$0.05  \\
        & S    &  519$\pm$10  & 0.53$\pm$0.01 &  81$\pm$1  & 0.17$\pm$0.03  \\
        & J4a? &   95$\pm$22  & 0.97$\pm$0.04 &  84$\pm$1  & 0.36$\pm$0.11  \\
        & J3   &   25$\pm$2   & 2.36$\pm$0.02 &  81$\pm$2  & 0.55$\pm$0.05  \\
        & J2   &    8$\pm$1   & 3.95$\pm$0.05 &  82$\pm$1  & \nodata        \\
2005.55 & C+S  & 1066$\pm$18  &  \nodata      &  \nodata   & \nodata        \\
        & C    &  587$\pm$10  & 0             &  \nodata   & 0.24$\pm$0.03  \\
        & S    &  481$\pm$12  & 0.55$\pm$0.01 &  81$\pm$1  & 0.17$\pm$0.05  \\
        & J4a? &  128$\pm$14  & 0.92$\pm$0.02 &  83$\pm$1  & 0.44$\pm$0.04  \\
        & J3   &   28$\pm$1   & 2.38$\pm$0.02 &  77$\pm$1  & 0.64$\pm$0.05  \\
        & J2   &   10$\pm$1   & 3.99$\pm$0.04 &  80$\pm$1  & \nodata        \\
  LT    & J1   &  2.4$\pm$0.4 & 6.87$\pm$0.25 &89.3$\pm$0.9& 1.24$\pm$0.11  \\
2005.70 & C+S  & 1105$\pm$27  &  \nodata      &  \nodata   & \nodata        \\
        & C    &  638$\pm$14  & 0             &  \nodata   & 0.30$\pm$0.03  \\
        & S    &  468$\pm$16  & 0.56$\pm$0.01 &  79$\pm$1  & 0.21$\pm$0.03  \\
        & J4a? &  116$\pm$24  & 0.92$\pm$0.04 &  86$\pm$1  & 0.45$\pm$0.07  \\
        & J3   &   29$\pm$1   & 2.44$\pm$0.02 &  73$\pm$1  & 0.55$\pm$0.06  \\
        & J2   &    9$\pm$1   & 3.92$\pm$0.04 &  80$\pm$2  & \nodata        \\
2005.86 & C+S  & 1017$\pm$15  &  \nodata      &  \nodata   & \nodata        \\
        & C    &  609$\pm$15  & 0             &  \nodata   & 0.24$\pm$0.03  \\
        & S    &  409$\pm$2   & 0.53$\pm$0.01 &  78$\pm$1  & 0.12$\pm$0.03  \\
        & J4a? &  195$\pm$12  & 0.85$\pm$0.01 &  83$\pm$1  & 0.39$\pm$0.04  \\
        & J3   &   33$\pm$2   & 2.42$\pm$0.03 &  75$\pm$2  & 0.78$\pm$0.08  \\
        & J2   &    9$\pm$2   & 4.07$\pm$0.10 &  79$\pm$2  & \nodata        \\
  LT    & J1   &  2.0$\pm$0.5 & 7.34$\pm$0.18 &87.6$\pm$0.3& $\leq$1.4      \\
2008.81 & C+S  &  395$\pm$8   &  \nodata      &  \nodata   & \nodata        \\
        & C    &  308$\pm$17  & 0             &  \nodata   & 0.32$\pm$0.06  \\
        & S    &   82$\pm$14  & 0.61$\pm$0.03 &  92$\pm$1  & $\leq$0.29     \\
        & J4b? &  133$\pm$10  & 1.18$\pm$0.02 &  81$\pm$1  & 0.49$\pm$0.09  \\
        & J4a? &   13$\pm$8   & 1.98$\pm$0.26 &  88$\pm$3  & $\leq$0.93     \\
        & J3   &   28$\pm$1   & 3.22$\pm$0.01 &  80$\pm$1  & 0.66$\pm$0.04  \\
        & J2   &    4$\pm$1   & 5.03$\pm$0.04 &  82$\pm$2  & $\leq$1.0      \\
2009.92 & C+S  &  466$\pm$12  &  \nodata      &  \nodata   & \nodata        \\
        & C    &  407$\pm$24  & 0             &  \nodata   & 0.20$\pm$0.03  \\
        & S    &   61$\pm$14  & 0.51$\pm$0.15 & 101$\pm$10 & $\leq$0.41     \\
        & J4c? &   58$\pm$17  & 1.12$\pm$0.21 &  82$\pm$2  & 0.30$\pm$0.11  \\
        & J4b? &   45$\pm$31  & 1.70$\pm$0.22 &  83$\pm$2  & 0.32$\pm$0.22  \\
        & J4a? &    5$\pm$3   & 2.73$\pm$0.13 &  84$\pm$3  & $\leq$0.74     \\
        & J3   &   25$\pm$3   & 3.48$\pm$0.06 &  81$\pm$1  & 0.67$\pm$0.13  \\
        & J2   &  3.7$\pm$0.7 & 5.66$\pm$0.15 &  78$\pm$3  & $\leq$1.3      \\
  HT    & J1   &  1.6$\pm$0.5 & 7.72$\pm$0.42 &  93$\pm$3  & $\leq$2.9      \\
  HT    & Jd   &  1.2$\pm$0.4 &11.49$\pm$0.39 &  96$\pm$4  & $\leq$2.6      \\
  HT    & Ja   &  1.5$\pm$0.5 &22.94$\pm$0.25 &100.8$\pm$0.6&$\leq$3.1      \\
2010.81 & C+S  &  716$\pm$12  &  \nodata      &  \nodata   & \nodata        \\
        & C    &  560$\pm$44  & 0             &  \nodata   & 0.18$\pm$0.05  \\
        & S    &  153$\pm$42  & 0.33$\pm$0.06 & 105$\pm$7  & $\leq$0.19     \\
        & J4c? &   47$\pm$27  & 1.04$\pm$0.36 &  86$\pm$4  & $\leq$0.55     \\
        & J4b? &   47$\pm$34  & 1.78$\pm$0.22 &  85$\pm$5  & $\leq$0.73     \\
        & J4a? &   10$\pm$7   & 2.60$\pm$0.34 &  89$\pm$6  & $\leq$0.87     \\
        & J3   &   26$\pm$7   & 3.65$\pm$0.14 &  82$\pm$1  & 0.70$\pm$0.25  \\
        & J2   &  2.9$\pm$1.2 & 5.87$\pm$0.13 &  82$\pm$2  & $\leq$0.55     \\
  HT    & Jd   &  1.4$\pm$0.3 &12.23$\pm$0.28 &  98$\pm$2  & $\leq$1.7      \\
\enddata
\tablenotetext{a}{The values and uncertainties of the component parameters were determined 
using the range of acceptable values for several modelfits to DIFMAP imaging runs with 
different approaches for starting models, initial sets of baselines, {\it a priori} 
amplitude corrections, windowing, phase and amplitude self-calibration, cleaning, etc., 
during the process of making the best final images. For cases in which two subcomponents 
(e.g., J2a and J2b) could be modeled, the single component entries here (e.g., J2) are 
determined from the brightness centroid and the sum of the flux densities of the two 
subcomponents and, in some cases, from additional results of single component fits.}
\tablenotetext{b}{Epoch of observation and, if used, Gaussian taper. 
The light taper (LT) and the heavy taper (HT) downweight the ({\it u,v}) data at a 
({\it u,v}) radius of 100 Mega-wavelengths by factors of two and ten, respectively. 
The only exception is at epoch 1991.16, for which the radius for the light taper is 
125 Mega-wavelengths.}
\tablenotetext{c}
{$S$ is the flux density.}
\tablenotetext{d}
{Here $d$ and P.A. give the separation and position angle of the jet 
component (J1, J2, etc., from the outermost to innermost knot), relative 
to the core component (C).} 
\tablenotetext{e}
{$D$ is the FWHM diameter of a circular gaussian component, unless an exception 
is noted. Upper limits are given in cases where point components 
are allowed.}
\tablenotetext{f}
{The ``C+S'' flux density and uncertainty were found from the range of values of   
this sum for the modelfits used to find the values and uncertainties of the 
individual flux densities; thus each ``C+S'' sum is not just the sum of the C and S 
flux densities.}
\tablenotetext{g}
{At epoch 2002.03, components C and S are blended as a single elliptical gaussian  
component. Here, $d$ gives the separation between the elliptical component's peak and 
an estimated core position determined from the average offset of 0.19$\pm$0.08 mas 
between the brightness centroid and component C at epochs 2001.86 and 2002.21. 
The values of P.A. and $D$ give the major axis position angle and FWHM 
of the elliptical component, which has axial ratio $\leq$0.30. For Figure 1, the 
flux density has been set to 543 mJy for both C and S, and the uncertainties have been 
set to the largest values -- 214 mJy and 216 mJy, respectively -- from 2001.86.}
\end{deluxetable}

\startlongtable
\begin{deluxetable}{cccccc}
\tablecaption{Model Fitting Results for 3C207: Double Subcomponents\tablenotemark{a} \label{tbl:mod_dbl}}
\tablecolumns{6}
\tablenum{3}
\tablewidth{0pt}
\tablehead{
\colhead{Epoch\tablenotemark{b}}          & \colhead{Subcomponent}                 &
\colhead{$S$ (mJy)\tablenotemark{c}}      & \colhead{$d$ (mas)\tablenotemark{d}}   &
\colhead{P.A. (\arcdeg)\tablenotemark{d}} & \colhead{$D$ (mas)\tablenotemark{e}}  
}
\colnumbers
\startdata
1995.67 & J2b  &   17$\pm$7   & 1.35$\pm$0.21 &  71$\pm$3  & $\leq$0.90     \\
        & J2a  &   19$\pm$6   & 2.01$\pm$0.08 &  73$\pm$2  & $\leq$0.55     \\
        & J1b  &    4$\pm$1   & 4.06$\pm$0.07 &  90$\pm$1  & $\leq$1.00     \\
        & J1a  &    5$\pm$2   & 4.89$\pm$0.05 &  91$\pm$1  & $<$0.1         \\
1998.08 & J2b  &   24$\pm$5   & 1.84$\pm$0.06 &  70$\pm$1  & 0.24$\pm$0.12  \\
        & J2a  &   33$\pm$6   & 2.51$\pm$0.04 &  74$\pm$1  & 0.36$\pm$0.16  \\
1999.64 & J2b  &   37$\pm$6   & 2.19$\pm$0.04 &  74$\pm$1  & 0.50$\pm$0.15  \\
        & J2a  &   28$\pm$5   & 2.91$\pm$0.04 &  77$\pm$1  & 0.36$\pm$0.09  \\
        & J1b  &  2.5$\pm$0.2 & 4.89$\pm$0.06 &  92$\pm$2  & $\leq$1.0      \\
        & J1a  &  2.8$\pm$0.1 & 5.69$\pm$0.05 &  91$\pm$1  & $\leq$0.78     \\ 
2001.14 & J2b  &   30$\pm$11  & 2.41$\pm$0.29 &  68$\pm$8  & $\leq$0.87     \\
        & J2a  &   19$\pm$13  & 3.21$\pm$0.33 &  78$\pm$4  & $\leq$0.66     \\
2001.39 & J2b  &   18$\pm$4   & 2.16$\pm$0.13 &  70$\pm$2  & $\leq$0.29     \\
        & J2a  &   28$\pm$5   & 2.95$\pm$0.10 &  77$\pm$1  & 0.53$\pm$0.16  \\
2001.63 & J2b  &   15$\pm$2   & 2.13$\pm$0.07 &  67$\pm$2  & $<$0.1         \\
        & J2a  &   30$\pm$3   & 2.92$\pm$0.05 &  76$\pm$1  & 0.64$\pm$0.05  \\
2001.86 & J2b  &   16$\pm$8   & 2.13$\pm$0.20 &  69$\pm$4  & 0.62$\pm$0.19  \\
        & J2a  &   28$\pm$7   & 3.04$\pm$0.13 &  78$\pm$1  & 0.75$\pm$0.14  \\
2002.03 & J2b  &   20$\pm$15  & 2.58$\pm$0.28 &  77$\pm$4  & $\leq$1.0      \\
        & J2a  &   18$\pm$14  & 3.29$\pm$0.20 &  78$\pm$1  & $\leq$0.91     \\
2002.21 & J2b  &   20$\pm$2   & 2.82$\pm$0.02 &  77$\pm$1  & 0.61$\pm$0.09  \\
        & J2a  &    9$\pm$1   & 3.28$\pm$0.03 &  79$\pm$1  & 0.77$\pm$0.01  \\  
2002.96 & J2b  &    9$\pm$6   & 2.66$\pm$0.35 &  73$\pm$6  & $\leq$0.48     \\
        & J2a  &   14$\pm$8   & 3.45$\pm$0.28 &  80$\pm$1  & $\leq$1.0      \\
2003.12 & J2b  &   13$\pm$8   & 3.00$\pm$0.13 &  81$\pm$3  & $\leq$0.83     \\
        & J2a  &   10$\pm$7   & 3.76$\pm$0.27 &  79$\pm$4  & 0.53$\pm$0.37  \\
2003.27 & J2b  &    9$\pm$3   & 2.99$\pm$0.03 &  79$\pm$1  & $<$0.1         \\
        & J2a  &   11$\pm$4   & 3.63$\pm$0.12 &  79$\pm$2  & $\leq$0.95     \\
2003.45 & J2b  &   11$\pm$1   & 3.03$\pm$0.02 &  79$\pm$1  & $<$0.1         \\
        & J2a  &    8$\pm$1   & 3.75$\pm$0.08 &  79$\pm$1  & 0.64$\pm$0.23  \\
2003.62 & J2b  &    9$\pm$2   & 3.00$\pm$0.05 &  79$\pm$1  & $<$0.1         \\
        & J2a  &    8$\pm$2   & 3.56$\pm$0.20 &  80$\pm$2  & $\leq$0.79     \\
2003.77 & J2b  &   12$\pm$3   & 3.09$\pm$0.07 &  80$\pm$1  & $\leq$0.66     \\
        & J2a  &    7$\pm$2   & 3.91$\pm$0.10 &  76$\pm$4  & $\leq$0.75     \\
2005.22 & J2b  &    8$\pm$1   & 3.52$\pm$0.04 &  77$\pm$1  & $\leq$0.61     \\
        & J2a  &    4$\pm$1   & 4.33$\pm$0.03 &  81$\pm$1  & $<$0.1         \\
2005.35 & J2b  &    5$\pm$1   & 3.64$\pm$0.04 &  81$\pm$1  & $\leq$0.33     \\
        & J2a  &    4$\pm$1   & 4.30$\pm$0.09 &  83$\pm$1  & $\leq$0.46     \\
2005.55 & J2b  &    5$\pm$1   & 3.66$\pm$0.02 &  77$\pm$2  & $<$0.1         \\
        & J2a  &    5$\pm$1   & 4.38$\pm$0.04 &  82$\pm$1  & $\leq$0.74     \\
2005.70 & J2b  &    5$\pm$1   & 3.60$\pm$0.06 &  75$\pm$2  & $<$0.1         \\
        & J2a  &    5$\pm$2   & 4.26$\pm$0.16 &  83$\pm$4  & $\leq$1.1      \\
2005.86 & J2b  &    4$\pm$1   & 3.65$\pm$0.03 &  76$\pm$1  & $<$0.1         \\
        & J2a  &    6$\pm$2   & 4.32$\pm$0.12 &  81$\pm$2  & $\leq$1.1      \\
2009.92 & J3b  &   15$\pm$3   & 3.39$\pm$0.02 &  80$\pm$1  & 0.35$\pm$0.12  \\
        & J3a  &    8$\pm$2   & 3.78$\pm$0.05 &  83$\pm$1  & 0.36$\pm$0.13  \\
        & J2b  &  1.4$\pm$0.2 & 5.25$\pm$0.15 &  73$\pm$2  & $\leq$0.61     \\
        & J2a  &  2.3$\pm$0.6 & 5.89$\pm$0.14 &  80$\pm$2  & $\leq$1.2      \\
2010.81 & J3b  &   13$\pm$4   & 3.41$\pm$0.25 &  82$\pm$1  & $\leq$0.61     \\
        & J3a  &   14$\pm$5   & 3.86$\pm$0.15 &  82$\pm$1  & $\leq$0.63     \\
        & J2b  &  1.9$\pm$0.1 & 5.36$\pm$0.01 &  83$\pm$1  & $\leq$0.12     \\
        & J2a  &  2.1$\pm$0.1 & 6.08$\pm$0.01 &  83$\pm$1  & $<$0.1         \\
\enddata
\tablenotetext{a}{The values and uncertainties of the subcomponent parameters were 
determined using the range of acceptable values for several modelfits to DIFMAP 
imaging runs with different approaches for starting models, initial sets of baselines, 
{\it a priori} amplitude corrections, windowing, phase and amplitude self-calibration, 
cleaning, etc., during the process of making the best final images.}
\tablenotetext{b}{Epoch of observation.}
\tablenotetext{c}{$S$ is the flux density.}
\tablenotetext{d}
{Here $d$ and P.A. give the separation and position angle of the jet subcomponent 
(J2b, J2a, etc.), from the outermost to innermost knot), relative to the core 
component (C).} 
\tablenotetext{e}
{$D$ is the FWHM diameter of a circular gaussian subcomponent. Upper limits are given 
in cases where point subcomponents are allowed.}
\end{deluxetable}

\bibliography{hough}{}
\bibliographystyle{aasjournal}

 



\end{document}